\documentclass[12pt]{article}
\usepackage[utf8]{inputenc}
\usepackage{epsf,amsmath,amssymb,graphicx,dcolumn}
\usepackage{caption}
\usepackage{subcaption}
\usepackage{scalefnt,ulem,pstricks}
\usepackage{booktabs,multirow,tabularx}
\usepackage{cite,hyperref}
\usepackage{cleveref}
\usepackage{color}
\usepackage{rotating}
\usepackage{microtype}
\usepackage{tablefootnote}

\definecolor{lightblue}{rgb}{.7,.8,1}
\def\refse#1{\mbox{Section~\ref{#1}}}
\renewcommand{\thefootnote}{\fnsymbol{footnote}}

\newcommand\f[2]{\frac{#1}{#2}} 
\newcommand{\abbrev}{\scalefont{1.}}

\newcommand{\sphid}[1]{}

\newcommand{\OpenLoops}{{\sc OpenLoops}}
\newcommand{\Matrix}{{\sc Matrix}}
\newcommand{\Munich}{{\sc Munich}}
\newcommand{\Collier}{{\sc Collier}}

\newcommand{\GINAC}{{\sc GiNaC}}

\def\refeq#1{\mbox{Eq.~\eqref{#1}}}

\def\reffi#1{\mbox{Figure~\ref{#1}}}
\def\reffitwo#1#2{\mbox{Figures~\ref{#1} and \ref{#2}}}
\def\reffis#1#2{\mbox{Figures~\ref{#1}--\ref{#2}}}
\def\refta#1{\mbox{Table~\ref{#1}}}

\def\refse#1{\mbox{Section~\ref{#1}}}

\def\citere#1{\mbox{Ref.~\cite{#1}}}
\def\citeres#1{\mbox{Refs.~\cite{#1}}}

\newcommand{\rcut}{\ensuremath{r_{\mathrm{cut}}}}

\newcommand{\ww}{\ensuremath{W^+W^-}}

\newcommand{\pt}{\ensuremath{p_T}}

\newcommand{\qt}{\ensuremath{q_T}}

\newcommand{\lo}{\text{\abbrev LO}}
\newcommand{\nlo}{\text{\abbrev NLO}}
\newcommand{\nloprime}{\text{\abbrev NLO$^\prime$}}
\newcommand{\nloplusgg}{\text{\abbrev NLO$^\prime$+$gg$}}
\newcommand{\nnlo}{\text{\abbrev NNLO}}

\newcommand{\fs}[1]{#1{\abbrev FS}}

\newcommand{\order}[1]{{\cal O}(#1)}

\newcommand{\sct}[1]{Section~\ref{#1}}

\newcommand{\muF}{\mu_{F}}
\newcommand{\muR}{\mu_{R}}

\newcommand{\bld}[1]{\boldmath{$#1$}}

\newcommand{\D}{\mathrm{d}}
\newcommand\as{\alpha_{\mathrm{S}}}

\newcommand{\Gt}{\Gamma_t}

\newcommand{\elle}{\ensuremath{l}}
\newcommand{\muenn}{\ensuremath{\mu^+e^-\nu_\mu {\bar \nu}_e}}
\newcommand{\emunn}{\ensuremath{e^+\mu^-\nu_e {\bar \nu}_\mu}}
\newcommand{\sfww}{\elle^+\elle^{\prime -}\nu_\elle{\bar \nu}_{\elle^\prime}}

\newcommand{\mmuenn}{\ensuremath{m_{\muenn}}}
\newcommand{\ptmuenn}{\ensuremath{p_{T,\muenn}}}

\def\efficiency{\epsilon}
\def\sigmafid{\sigma_{\rm fiducial}}
\def\sigmainc{\sigma_{\rm inclusive}}

\Crefname{figure}{Fig.}{Figs.}
\usepackage{datetime}

\newdateformat{monthyeardate}{%
  \monthname[\THEMONTH] \THEYEAR}

\begin{document}
\begin{titlepage}
\renewcommand{\thefootnote}{\fnsymbol{footnote}}
\begin{flushright}
ZU-TH 16/16\\
MITP/16-038\\
NSF-KITP-16-047\\
DESY 16-075
\end{flushright}
\vspace*{1cm}

\begin{center}
{\Large \bf \bld{\ww{}} production at the LHC:\\[0.4cm]
fiducial cross sections and distributions in NNLO QCD}
\end{center}

\par \vspace{2mm}
\begin{center}
{\bf Massimiliano Grazzini$^{(a)}$},
{\bf Stefan Kallweit$^{(b,c)}$},
{\bf Stefano Pozzorini$^{(a,c)}$},\\[0.2cm]
{\bf Dirk Rathlev$^{(d)}$} and {\bf Marius Wiesemann$^{(a)}$}

\vspace{5mm}

$^{(a)}$ Physik-Institut, Universit\"at Z\"urich, 
CH-8057 Z\"urich, Switzerland 

$^{(b)}$ PRISMA Cluster of Excellence, Institute of Physics,\\[0.1cm]
Johannes Gutenberg University, D-55099 Mainz, Germany

$^{(c)}$Kavli Institute for Theoretical Physics,\\[0.1cm]
University of California, Santa Barbara, CA 93106, USA

$^{(d)}$Theory Group, Deutsches Elektronen-Synchrotron (DESY), D-22607 Hamburg, Germany

\end{center}

\par \vspace{2mm}
\begin{center} {\large \bf Abstract} \end{center}
\begin{quote}
\pretolerance 10000

We consider QCD radiative corrections to 
\ww{} production at the LHC and present 
the first fully differential predictions 
for this process at next-to-next-to-leading 
order (NNLO) in perturbation theory.
Our computation consistently includes the leptonic decays of the $W$ bosons, taking into account spin correlations, off-shell effects and non-resonant contributions. 
Detailed predictions are presented for the different-flavour channel 
$pp\to\muenn+X$ at $\sqrt{s}=8$ and $13$\,TeV.  In particular, we discuss
fiducial cross sections and distributions in the presence of standard selection
cuts used in  
experimental \ww{} and $H\to \ww$ analyses at the LHC.
The inclusive \ww{} cross section receives large NNLO corrections,
and, due to the presence of a jet veto, typical fiducial cuts have a
sizeable influence on the behaviour of 
the perturbative expansion.
The availability of differential NNLO predictions, both for inclusive
and fiducial observables, will play an important role in the rich physics
programme that is based on precision studies of \ww{} signatures at the LHC.

\end{quote}

\vspace*{\fill}
\begin{flushleft}
\monthyeardate\today

\end{flushleft}
\end{titlepage}

\newcommand{\mll}{\ensuremath{m_{ll}}}
\newcommand{\mww}{\ensuremath{m_{WW}}}
\newcommand{\mtatlas}{\ensuremath{m_T^{\rm ATLAS}}}
\newcommand{\ptww}{\ensuremath{p_{T,WW}}}
\newcommand{\ptw}{\ensuremath{p_{T,W}}}
\newcommand{\ptwone}{\ensuremath{p_{T,W_1}}}
\newcommand{\ptwtwo}{\ensuremath{p_{T,W_2}}}
\newcommand{\ptwp}{\ensuremath{p_{T,W^-}}}
\newcommand{\ptwm}{\ensuremath{p_{T,W^+}}}
\newcommand{\ptlone}{\ensuremath{p_{T,l_1}}}
\newcommand{\ptltwo}{\ensuremath{p_{T,l_2}}}
\newcommand{\Etlone}{\ensuremath{E_{T,l_1}}}
\newcommand{\Etltwo}{\ensuremath{E_{T,l_2}}}
\newcommand{\ptli}{\ensuremath{p_{T,l_i}}}
\newcommand{\ptll}{\ensuremath{p_{T,ll}}}
\newcommand{\ptjet}{\ensuremath{p_{T,j}}}
\newcommand{\ptmiss}{\ensuremath{p_{T}^{\text{miss}}}}
\newcommand{\ptmissrel}{\ensuremath{p_{T}^{\text{miss,rel}}}}
\newcommand{\ymu}{\ensuremath{y_\mu}}
\newcommand{\ye}{\ensuremath{y_e}}
\newcommand{\yj}{\ensuremath{y_j}}
\newcommand{\dphillnunu}{\ensuremath{\Delta\phi_{ll,\nu\nu}}}
\newcommand{\dphill}{\ensuremath{\Delta\phi_{ll}}}
\newcommand{\dRll}{\ensuremath{\Delta R_{ll}}}
\newcommand{\dRej}{\ensuremath{\Delta R_{e,j}}}

\section{Introduction}

The production of $W$-boson pairs is one of the most important electroweak~(EW)
processes at hadron colliders.  
Experimental studies of \ww{} production 
play a central role in precision tests 
of the gauge symmetry structure
of EW interactions and of the mechanism of EW symmetry
breaking.
The \ww{} cross section has been measured at the
Tevatron~\cite{Aaltonen:2009aa,Abazov:2011cb} and at the LHC, both at
7\,TeV~\cite{ATLAS:2012mec,Chatrchyan:2013yaa} and
8\,TeV~\cite{ATLAS-CONF-2014-033,Aad:2016wpd,Chatrchyan:2013oev,Khachatryan:2015sga}. 
The dynamics of $W$-pair production is of great interest,
not only in
the context of precision tests of the Standard Model, but also in searches
of physics beyond the Standard Model (BSM).
Any small anomaly in the 
production rate or in the shape
of distributions could be a signal of new physics. 
In particular, due to the high sensitivity to 
modifications of the Standard Model trilinear gauge couplings, 
\ww{} measurements are a powerful tool for 
indirect BSM searches via anomalous
couplings~\cite{ATLAS:2012mec,Chatrchyan:2013yaa,Wang:2014uea,Khachatryan:2015sga,Aad:2016wpd}. 
Thanks to the increasing reach in 
transverse momentum, Run\,2 of
the LHC will considerably tighten the present bounds on anomalous couplings.
Final states with $W$-boson pairs are widely studied also
in the context of direct BSM searches \cite{Morrissey:2009tf}.

In Higgs-boson studies~\cite{Aad:2012tfa,ATLAS:2014aga,Aad:2016lvc,Chatrchyan:2012xdj,Chatrchyan:2013lba,Chatrchyan:2013iaa},
\ww{} production plays an important role 
as irreducible
background
in the $H\to \ww$ channel. Such measurements are mostly based on
final states with two leptons and two neutrinos, which provide a clean
experimental signature, but do not allow for a full reconstruction of the
$H\to \ww$ resonance.  As a consequence, it is not possible to
extract the irreducible \ww{} background
from data with a simple side-band approach. Thus,
the availability of precise theory predictions for the \ww{} background is
essential for the sensitivity 
to $H\to \ww$ and to any 
BSM particle that decays into $W$-boson pairs.In the context of Higgs studies, the
off-shell treatment of $W$-boson decays is of great relevance,
both for the description of the $H\to \ww$ signal region below the \ww{} threshold, 
and for indirect determinations of the Higgs-boson width through
signal--background interference effects at high invariant
masses~\cite{Kauer:2012hd,Caola:2013yja,Campbell:2013wga}.

The accurate description of the jet activity 
is another critical aspect of Higgs measurements, and of \ww{}
measurements in general.  Such analyses
typically rely on a rather strict jet
veto, which suppresses the severe signal contamination due to the 
$t\bar t$ background, 
but induces potentially large logarithms that challenge the
reliability of fixed-order predictions in perturbation
theory.
All these requirements, combined with the ever increasing accuracy of
experimental measurements, call for continuous improvements in the
theoretical description of \ww{} production.

Next-to-leading order (NLO) QCD predictions for \ww{} production at
hadron colliders have been available for a long time, both for the case of
stable $W$-bosons~\cite{Ohnemus:1991kk,Frixione:1993yp} and with
spin-correlated decays of vector bosons into
leptons~\cite{Campbell:1999ah,Dixon:1999di,Dixon:1998py,Campbell:2011bn}. 
Recently, also the NLO EW corrections have been
computed~\cite{Bierweiler:2012kw,Baglio:2013toa,Billoni:2013aba}. 
Their impact on inclusive cross sections hardly exceeds
a few percent, but can be strongly enhanced up to 
several tens of percent at transverse momenta of about 1\,TeV.

Given the sizeable impact of ${\cal O}(\as)$ corrections, the
calculation of higher-order QCD effects is indispensable in order to reach
high precision.  The simplest ingredient of $pp\to\ww+X$ at
${\cal O}(\as^2)$ is given by the 
loop-induced gluon-fusion contribution.
Due to the strong
enhancement of the gluon luminosity, the $gg$ channel 
was generally regarded as the dominant source of NNLO QCD corrections 
to $pp\to\ww+X$ in the literature.
Predictions for $gg\to \ww$ at LO have been widely
studied~\cite{Dicus:1987dj,Glover:1988fe,Binoth:2005ua,Binoth:2006mf,Campbell:2011bn},
and squared quark-loop contributions at LO are known also for 
$gg\to \ww{} g$~\cite{Melia:2011tj,Melia:2012zg}.
Two-loop helicity amplitudes for $gg\to VV'$ became available
in~\citeres{Caola:2015ila,vonManteuffel:2015msa}, and have been used to compute the
NLO QCD corrections to $gg\to W^+W^-$~\cite{Caola:2015rqy},
including all partonic processes with external gluons,
while the ones with external quarks are still unknown to date.
Calculations at NLO QCD for \ww{} production in association with
one~\cite{Dittmaier:2007th,Campbell:2007ev,Dittmaier:2009un,Campbell:2015hya}
and two~\cite{Melia:2011dw,Greiner:2012im} jets are also important ingredients 
of inclusive \ww{} production at NNLO QCD and beyond.  
The merging of NLO QCD predictions for $pp\to\ww+0,1$\,jets\footnote{See also~\cite{Campanario:2013wta} for a
combination of fixed-order NLO predictions for \ww{}+0,1\,jet production.}
has been presented in~\citere{Cascioli:2013gfa}. 
This merged calculation also consistently includes squared
quark-loop contributions to $pp\to\ww+0,1$\,jets 
in all gluon- and quark-induced channels.

First NNLO QCD predictions for the inclusive \ww{} cross section became
available in~\citere{Gehrmann:2014fva}.  This calculation was based on
two-loop scattering amplitudes for on-shell \ww{} production, while two-loop
helicity amplitudes are now available for all vector-boson pair production
processes, including off-shell leptonic
decays~\cite{Caola:2014iua,Gehrmann:2015ora}.
In the energy range from 7 to 14\,TeV, NNLO corrections shift the NLO
predictions for the total cross section by about
9\% to 12\%~\cite{Gehrmann:2014fva}, which is 
around three times as large as
the $gg\to \ww$ contribution alone.  Thus, contrary to what was widely
expected, gluon--gluon fusion is not the dominant source of radiative
corrections beyond NLO.  Moreover, the relatively large size of NNLO effects
turned out to alleviate the tension that was observed between earlier
experimental measurements~\cite{Chatrchyan:2013oev,ATLAS-CONF-2014-033} and NLO QCD
predictions supplemented with the loop-induced gluon fusion contribution~\cite{Campbell:2011bn}.
In fact, NNLO QCD predictions are in good agreement with the latest measurements of
the \ww{} cross section~\cite{Khachatryan:2015sga,Aad:2016wpd}.

Besides perturbative calculations for the
inclusive cross section, the modelling of 
the jet-veto efficiency is another theoretical ingredient that plays a critical role in the
comparison of data with Standard Model predictions.
In particular, it was pointed out that 
a possible underestimate of the jet-veto efficiency through the
{\sc Powheg} Monte Carlo~\cite{Nason:2013ydw}, which is used 
to extrapolate the measured cross section from the fiducial region to the full phase space,
would lead to an artificial excess in the total cross section~\cite{Monni:2014zra}.
The relatively large size of higher-order effects
and the large intrinsic uncertainties of NLO+PS Monte Carlo simulations call for improved 
theoretical predictions for the jet-veto efficiency.
The resummation of logarithms of the jet-veto scale at next-to-next-to-leading logarithmic (NNLL) accuracy
was presented in~\citeres{Jaiswal:2014yba,Becher:2014aya}.
Being matched to the $pp\to\ww+X$ cross sections at NLO, these predictions
cannot describe the vetoing of hard jets beyond LO accuracy.  In order to
reach higher theoretical accuracy, NNLL resummation needs to be matched to
differential NNLO calculations.  Such NNLL+NNLO predictions have been
presented in~\citere{Grazzini:2015wpa} for the distribution in the
transverse momentum of the \ww{} system, and could be used to obtain
accurate predictions for the jet-veto efficiency through a reweighting
of Monte Carlo samples.


In this paper we present, for the first time, fully differential predictions
for \ww{} production with leptonic decays at NNLO.  More precisely, the full process
that leads to a final state with two leptons and two neutrinos is 
considered, including all relevant off-shell and interference effects in the
complex-mass scheme~\cite{Denner:2005fg}.
The calculation is carried out with \Matrix{}~\cite{matrix}, a
new tool that is based on the
\Munich{} Monte Carlo program~\cite{munich} interfaced with the {\sc OpenLoops}
generator of one-loop scattering amplitudes~\cite{Cascioli:2011va,hepforge:OpenLoops}, and
includes an automated implementation of the \qt{}-subtraction~\cite{Catani:2007vq} and
-resummation~\cite{Bozzi:2005wk} formalisms.
This widely automated framework has already been used, in combination with
the two-loop scattering amplitudes of~\citeres{Gehrmann:2011ab,Gehrmann:2015ora}, for the
calculations of 
$Z\gamma$~\cite{Grazzini:2013bna,Grazzini:2015nwa},
$ZZ$~\cite{Cascioli:2014yka,Grazzini:2015hta}, 
\ww{}~\cite{Gehrmann:2014fva}, 
$W^\pm\gamma$~\cite{Grazzini:2015nwa} and 
$W^\pm Z$~\cite{Grazzini:2016swo} 
production at NNLO QCD as well as in 
the resummed computations of the $ZZ$ and \ww{} transverse-momentum 
spectra~\cite{Grazzini:2015wpa} at NNLL+NNLO.
The present calculation relies on the two-loop amplitudes
of~\citere{Gehrmann:2015ora}.  Their implementation in
\Matrix{}~\cite{matrix} is applicable to any final state with two
charged leptons and two neutrinos, but in this paper we will focus on
the different-flavour signature $\muenn$.
The impact of QCD corrections on cross sections and distributions will be studied 
both at inclusive level and in presence of typical experimental selection cuts for \ww{} measurements and $H\to \ww$ studies.  The presented NNLO
results for fiducial cross sections and for the efficiencies of the
corresponding acceptance cuts provide first insights into acceptance efficiencies and
jet-veto effects at NNLO.

As pointed out in~\citere{Gehrmann:2014fva}, radiative QCD
corrections resulting from real bottom-quark emissions lead to a severe
contamination of $W$-pair production through top-quark resonances in the
$\ww{} b$ and $\ww{} b\bar b$ channels.  The enhancement of the $\ww$ cross
section that results from the opening of the $t\bar t$ channel at NNLO can
exceed a factor of five.  It is thus clear that a careful subtraction of $t\bar t$ and single-top
contributions is indispensable in order to ensure a decent 
convergence of the perturbative series.  To this end,
we adopt a top-free
definition of the \ww{} cross section based on a complete bottom-quark veto in the
four-flavour scheme. 
The uncertainty related with
this prescription will be assessed
by means of an alternative top-subtraction approach
based on the top-quark-width dependence of the \ww{} cross section in the
five-flavour scheme~\cite{Gehrmann:2014fva}.

The manuscript is organized as follows.
In \sct{sec:calculation} we describe technical aspects of the computation, 
including 
the subtraction of resonant top-quark contributions (\sct{subsec:top}),
\mbox{\qt{} subtraction} (\sct{subsec:formalism}), 
the \Matrix{} framework (\sct{subsec:matrix}),
and the stability of (N)NLO predictions 
based on \qt{} subtraction (\sct{subsec:stability}).
\sct{sec:results} describes our numerical results for $pp\to\muenn+X$:
We present the input parameters (\sct{sec:results-setup}),
cross sections and distributions without acceptance cuts (\sct{sec:results-incl}) and
with cuts corresponding to \ww{} signal (\sct{sec:results-ww}) and Higgs analyses
(\sct{sec:results-higgs}).
The main results are summarized in \sct{sec:summary}.

\section{Description of the calculation}
\label{sec:calculation}

We study the process 
\begin{equation}
\label{eq:process}
pp\to \elle^+\elle^{\prime\, -}\nu_{\elle}{\bar\nu}_{\elle^\prime}+X, 
\end{equation}
including all resonant and non-resonant Feynman
diagrams that contribute to the production of two charged leptons and two
neutrinos.

\begin{figure}
\begin{center}
\begin{tabular}{ccccc}
\includegraphics[width=.25\textwidth]{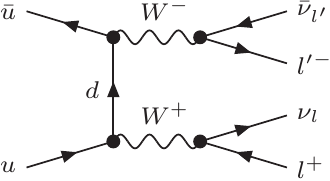} & &
\includegraphics[width=.25\textwidth]{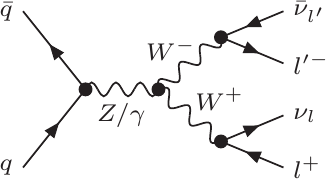} & &
\includegraphics[width=.25\textwidth]{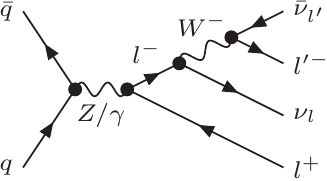} \\[0ex]
(a) & & (b) & & (c)
\end{tabular}
\caption[]{\label{fig:Borndiagrams}{Sample of Born diagrams contributing to \ww{} production both in the different-flavour case ($\elle\neq \elle^\prime$) and in the same-flavour case ($\elle=\elle^\prime$).}}
\end{center}
\vspace*{.5ex}
\begin{center}
\begin{tabular}{ccc}
\includegraphics[width=.25\textwidth]{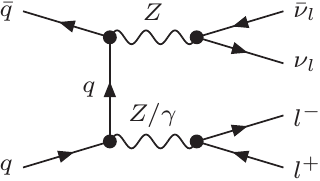} & &
\includegraphics[width=.25\textwidth]{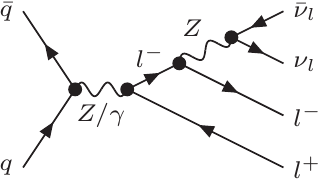} \\[0ex]
(d) & & (e)
\end{tabular}
\caption[]{\label{fig:BorndiagramsSF}{Sample of Born diagrams contributing to \ww{} production 
only in the same-flavour case. In the different-flavour case, they would describe $ZZ$ production in the $2\elle2\nu'$ channel.}}
\end{center}
\end{figure}

Depending on the flavour of the final-state leptons,
the generic reaction in \refeq{eq:process} can involve different combinations of
vector-boson resonances.
The different-flavour final state $\elle^+\elle^{\prime\, -}\nu_{\elle}{\bar\nu}_{\elle^\prime}$ 
is generated, as shown in \reffi{fig:Borndiagrams} for the $q\bar q$ process at LO,
\setlength\parskip{0em}
\begin{itemize}\setlength\itemsep{0em}
\item[(a)] via resonant $t$-channel \ww{} production with subsequent \mbox{$W^+\to\elle^+\nu_\elle$} 
and \mbox{$W^-\to \elle^{\prime\, -}\bar\nu_{\elle^\prime}$} decays; 
\item[(b)] via $s$-channel production in \mbox{$Z^{(}\hspace{-0.1em}^\ast\hspace{-0.1em}^{)}/\gamma^\ast\to WW^{(}\hspace{-0.1em}^\ast\hspace{-0.1em}^{)}$} topologies through a triple-gauge-boson vertex 
with subsequent \mbox{$W^+\to\elle^+\nu_\elle$} 
and \mbox{$W^-\to \elle^{\prime\, -}\bar\nu_{\elle^\prime}$} decays,  
where either both $W$ bosons, or the $Z$ boson and one of the $W$ bosons can 
become simultaneously resonant;
\item[(c)] via $Z/\gamma^\ast$ production with a subsequent decay 
\mbox{$Z/\gamma^\ast\to\elle\nu_{\elle} W\to\elle\elle^\prime\nu_{\elle}\nu_{\elle^\prime}$}.
Note that kinematics again allows for a resonant $W$ boson in the decay chain 
of a resonant $Z$ boson.
\end{itemize}
\setlength\parskip{2ex}
Additionally, in the case of equal lepton flavours, $\elle=\elle^\prime$, 
off-shell $ZZ$ production diagrams are involved, as shown in \reffi{fig:BorndiagramsSF}, 
where the $\elle^+\elle^-\nu_{\elle}{\bar\nu}_{\elle}$ final state is generated
\setlength\parskip{0em}
\begin{itemize}\setlength\itemsep{0em}
\item[(d)] via resonant $t$-channel $ZZ$ production with $Z\to l^+l^-$ and $Z\to \nu_\elle\bar\nu_\elle$ decays; 
\item[(e)] via further $Z\to4$ leptons topologies, 
$Z/\gamma^\ast\to\elle\elle Z\to\elle\elle\nu_{\elle}\nu_{\elle}$ or $Z\to\nu_{\elle}\nu_{\elle} Z\to\elle\elle\nu_{\elle}\nu_{\elle}$. 
Any double-resonant configurations are kinematically suppressed or excluded by phase-space cuts.
\end{itemize}
\setlength\parskip{2ex}
Note that the appearance of infrared (IR) divergent $\gamma^\ast\to \elle^+\elle^-$ splittings in the case of equal lepton flavours would prevent a fully inclusive phase-space integration.

Our calculation is performed in the
complex-mass scheme~\cite{Denner:2005fg}, and
besides resonances, it includes also 
contributions from off-shell EW bosons
and all relevant interferences; no resonance approximation is applied.
Our implementation can deal with any combination of leptonic flavours,
$\elle,\elle^\prime\in \{e,\mu,\tau\}$.  However, in this paper we will focus on the
different-flavour channel \mbox{$pp\to\muenn+X$}.  For the sake of brevity,
we will often denote this process as \ww{}
production though.

The NNLO computation requires the following scattering amplitudes at $\mathcal{O}(\as^2)$:
\setlength\parskip{0em}
\begin{itemize}\setlength\itemsep{0em}
\item tree amplitudes for 
$q\bar q \to \sfww\, gg$,\; $q\bar q^{(\prime)} \to \sfww\,q^{(\prime\prime)}\bar q^{(\prime\prime\prime)}$, 
and crossing-related processes;
\item one-loop amplitudes for $q\bar q \to \sfww\, g$, and crossing-related processes; 
\item squared one-loop amplitudes for $q\bar q \to \sfww$ and $gg\to \sfww$;
\item two-loop amplitudes for $q\bar q \to \sfww$.
\end{itemize}
\setlength\parskip{2ex}
All required tree-level and one-loop amplitudes are
obtained from the {\sc OpenLoops} generator~\cite{Cascioli:2011va,hepforge:OpenLoops}, which
implements a fast numerical recursion for the calculation of NLO scattering
amplitudes within the Standard Model.  For the numerically stable evaluation of tensor
integrals we employ the {\sc Collier} library~\cite{Denner:2014gla,Denner:2016kdg,hepforge:COLLIER}, which
is based on the Denner--Dittmaier reduction
techniques~\cite{Denner:2002ii,Denner:2005nn} and the scalar integrals
of~\citere{Denner:2010tr}. 
For the two-loop helicity
amplitudes we rely on a public C++ library~\cite{hepforge:VVamp} that implements 
the results of~\citere{Gehrmann:2015ora},
and for the numerical evaluation of the relevant 
multiple polylogarithms we use the implementation \cite{Vollinga:2004sn} in
the \GINAC\cite{Bauer:2000cp} library.
The contribution of the massive-quark loops
is neglected in the two-loop amplitudes, but accounted for anywhere else,
in particular in the loop-induced $gg$ channel.

\subsection{\bld\ww{} contamination through single-top and $t\bar t$ production}
\label{subsec:top}

The theoretical description of \ww{} production at higher orders in QCD is
complicated by a subtle interplay with top-production processes, which
originates from real-emission channels with final-state
bottom quarks~\cite{Dittmaier:2007th,Cascioli:2013gfa,Gehrmann:2014fva}.  In
the five-flavour scheme (\fs{5}), where bottom quarks are included in the 
parton-distribution
functions and the bottom-quark mass is set to zero, the presence of real bottom-quark
emission is essential to cancel collinear singularities that arise from
$g\to b\bar b$ splittings in the virtual corrections.  At the same time, the
occurrence of $Wb$ pairs in the real-emission matrix elements induces $t\to
Wb$ resonances that lead to a severe contamination of $W^+W^-$
production.
The problem starts with the NLO cross section, which receives a single-resonant
$tW\to W^+W^-b$ contribution
of about $30\%\,(60\%)$ at $7\,(14)$\,TeV.
At NNLO, the appearance of double-resonant 
$t{\bar t}\to W^+W^-b\bar b$ production
channels enhances the $W^+W^-$ cross section by about a factor of
four~(eight)~\cite{Gehrmann:2014fva}. 
Such single-top and $t\bar t$ contributions arise through the couplings of
$W$ bosons to external bottom quarks and enter at the same orders in $\alpha$ and
$\as$ as (N)NLO QCD contributions from light quarks.  Their huge impact
jeopardises the convergence of the perturbative expansion. 
Thus, precise theoretical predictions for \ww{} production 
require a consistent prescription to subtract the top contamination.

In principle, resonant
top contributions can be suppressed by imposing a $b$-jet veto, similarly as
in experimental analyses.  However,
for a $b$-jet veto with typical \pt{} values 
of $20-30$\,GeV, the top contamination remains
as large as about $10\%$~\cite{Gehrmann:2014fva}, while
in the limit of a vanishing $b$-jet veto $p_T$'s the NLO and NNLO \ww{} cross
sections suffer from collinear singularities
associated with massless bottom quarks in the \fs{5}.

To circumvent this problem, throughout this paper we use the four-flavour
scheme (\fs{4}), where the bottom mass renders 
all partonic subprocesses
with bottom quarks in the final state separately finite.  In this scheme,
the contamination from $t\bar{t}$ and single-top production is easily avoided
by omitting bottom-quark emission subprocesses.
However, this prescription generates
logarithms of the bottom mass 
that could have a non-negligible impact on the \ww{} cross section.
In order to assess the related uncertainty, results
in the \fs{4} are compared against a second calculation in the \fs{5}.
In that case, the contributions that are free from top resonances are isolated
with a gauge-invariant approach that exploits the scaling behaviour of the
cross sections in the limit of a vanishing top-quark
width~\cite{Gehrmann:2014fva}. 
The idea is that double-resonant (single-resonant) contributions depend
quadratically (linearly) on $1/\Gt$, while top-free \ww{} contributions are
not enhanced at small $\Gt$.  Exploiting this scaling property, the $t\bar
t$, $tW$ and (top-free) \ww{} components in the \fs{5} are separated
from each other through a numerical fit based on multiple high-statistics
evaluations of the cross section for increasingly small values of $\Gt$.
The subtracted result
in the \fs{5} can then be
understood as a theoretical prediction of the genuine \ww{} cross section
and directly compared to the \fs{4} result.
The difference should
be regarded as an ambiguity 
in the definition of a top-free \ww{} cross section
and includes, among other contributions,
the quantum interference between \ww{} production (plus
unresolved bottom quarks) and $t\bar t$ or single-top production.
This ambiguity was shown to be around $1\%-2\%$ 
for the inclusive \ww{} cross section at NNLO~\cite{Gehrmann:2014fva},
and turns out to be of the same size or even smaller
in presence of a jet veto (see \refse{sec:results}).

\subsection{The $\boldsymbol{q_T}$-subtraction formalism}
\label{subsec:formalism}

The implementation of the various IR-divergent amplitudes 
into a numerical code that provides finite
NNLO predictions for physical observables 
is a highly non-trivial task.  
In particular, 
the numerical computations need to be arranged in a way that guarantees the
cancellation of IR singularities 
across subprocesses with different parton multiplicities.
To this end various methods have been developed. They
can be classified in two broad categories.
In the first one, the NNLO calculation is organized
so as to cancel IR singularities of  both NLO and NNLO type at the same time.
The formalisms of antenna subtraction 
\cite{Kosower:1997zr, GehrmannDeRidder:2005cm,Daleo:2006xa,Currie:2013vh},
colourful subtraction \cite{Somogyi:2005xz,DelDuca:2015zqa,DelDuca:2016csb}
and Stripper \cite{Czakon:2010td,Czakon:2011ve,Czakon:2014oma}
belong to this category. Antenna subtraction and colourful subtraction
can be considered as extensions of the \nlo{} subtraction methods of \citeres{Frixione:1995ms,Frixione:1997np,Catani:1996jh,Catani:1996vz} to NNLO.
Stripper, instead, is a combination of the 
FKS subtraction method~\cite{Frixione:1995ms}
with numerical techniques based on 
sector decomposition \cite{Anastasiou:2003gr,Binoth:2000ps}.
The methods in the second category start from an NLO calculation with 
one additional parton (jet) in the final state 
and devise suitable subtractions to make the cross section
finite in the region in which the additional parton (jet) leads
to further divergences.
The \qt{}-subtraction method~\cite{Catani:2007vq}
as well as  
$N$-jettiness subtraction 
\cite{Boughezal:2015dva,Boughezal:2015eha,Gaunt:2015pea}, and
the Born-projection
method of \citere{Cacciari:2015jma}
belong to this class.

The \qt{}-subtraction formalism~\cite{Catani:2007vq} has been conceived in order to 
deal  with the production of any colourless\footnote{The extension to heavy-quark production
has been discussed in \citere{Bonciani:2015sha}.} high-mass system $F$ 
at hadron colliders. This method has already been applied in several 
NNLO calculations~\cite{Catani:2007vq,Catani:2009sm,Ferrera:2011bk,Catani:2011qz,Grazzini:2013bna,Ferrera:2014lca,Cascioli:2014yka,Gehrmann:2014fva,Grazzini:2015nwa,Grazzini:2015hta,Grazzini:2016swo},
and we have employed it also to obtain the results presented in this paper.
In the \qt{}-subtraction framework,
the $pp\to F+X$ cross section at (N)NLO can be written as
\begin{equation}
\label{eq:main}
\D{\sigma}^{\mathrm{F}}_{\mathrm{(N)NLO}}={\cal H}^{\mathrm{F}}_{\mathrm{(N)NLO}}\otimes \D{\sigma}^{\mathrm{F}}_{\mathrm{LO}}
+\left[ \D{\sigma}^{\mathrm{F + jet}}_{\mathrm{(N)LO}}-
\D{\sigma}^{\mathrm{CT}}_{\mathrm{(N)NLO}}\right].
\end{equation}
The term $d{\sigma}^{\mathrm{F + jet}}_{\mathrm{(N)LO}}$ represents the cross
section for the production of the system $F$ plus one jet at (N)LO accuracy
and can be evaluated with any available NLO subtraction formalism.  The
counterterm $\D{\sigma}^{\mathrm{CT}}_{\mathrm{(N)NLO}}$ guarantees the cancellation of 
the remaining IR divergences of the $F+$jet cross section.  It is obtained via fixed-order expansion
from the resummation formula for logarithmically enhanced contributions at small
transverse momenta~\cite{Bozzi:2005wk}.  The practical implementation of the
contributions in the square bracket in~\refeq{eq:main} is described in more
detail in \refse{subsec:matrix}.

The hard-collinear coefficient ${\cal H}^{\mathrm{F}}_{\mathrm{(N)NLO}}$ encodes the loop corrections to the Born-level 
process and compensates\footnote{More precisely, while the behaviour of $\D{\sigma}^{\mathrm{CT}}_{\mathrm{(N)NLO}}$
for $\qt\to 0$ is dictated by
the singular structure of $d{\sigma}^{\mathrm{F + jet}}_{\mathrm{(N)LO}}$, its non-divergent
part in the same limit is to some extent arbitrary, and its choice determines the explicit form of ${\cal H}^{\mathrm{F}}_{\mathrm{(N)NLO}}$.}
for the subtraction of $\D{\sigma}^{\mathrm{CT}}_{\mathrm{(N)NLO}}$.
It is obtained from the (N)NLO truncation of the process-dependent perturbative function
\begin{equation}
{\cal H}^{\mathrm{F}}=1+\f{\as}{\pi}\,
{\cal H}^{\mathrm{F}(1)}+\left(\f{\as}{\pi}\right)^2
{\cal H}^{\mathrm{F}(2)}+ \dots \;\;.
\end{equation}
The NLO calculation  of $\D{\sigma}^{\mathrm{F}}$ 
requires the knowledge
of ${\cal H}^{\mathrm{F}(1)}$, and the NNLO calculation also requires ${\cal H}^{\mathrm{F}(2)}$.
The general structure of ${\cal H}^{\mathrm{F}(1)}$
has been known for a long time~\cite{deFlorian:2001zd}. 
Exploiting the explicit results of ${\cal H}^{\mathrm{F}(2)}$ for Higgs
\cite{Catani:2011kr} and vector-boson~\cite{Catani:2012qa} 
production,
the result of \citere{deFlorian:2001zd} has been extended to the calculation of the NNLO coefficient 
${\cal H}^{\mathrm{F}(2)}$~\cite{Catani:2013tia}.
These results have been confirmed through an independent calculation in
the framework of Soft--Collinear Effective Theory~\cite{Gehrmann:2012ze,Gehrmann:2014yya}.
The counterterm $\D{\sigma}^{\mathrm{CT}}_{\mathrm{(N)NLO}}$ 
only depends on ${\cal H}^{\mathrm{F}}_{\mathrm{(N)LO}}$, i.e.\ for an NNLO computation it requires only 
${\cal H}^{\mathrm{F}(1)}$ as input, which can be derived from the one-loop amplitudes 
for the Born subprocesses.

\subsection{Organization of the calculation in M{\small ATRIX}}
\label{subsec:matrix}

Our calculation of \ww{} production is based on \Matrix{}~\cite{matrix}, a 
widely automated program for NNLO calculations at hadron colliders. 
This new tool is based on \qt{} subtraction, and is thus applicable to any process
with a colourless high-mass final state, provided that the two-loop amplitudes for the Born subprocess are available.
Moreover, besides fixed-order calculations,
it supports also the resummation of logarithmically enhanced terms at NNLL
accuracy (see \citere{Grazzini:2015wpa}, and \citere{Wiesemann:2016tae} for more details).

\Matrix{} is based on \Munich{}~\cite{munich},
a general-purpose Monte Carlo program that includes 
a fully automated implementation of the Catani--Seymour dipole 
subtraction method~\cite{Catani:1996jh,Catani:1996vz},
an efficient phase-space integration,
as well as an interface to the one-loop generator \OpenLoops{}~\cite{Cascioli:2011va,hepforge:OpenLoops} 
to obtain all required (spin- and colour-correlated) 
tree-level and one-loop amplitudes.
\Munich{} takes care of the bookkeeping of all relevant partonic subprocesses.
For each subprocess it automatically generates adequate phase-space parameterizations 
based on the resonance structure of the underlying (squared) tree-level Feynman diagrams.
These parameterizations are combined using a multi-channel approach 
to simultaneously flatten the resonance structure of the 
amplitudes, and thus guarantee a fast convergence of the numerical integration.
Several improvements like an adaptive weight-optimization procedure are implemented as well.

Supplementing the fully automated NLO framework of \Munich{} with a generic
implementation of the \qt{}-subtraction and -resummation techniques,
\Matrix{} achieves NNLL+NNLO accuracy in a way that limits the additionally
introduced dependence on the process to the two-loop amplitudes that enter
${\cal H}^{\mathrm{F}}_{\mathrm{NNLO}}$
in \refeq{eq:main}.  All other process-dependent
information entering the various ingredients in \refeq{eq:main} are
expressed in terms of NLO quantities already available within 
\Munich{}+\OpenLoops{}.

All NNLO contributions with vanishing total transverse momentum \qt{} of the
final-state system $F$ are collected in the coefficient ${\cal
H}^{\mathrm{F}}_{\mathrm{NNLO}}$.
The remaining part of the NNLO cross section, namely the
difference in the square bracket in \refeq{eq:main}, is formally finite in
the limit $\qt\to 0$, but each term separately exhibits logarithmic
divergences in this limit.
Since the subtraction is non-local, a technical cut on $\qt$ is introduced 
in order to render both terms separately finite. 
In this way, the \qt{}-subtraction method works very similarly to a 
phase-space slicing method.
In practice, it turns out to be more convenient to use a cut, $r_{\mathrm{cut}}$, on the 
dimensionless quantity $r=\qt/M$, where $M$ denotes the invariant mass
 of the final-state system $F$.

The counterterm $\D{\sigma}^{\mathrm{CT}}_{\mathrm{(N)NLO}}$ cancels all 
divergent terms from the real-emission contributions at small $\qt$, 
implying that the $r_{\mathrm{cut}}$ dependence of their difference 
should become numerically negligible for sufficiently small values of $r_{\mathrm{cut}}$.
In practice, as both the counterterm and the real-emission contribution 
grow arbitrarily large for $r_{\mathrm{cut}}\to0$, the statistical accuracy of the Monte Carlo integration degrades, preventing one from pushing $r_{\mathrm{cut}}$ too low.
In general, the absence of any strong residual $r_{\mathrm{cut}}$ dependence 
provides a stringent check on the correctness of the computation since any significant mismatch between the 
contributions would result in a divergent cross section in the limit $r_{\mathrm{cut}}\to0$.
To monitor the $r_{\mathrm{cut}}$ dependence without the need of repeated CPU-intensive runs,
\Matrix{} allows for simultaneous cross-section evaluations at variable $r_{\mathrm{cut}}$ 
values.
The numerical information on the $r_{\mathrm{cut}}$ dependence of the cross
section can be used to quantify the uncertainty due to finite
$r_{\mathrm{cut}}$ values (see \refse{subsec:stability}).

\subsection{Stability of $\boldsymbol{q_T}$ subtraction for $\boldsymbol{\muenn}$ production}
\label{subsec:stability}

\begin{figure}[t]
\begin{center}
\includegraphics[width=0.48\textwidth]{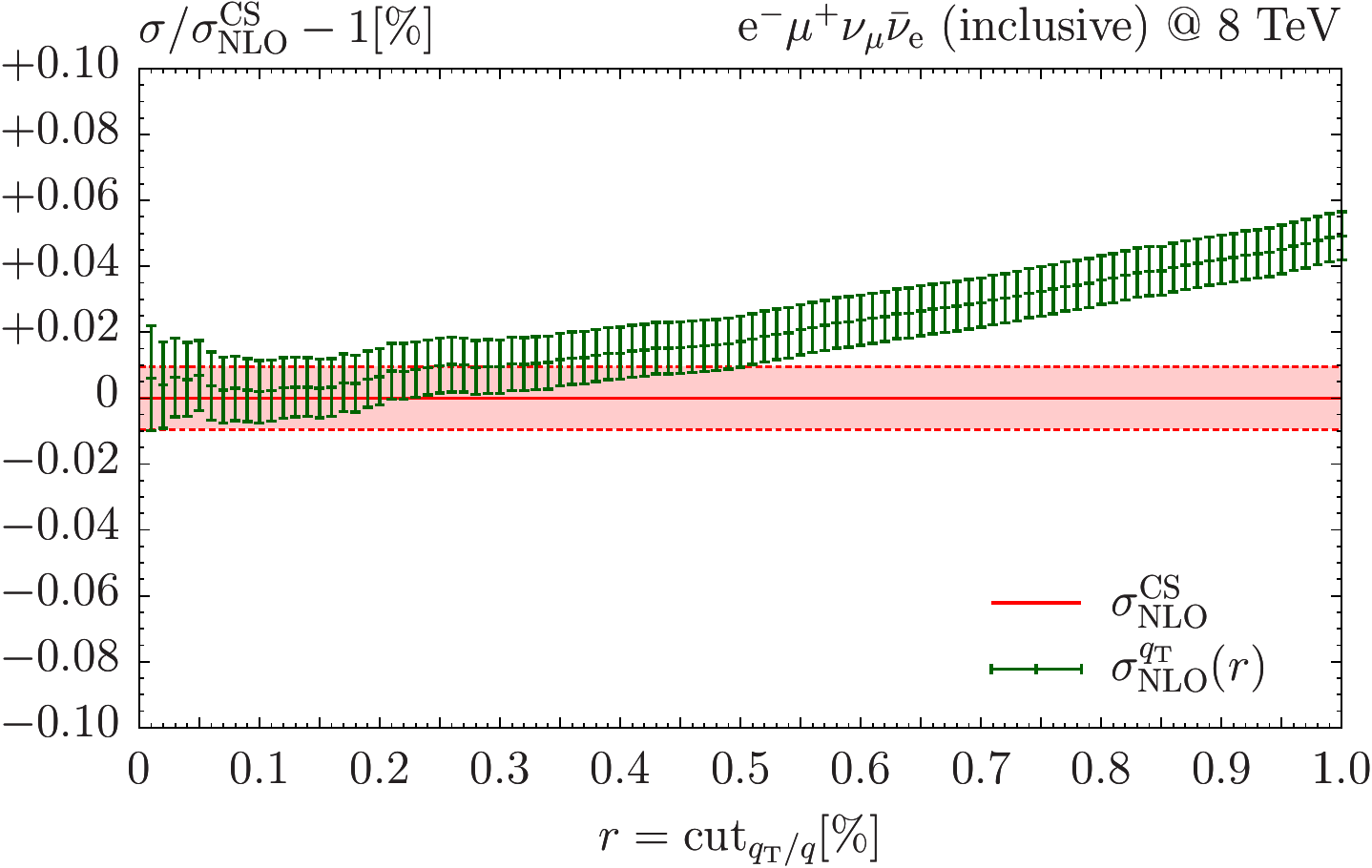}\hfill
\includegraphics[width=0.48\textwidth]{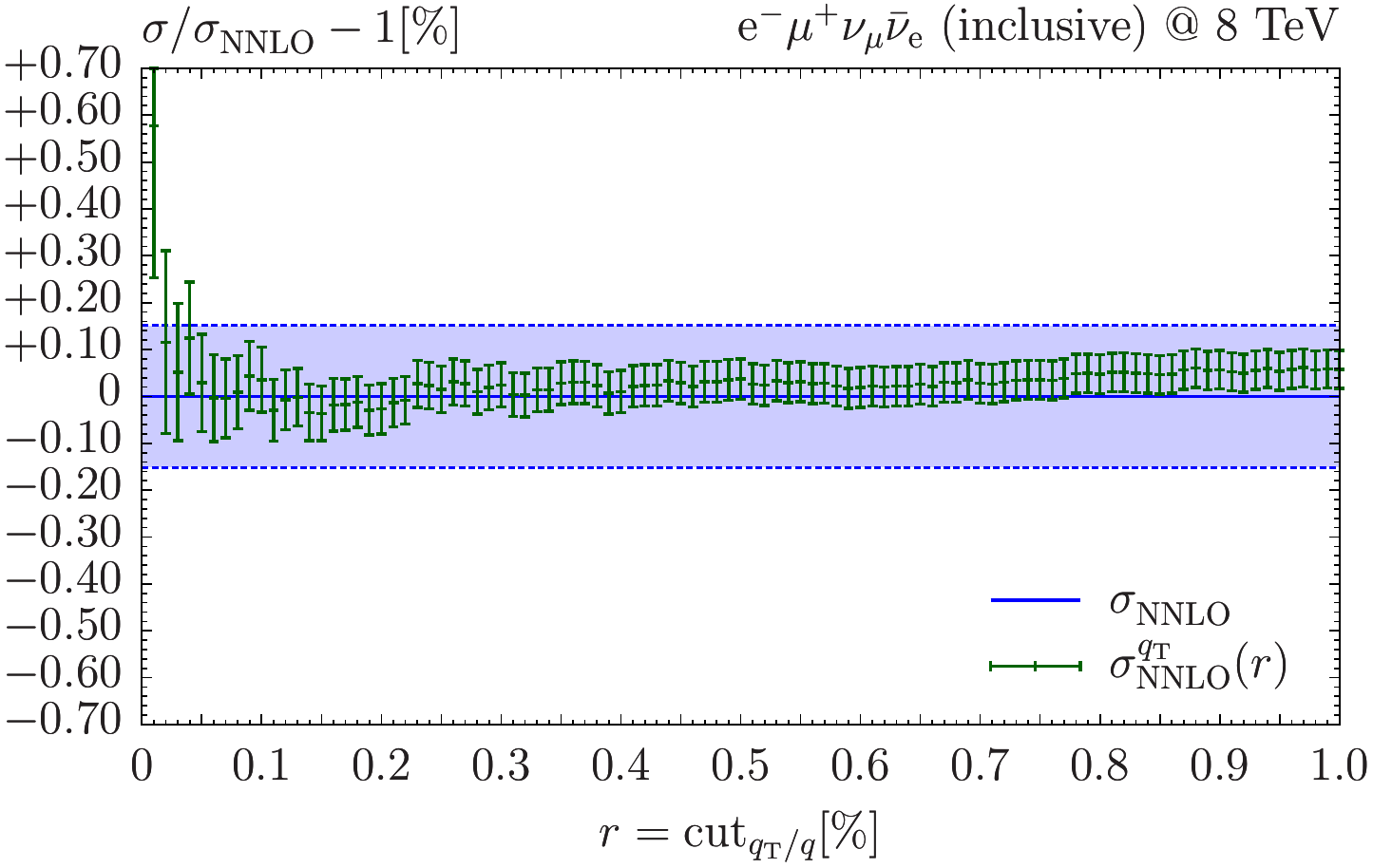}\\[1ex]
\includegraphics[width=0.48\textwidth]{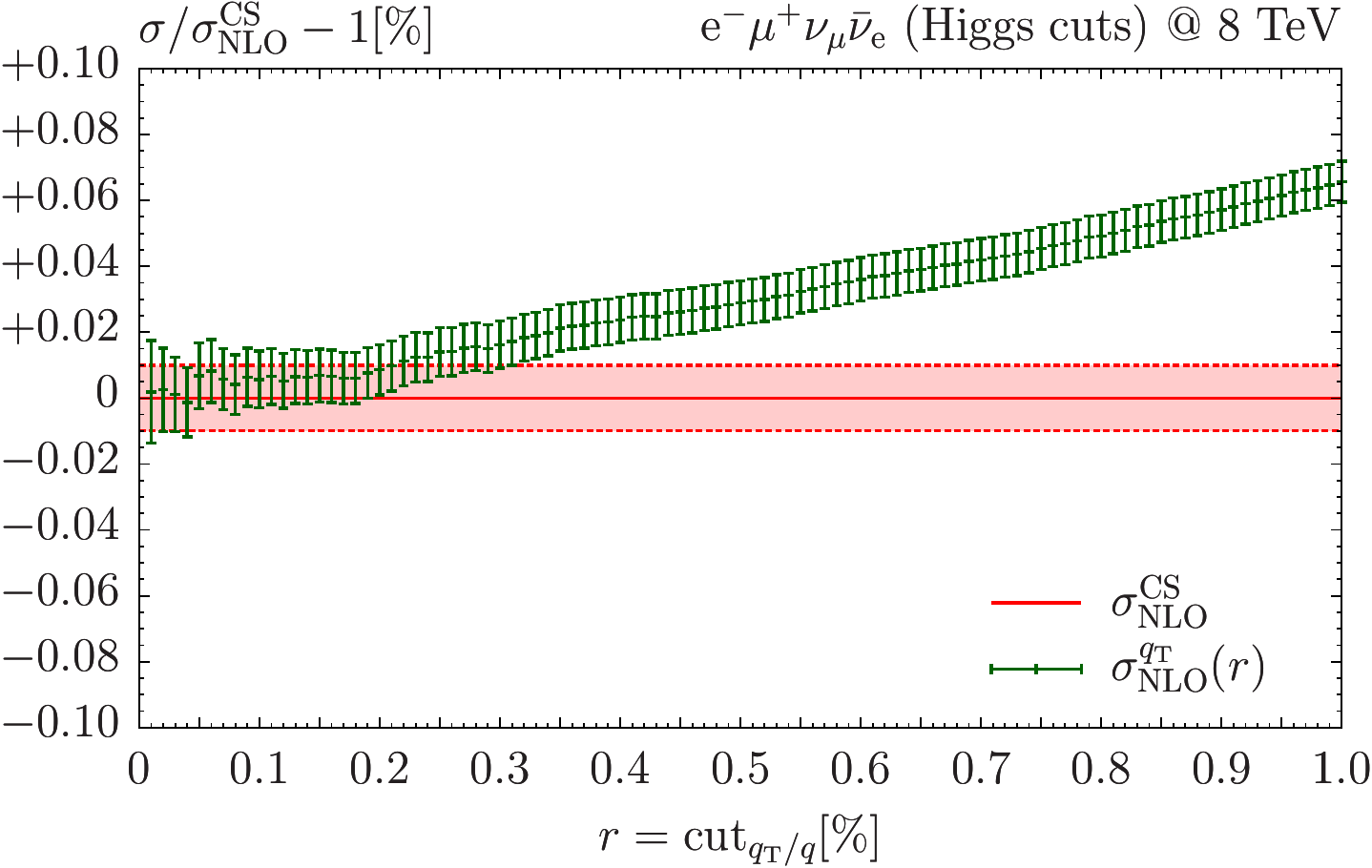}\hfill
\includegraphics[width=0.48\textwidth]{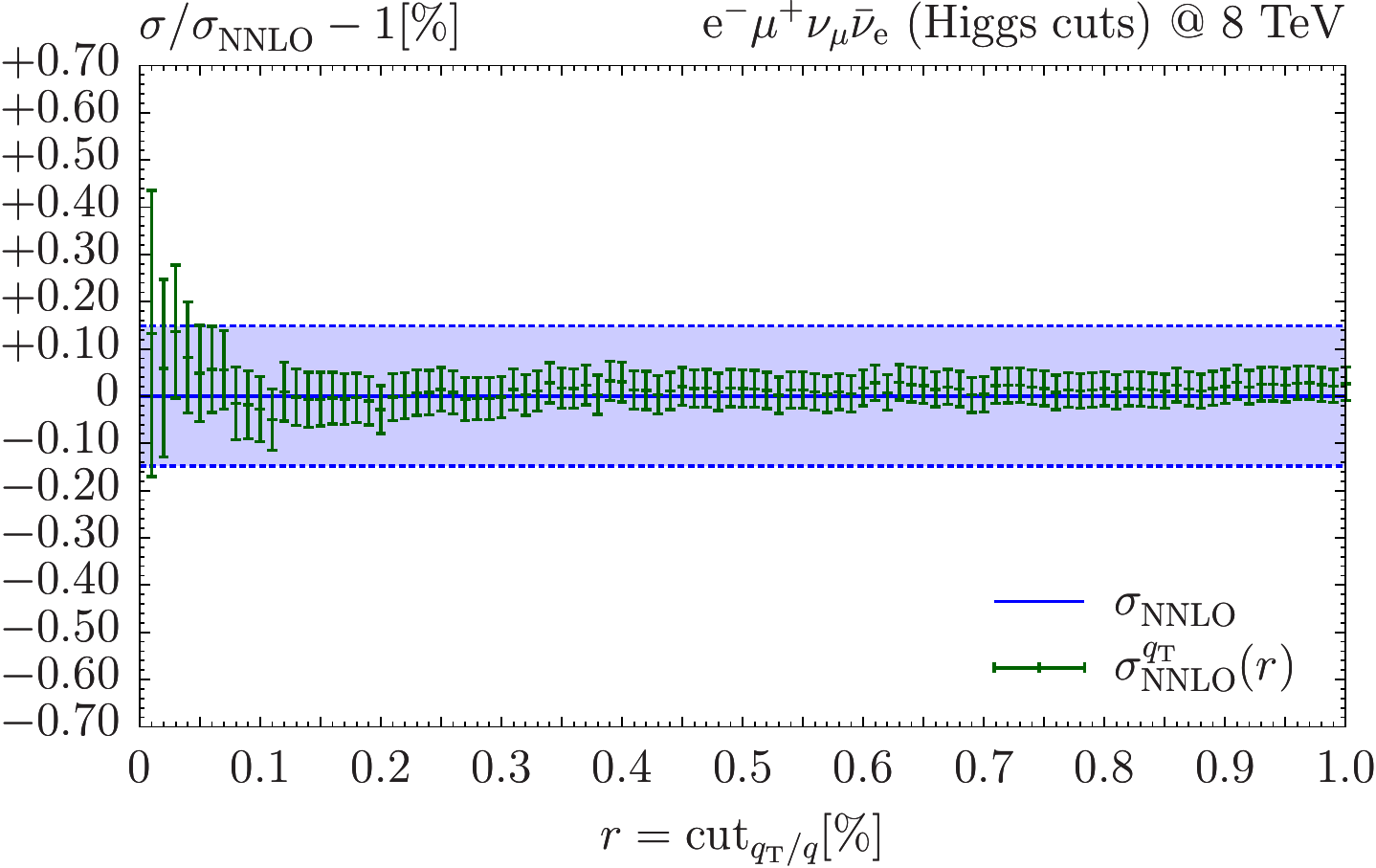}\\[1ex]
\caption[]{\label{fig:stability}{
Dependence of the $pp\to\muenn+X$ cross sections at 8\,TeV on the \qt{}-subtraction cut, $\rcut{}$,
for both NLO (left plots) and NNLO (right plots) results in the inclusive phase space (upper plots) and with Higgs cuts (lower plots). 
 NLO results are normalized to the \rcut{}-independent NLO cross section computed with Catani--Seymour subtraction, 
and the NNLO results are normalized to their values at $\rcut\to0$, with a conservative extrapolation-error indicated by the blue bands.}
}
\end{center}
\end{figure}

In the following we investigate the stability of the \qt{} subtraction
approach for $pp\to\muenn+X$.  To this end, in~\reffi{fig:stability}
we plot the NLO and NNLO cross sections as functions of the
\qt{}-subtraction cut, $\rcut$, which acts on the dimensionless variable
$r=\ptmuenn/\mmuenn$.  
Validation plots are presented at 8\,TeV both for the fully inclusive cross
section (see \refse{sec:results-incl}) and for the most exclusive case we 
have investigated, i.e.\ the cross section in presence of standard fiducial cuts 
for Higgs background analyses (see \refse{sec:results-higgs}). 
All considered scenarios at 8 and 13\,TeV lead essentially to the same conclusions.

At NLO the \rcut-independent cross section obtained with Catani--Seymour subtraction 
is used as a reference for the validation of the \qt{}-subtraction result.  
The comparison of the NLO cross sections in the left panels of \reffi{fig:stability} 
demonstrates
that \qt{} subtraction reaches about half-permille accuracy already at the moderate
value of $\rcut=1\%$, where we can, however, still resolve a
difference, which is slightly larger than the respective numerical 
uncertainties, with respect to the \rcut{}-independent result achieved using Catani--Seymour
subtraction. This difference is due to the well-known power-suppressed contributions
that are left after the cancellation of the logarithmic singularity at small \rcut{}. 
Going to even smaller values of \rcut{}, we observe a perfect convergence 
within statistical uncertainties towards the Catani--Seymour-subtracted result
in the limit $\rcut\to0$. 

At NNLO, where an \rcut{}-independent control result is not available, 
we observe no significant, i.e.\ beyond the numerical uncertainties, \rcut{} dependence below about $\rcut=1\%$;
we thus use the finite-\rcut{} results to extrapolate to $\rcut=0$, 
taking into account the breakdown of predictivity for very low $\rcut$ values,
and conservatively assign an additional numerical error to our results
due to this extrapolation. This procedure allows us
to control all NNLO predictions to inclusive and fiducial cross sections presented 
in \refse{sec:results} well below the level of two per mille. 
The increasing error bars indicate 
that arbitrarily low \rcut{} values cannot be tested as the contributions cancelling in
the limit are separately divergent.

Based on the observation that no significant $\rcut$ dependence is found below $\rcut=1\%$, 
the value $\rcut=0.25\%$ was adopted for the
calculation of the differential observables presented in~\refse{sec:results}. 
We have checked that the total rates for that value are fully consistent 
within numerical uncertainties with our extrapolated results and 
that a smaller value $\rcut=0.1\%$ leads to distributions in full statistical
agreement, thus confirming the robustness of our results also at the differential level.

\section{Results}
\label{sec:results}

We present numerical results for the different-flavour
process $pp\to \muenn+X$ at $\sqrt{s}=8$\,TeV and $13$\,TeV.
Cross sections and distributions are studied both in the
inclusive phase space and in presence of typical selection cuts for \ww{}
and $H\to\ww{}$ analyses.

Different-flavour final states provide the highest sensitivity both in \ww{}
measurements and Higgs studies.  We note that, due to the charge asymmetry
of \ww{} production in proton--proton collisions and the differences in the muon and electron acceptance
cuts (in particular regarding the rapidity cuts), the two different-flavour channels, 
$\muenn$ and $\emunn$, do not yield identical cross sections. 
However, we have checked that the absolute differences are not resolved on the level of our statistical errors.
Thus (N)NLO predictions and $K$-factors for $\muenn$
production can be safely applied also to $pp\to\emunn+X$.

\subsection{Input parameters, PDFs and selection cuts}
\label{sec:results-setup}

\def\hatm{\mu}
\def\hatthetaw{\hat\theta_\mathrm{w}}
\def\thetaw{\theta_\mathrm{w}}

Results in this paper 
are based on the EW input parameters 
$G_\mu = 1.1663787\times 10^{-5}$~GeV$^{-2}$, $m_W=80.385$ GeV
and $m_Z = 91.1876$~GeV. The other couplings in the EW
sector are derived in the \mbox{$G_\mu$-scheme}, where $\cos{\thetaw}=m_W/m_Z$ and
$\alpha=\sqrt{2} G_\mu m_W^2\sin^2\thetaw/\pi$.
In the complex-mass scheme, the physical gauge-boson masses and the weak mixing angle are replaced by 
$\hatm_V=\sqrt{m_V^2-\mathrm{i}\Gamma_Vm_V}$ and 
$\cos{\hatthetaw}=\hatm_W/\hatm_Z$, 
while  for $\alpha$ the above real-valued expression is used.
For the vector-boson widths we employ 
$\Gamma_W=2.085$~GeV and $\Gamma_Z=2.4952$~GeV~\cite{Agashe:2014kda},
and for the heavy quarks we set $m_b=4.92$~GeV and
$m_t=172.5$~GeV.
These input parameters result in a 
branching fraction  
$\mathrm{BR}(W^\pm\to\elle^\pm\nu_\elle)=0.1090040$ 
for each massless lepton generation, i.e.\  $l=e,\mu$.
Contributions from resonant Higgs bosons and their interference
with the \ww{} continuum are fully supported in our implementation. However, 
since this study is focused on \ww{} production as EW signal or 
as background to $H\to\ww$, Higgs contributions 
have been decoupled by taking the $m_H\to\infty$ limit.

To compute hadronic cross sections,
we use NNPDF3.0 parton-distribution functions (PDFs)~\cite{Ball:2014uwa},
and, unless stated otherwise, we work in the \fs{4},
while removing all contributions with final-state bottom quarks in order to avoid any contamination from top-quark resonances.
In the NNPDF framework, \fs{4} PDFs are 
derived from the standard variable-flavour-number PDF set with
$\alpha_s^{(5\mathrm{F})}(M_Z)=0.118$ 
via appropriate backward and forward evolution with five and four active flavours, 
respectively.
The resulting values of the strong coupling 
$\alpha_s^{(4\mathrm{F})}(M_Z)$ at 
LO, NLO and NNLO are 
0.1136, 0.1123 and 0.1123, respectively.
Predictions at N$^n$LO are obtained using
PDFs at the corresponding perturbative order
and the evolution of $\as$ at $(n+1)$-loop order, 
as provided by the PDF set.
The central values of the factorization and renormalization scales 
are set to $\muF=\muR=m_W$. Scale uncertainties are estimated by varying 
$\muF$ and $\muR$ in the range $0.5\, m_W\leq \muF, \muR \leq 2\, m_W$ 
with the restriction $0.5\leq \muF/\muR \leq 2$.

\renewcommand{\baselinestretch}{1.2}
\begin{table}[t]
\begin{center}
\begin{tabular}{|c|c|c|}
\hline
cut variable & \ww{} cuts & Higgs cuts \\
\hline
\multicolumn{3}{|c|}{lepton definition}\\
\hline
$\ptlone$ & $>25$\,GeV & $>22$\,GeV \\
$\ptltwo$ & $>20$\,GeV & $>10$\,GeV \\
$|\ymu|$ & $<2.4$ & $<2.4$ \\
$|\ye|$ & $<2.47$ and $\notin[1.37;1.52]$ & $<2.47$ and $\notin[1.37;1.52]$ \\
\hline
\multicolumn{3}{|c|}{leptonic cuts}\\
\hline
$\ptmiss$ & $>20$\,GeV & $>20$\,GeV\\
$\ptmissrel$ & $>15$\,GeV & --- \\
$\ptll$ & --- & $>30$\,GeV \\
$\mll$ & $>10$\,GeV & $\in[10$\,GeV$;55$\,GeV$]$\\
$\dRll$ & $>0.1$ & --- \\
$\dphill$ & --- & $<1.8$\\
$\dphillnunu$ & --- & $>\pi/2$\\
\hline
\multicolumn{3}{|c|}{anti-$k_T$ jets with $R=0.4$, $\;\ptjet>25$\,GeV, $\;|\yj|<4.5$}\\
\hline
$N_{\mathrm{jets}}$ & $0$ & $0$ \\\hline
\end{tabular}
\end{center}
\renewcommand{\baselinestretch}{1.0}
\caption{\label{tablecuts} 
Selection cuts targeted at \ww{} signal measurements (central
column) and $H\to\ww$~studies (right column).  The hardest and second hardest lepton are denoted as $l_1$
and $l_2$, respectively.
The missing transverse momentum, $\ptmiss$, is identified with the total transverse momentum of the
$\nu\bar \nu$ pair, while the relative missing transverse
momentum $\ptmissrel$ is defined as $\ptmiss\times \sin|\Delta\phi|$, where
$\Delta\phi$ is the azimuthal separation between $\mathbf\ptmiss$
and the momentum of the closest lepton; 
$\dphillnunu$ is the azimuthal angle between
the vectorial sum of the leptons'
transverse momenta, $\mathbf\ptll$,
and $\mathbf\ptmiss$.}
\end{table}

\renewcommand{\baselinestretch}{1.0}

In the following subsections we investigate $\muenn$ production in the
inclusive phase space (\refse{sec:results-incl}) and in presence of typical selection cuts that are designed
for measurements of \ww{} production (\refse{sec:results-ww}) and for $H\to\ww$ studies 
(\refse{sec:results-higgs})  at the LHC.
The detailed list of cuts is specified in~\refta{tablecuts}.
Besides the requirement  
of two charged leptons within a certain transverse-momentum and rapidity region,
they involve additional restrictions on the missing transverse momentum
($\ptmiss=p_{T,\nu\bar\nu}$), the transverse momentum ($p_{T,ll}$) and
invariant mass ($m_{ll}$) of the dilepton system, the combined rapidity--azimuth ($\Delta R_{ll}$) and
azimuthal ($\Delta \phi_{ll}$) separation of the charged leptons, as well as on the
relative missing transverse momentum ($\ptmissrel$) and the
azimuthal angle between $\mathbf\ptll$,
and $\mathbf\ptmiss$ ($\dphillnunu$), as
defined in~\refta{tablecuts}. Moreover, the \ww{} and Higgs selection criteria
involve a veto against anti-$k_T$ jets~\cite{Cacciari:2008gp} 
with $R=0.4$, $p_T>25$\,GeV and $|y|<4.5$.

\subsection{Analysis of inclusive \bld{\muenn} production}
\label{sec:results-incl}

\renewcommand{\baselinestretch}{1.2}
\begin{table}
\begin{minipage}{\textwidth}  
\begin{center}
\begin{tabular}{|l|ccc|ccc|}
\hline
& \multicolumn{3}{c|}{$\sigmainc$\,[fb]} &\multicolumn{3}{c|}{$\sigma/\sigma_{\rm NLO}-1$} \\
\hline
\multicolumn{1}{|c|}{$\sqrt{s}$} & 8\,TeV  & & 13\,TeV & 8\,TeV & & 13\,TeV  \\
\hline
LO       
&    425.41(4)\,$^{+2.8\%}_{-3.6\%}$ 
& & \phantom{0}778.99\phantom{0}(8)\,$^{+5.7\%}_{-6.7\%}$ 
& $-31.8\%$ 
& & $-35.4\%$ \\
NLO      
&    623.47(6)\,$^{+3.6\%}_{-2.9\%}$ 
& & 1205.11(12)\,$^{+3.9\%}_{-3.1\%}$ 
& 0 
& & 0 \\
\nloprime   
&    635.95(6)\,$^{+3.6\%}_{-2.8\%}$ 
& & 1235.82(13)\,$^{+3.9\%}_{-3.1\%}$ 
& $+\phantom{0}2.0\%$ 
& & $+\phantom{0}2.5\%$\\
\nloplusgg 
&    655.83(8)\,$^{+4.3\%}_{-3.3\%}$ 
& & 1286.81(13)\,$^{+4.8\%}_{-3.7\%}$ 
& $+\phantom{0}5.2\%$ 
& & $+\phantom{0}6.8\%$ \\ 
NNLO     
&    690.4(5)\phantom{0}\,$^{+2.2\%}_{-1.9\%}$ 
& & 1370.9(11)\phantom{0}\,$^{+2.6\%}_{-2.3\%}$ 
& $+10.7\%$ 
& & $+13.8\%$\\
\hline
\end{tabular}
\end{center}
\renewcommand{\baselinestretch}{1.0}
\caption{\label{tableincl} Total inclusive cross sections at different perturbative orders and
relative differences with respect to NLO. The quoted uncertainties correspond to scale variations as described in the text, and the numerical integration errors on the previous digit(s) are stated in brackets; for the NNLO results, 
the latter include the uncertainty due the \rcut{} extrapolation (see \refse{subsec:stability}).}
\end{minipage}
\end{table}

\renewcommand{\baselinestretch}{1.0}

In this Section we study $\muenn$ production in absence of 
acceptance cuts.  Predictions for the total inclusive 
cross section at  LO, NLO and NNLO are listed in \refta{tableincl}.
The NLO cross section computed with NNLO PDFs, denoted by NLO$^\prime$, and NLO$^\prime$
supplemented with the loop-induced gluon-fusion contribution (\nloplusgg) are provided as well.

At $\sqrt{s}=8\,(13)$\,TeV the NLO corrections increase the LO cross section by $47\%\,(55\%)$, and the 
NNLO corrections result in a further sizeable shift of $+11\%\,(+14\%)$ with respect to NLO. 
The total NNLO correction can be understood as the sum of three contributions that can be read off \refta{tableincl}:
Evaluating the cross section up to ${\cal O}(\as)$ with NNLO PDFs 
increases the NLO result by about $2\%\,(3\%)$. 
The loop-induced gluon-fusion channel, which used to be considered the dominant part of the NNLO corrections,
further raises the cross section by only $3\%\,(4\%)$, while the genuine ${\cal O}(\as^2)$ corrections 
to the $q\bar q$ channel\footnote{Here and in what follows, 
all NNLO corrections that do not stem from the loop-induced $gg\to \ww$ channel
are denoted as genuine 
${\cal O}(\as^2)$  corrections or NNLO corrections to the $q\bar q$ channel. 
Besides $q\bar q$-induced partonic processes, they actually 
contain also $gq$ and $g\bar q$ channels with one extra final-state parton 
as well as $gg$, $qq^{(}\hspace{-0.1em}'\hspace{-0.1em}^{)}$, $\bar q\bar q^{(}\hspace{-0.1em}'\hspace{-0.1em}^{)}$
and $q{\bar q}'$ channels
with two extra final-state partons.} 
amount to about $+6\%\,(+7\%)$.
Neglecting PDF effects,  we find that the 
loop-induced $gg$ contribution corresponds to only $37\%\,(38\%)$ of the total
${\cal O}(\as^2)$ effect, i.e.\ of $\sigma_\nnlo -\sigma_\nloprime$,  
with the remaining  $63\%\,(62\%)$ being due to genuine NNLO 
corrections. 

These results are in line with the inclusive on-shell predictions of \citere{Gehrmann:2014fva}, 
where the relative weight of the $gg$ contribution was found to be $35\%\,(36\%)$, 
and the small difference is due to the chosen PDFs.
We also find by up to about $2\%$ larger NNLO corrections than stated in \citere{Gehrmann:2014fva}, which can
also be attributed to the chosen PDF sets.
Indeed, repeating the on-shell calculation of \citere{Gehrmann:2014fva} using the input parameters 
of \refse{sec:results-setup} (with $\Gamma_W=\Gamma_Z=0$),
we find that the relative corrections agree on the level of the statistical error when the same PDF sets are applied.
Moreover, comparing the results of \refta{tableincl} with this on-shell calculation allows us
to quantify the size of off-shell effects, which turn out to reduce the on-shell result by about $2\%$ with a 
very mild dependence (at the permille level) on the perturbative order and the collider energy.
The results for the two considered collider energies confirm 
that
the size of relative corrections slightly increases with the centre-of-mass energy, as in the on-shell case.

We add a few comments on the theoretical uncertainties of the above results. As is well known,
scale variations do not give a reliable estimate of
the size of missing higher-order contributions
at the first orders of the perturbative expansion. In fact, LO and NLO predictions are not
consistent within scale uncertainties, and the same conclusion can be drawn
by comparing NLO or \nloplusgg{} predictions with their respective scale uncertainties to the central NNLO result.
This can be explained by the fact that the $qg$ (as well as $\bar qg$) and $gg$ 
(as well as $qq^{(}\hspace{-0.1em}'\hspace{-0.1em}^{)}$, $\bar q\bar q^{(}\hspace{-0.1em}'\hspace{-0.1em}^{)}$
and $q{\bar q}'$) channels 
open up only at NLO and NNLO, respectively.
Since the NNLO is the first order where all the partonic channels are contributing, 
the NNLO scale dependence
should provide a realistic estimate of the uncertainty from missing higher-order corrections.
The loop-induced gluon--gluon channel, which contributes only at its leading order at $\order{\as^2}$ and thus 
could receive large relative corrections, 
was not expected to break this picture due to its overall smallness already in \citere{Gehrmann:2014fva}.
That conclusion is supported by the recent calculation of the NLO corrections 
to the loop-induced $gg$ channel \cite{Caola:2015rqy}.

\begin{figure}[t]
\begin{center}
\begin{tabular}{cc}
\hspace*{-0.17cm}
\includegraphics[trim = 7mm -7mm 0mm 0mm, width=.33\textheight]{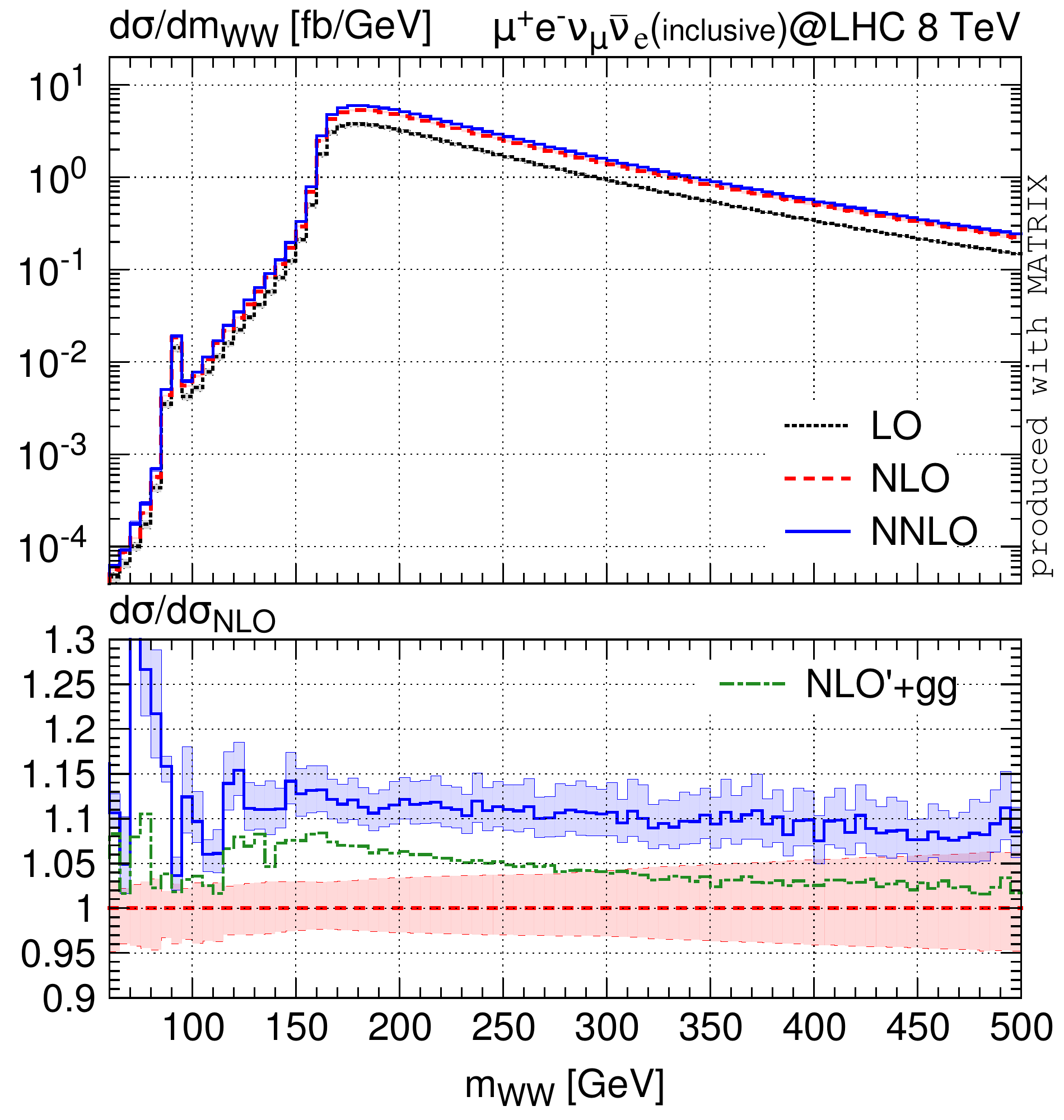} &
\includegraphics[trim = 7mm -7mm 0mm 0mm, width=.33\textheight]{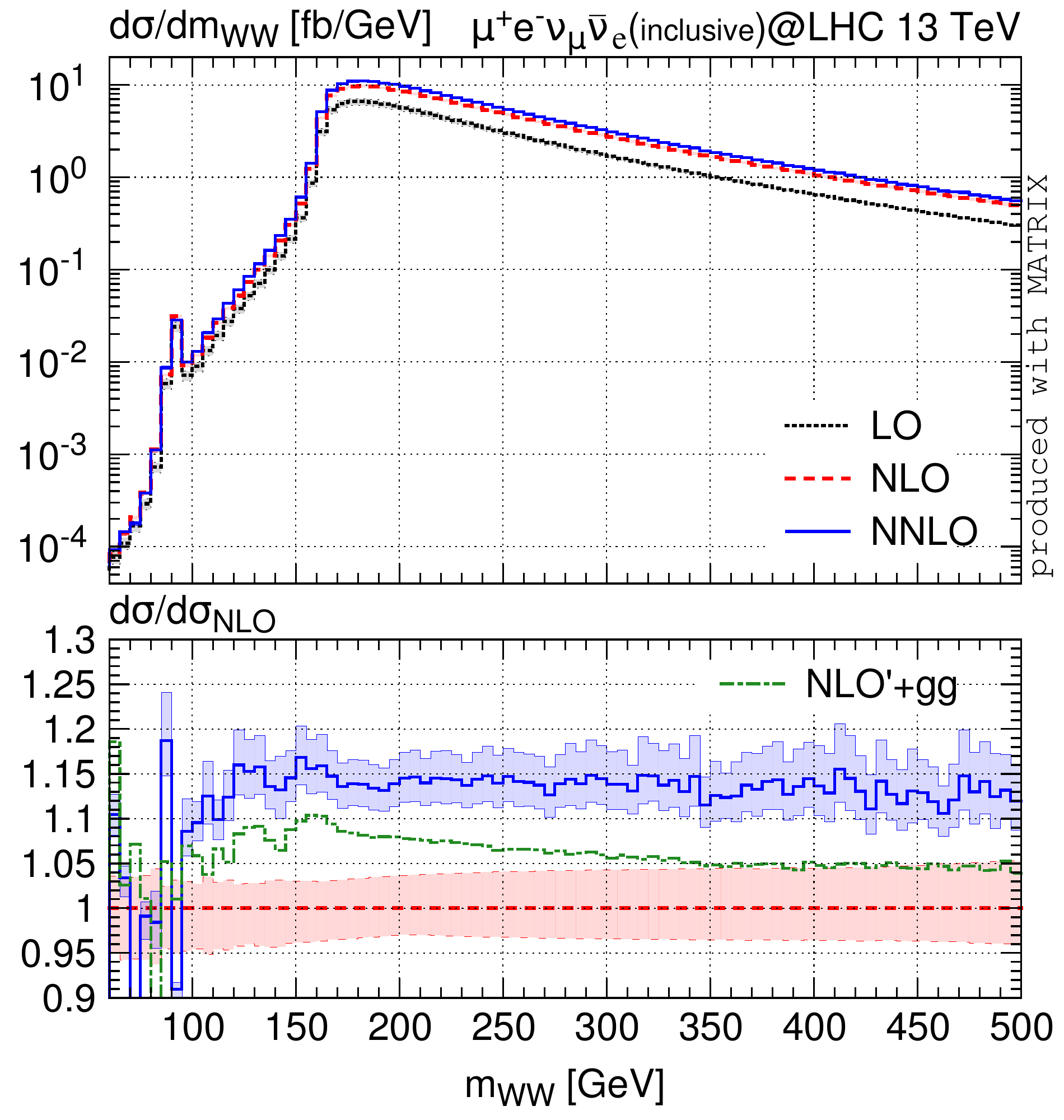} \\[-1em]
\hspace{0.6em} (a) & \hspace{1em}(b)
\end{tabular}
\caption[]{\label{fig:mWWinclusive}{Distribution in the invariant mass of the \ww{} pair,
$m_{\ww}=m_{\muenn}$. No acceptance cuts are applied. Absolute \lo{} (black, dotted), \nlo{} (red, dashed) and 
\nnlo{} (blue, solid) predictions at $\sqrt{s}=8$\,TeV (left) 
and $\sqrt{s}=13$\,TeV (right) are plotted in the upper frames. 
The lower frames display \nloplusgg{} (green, dot-dashed) 
and NNLO predictions normalized to NLO.
The bands illustrate
the scale dependence of the NLO and NNLO predictions.  In the case of ratios,
scale variations are applied only to the numerator, while the NLO prediction
in the denominator corresponds to the central scale.
}}
\end{center}
\end{figure}

In~\reffis{fig:mWWinclusive}{fig:pTW2inclusive} we present 
distributions that characterize the kinematics of the 
reconstructed $W$ bosons\footnote{The various kinematic variables are 
defined in terms of the off-shell $W$-boson momenta, 
$p_{W^+}=p_{\mu^+}+p_{\nu_\mu}$ and $p_{W^-}=p_{e^-}+p_{\bar\nu_e}$.}.
Absolute predictions at the various perturbative orders 
are complemented by ratio plots that illustrate the relative
differences with respect to NLO. 
In order to
assess the importance of genuine NNLO corrections, full NNLO results are
compared to \nloplusgg{} predictions in the ratio plots.

In \reffi{fig:mWWinclusive} we show the distribution in the total invariant mass, $m_{\ww}=m_{\muenn}$.
This observable features the characteristic threshold behaviour around $2\,m_W$,
with a rather long tail and a steeply falling cross section in the off-shell region
below threshold.
Although suppressed by two orders of magnitude, the
$Z$-boson resonance that 
originates from topologies of type (b) and (c) in \reffi{fig:Borndiagrams}
is clearly visible at $m_{\muenn}=m_Z$.
Radiative QCD effects turn out to be largely insensitive to the
EW dynamics that governs 
off-shell $W$-boson decays and dictates the shape of the $m_{\muenn}$ distribution.
In fact, the $\sigma_\nnlo{}/\sigma_\nlo{}$ ratio is rather flat, and shape distortions do not exceed about $5\%$, apart from the strongly suppressed region far below the $2\,m_W$ threshold.

\begin{figure}[t]
\begin{center}
\begin{tabular}{cc}
\hspace*{-0.17cm}
\includegraphics[trim = 7mm -7mm 0mm 0mm, width=.33\textheight]{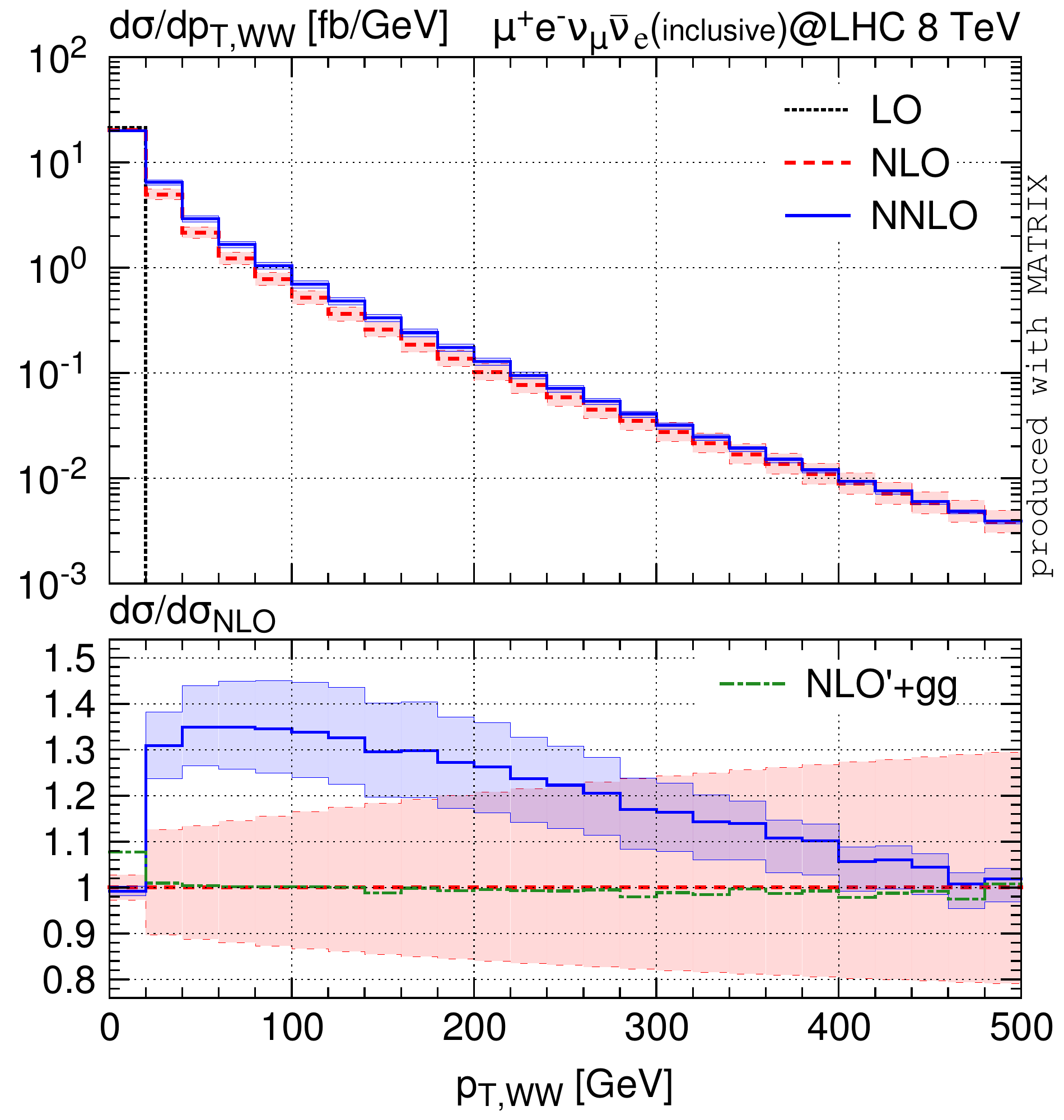} &
\includegraphics[trim = 7mm -7mm 0mm 0mm, width=.33\textheight]{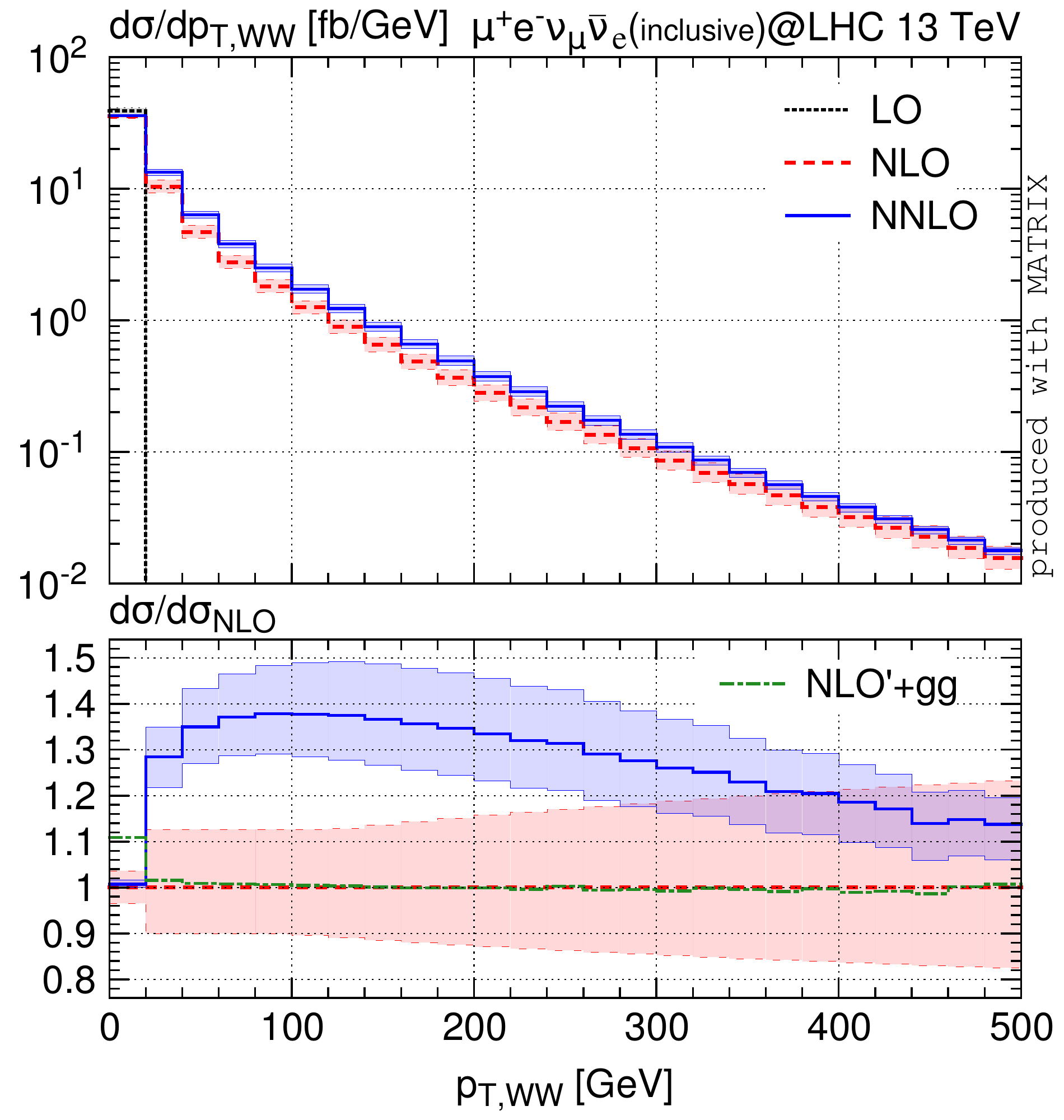} \\[-1em]
\hspace{0.6em} (a) & \hspace{1em}(b)
\end{tabular}
\caption[]{\label{fig:pTWWinclusive}{
Distribution in the  transverse momentum of the \ww{} pair. 
No acceptance cuts are applied. 
Absolute predictions and relative corrections as in~\reffi{fig:mWWinclusive}.
}}
\end{center}
\end{figure}

The distribution in the transverse momentum of the \ww{} pair, shown in \reffi{fig:pTWWinclusive}, 
vanishes at \lo{}. Thus, at non-zero transverse momenta
\nlo{} (\nnlo{}) results are formally only \lo{} (\nlo{}) accurate.
Moreover, the loop-induced $gg$ channel contributes only at $\ptww=0$.
The relative NNLO corrections are consistent with the results discussed 
in \citere{Grazzini:2015wpa}: they are large and exceed 
the estimated scale uncertainties in the small and intermediate 
transverse-momentum regions, while
the \nlo{} and \nnlo{} uncertainty bands overlap  
at large transverse momenta. At very low $\pt$, the fixed-order NNLO calculation 
diverges, but NNLL+NNLO resummation~\cite{Grazzini:2015wpa} 
can provide accurate predictions also in that region.

\begin{figure}[tp]
\begin{center}
\begin{tabular}{cc}
\hspace*{-0.17cm}
\includegraphics[trim = 7mm -7mm 0mm 0mm, width=.33\textheight]{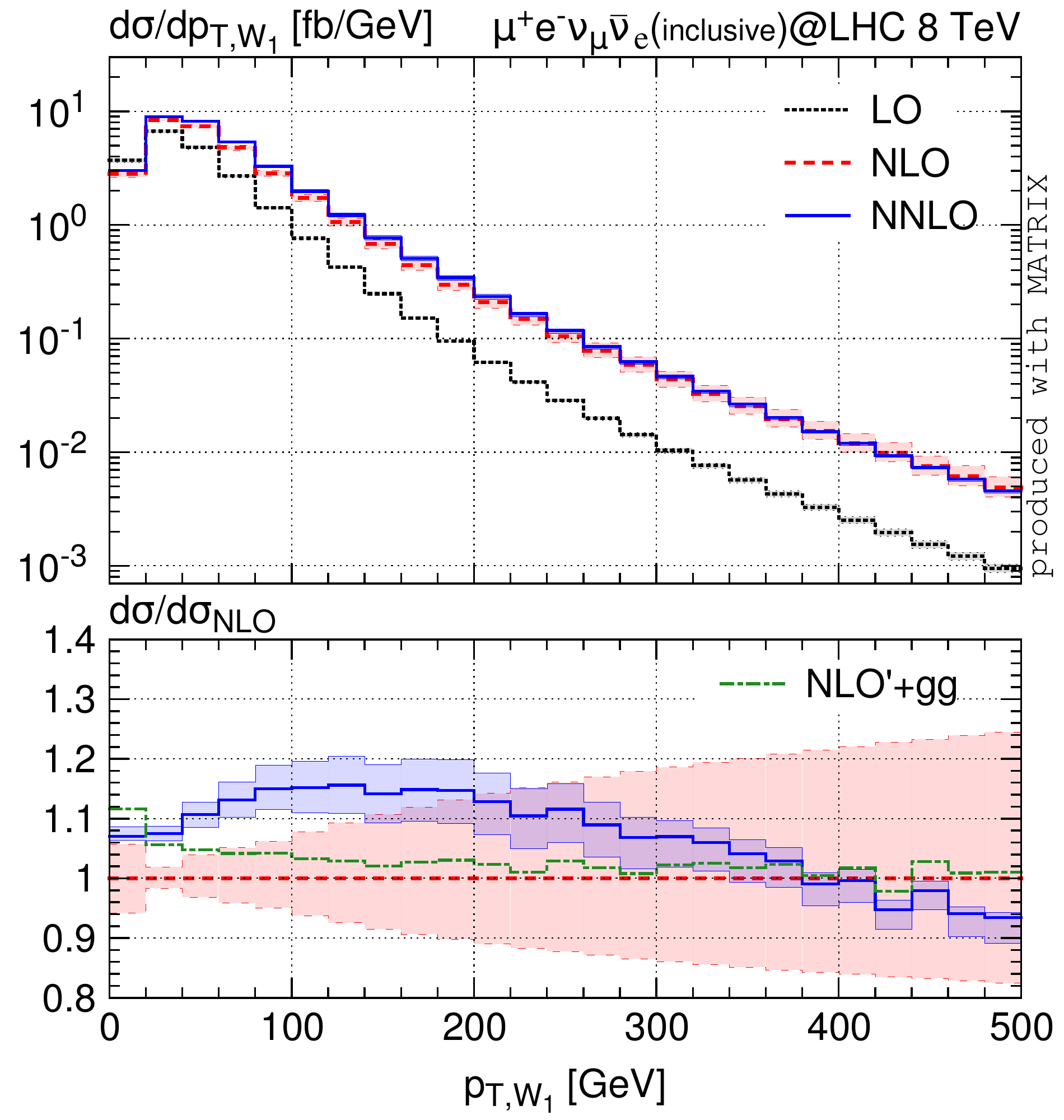} &
\includegraphics[trim = 7mm -7mm 0mm 0mm, width=.33\textheight]{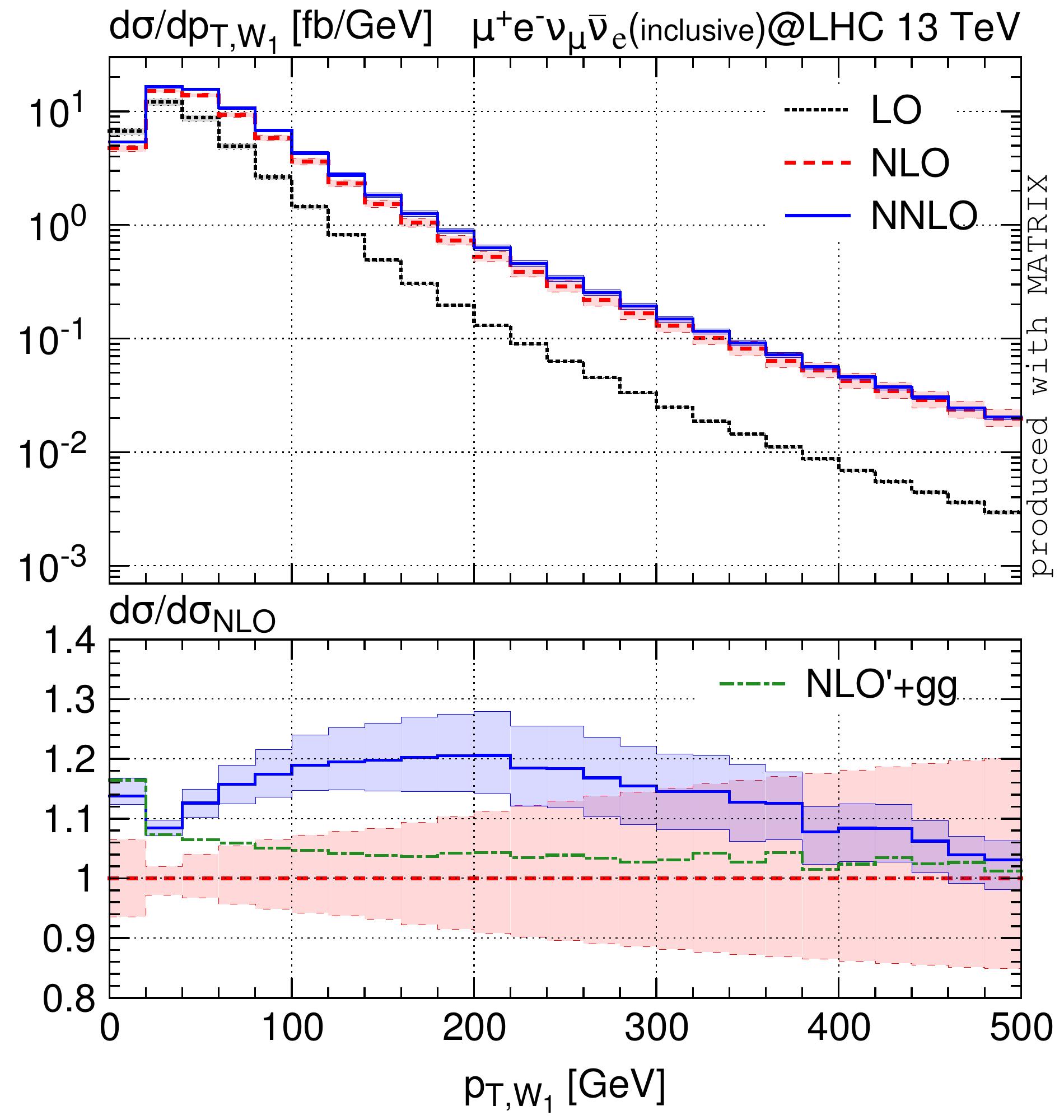} \\[-1em]
\hspace{0.6em} (a) & \hspace{1em}(b)
\end{tabular}
\caption[]{\label{fig:pTW1inclusive}{
Distribution in the 
transverse momentum of the harder reconstructed $W$ boson. 
No acceptance cuts are applied. 
Absolute predictions and relative corrections as in~\reffi{fig:mWWinclusive}.
}}
\end{center}
\vspace{1cm}
\begin{center}
\begin{tabular}{cc}
\hspace*{-0.17cm}
\includegraphics[trim = 7mm -7mm 0mm 0mm, width=.33\textheight]{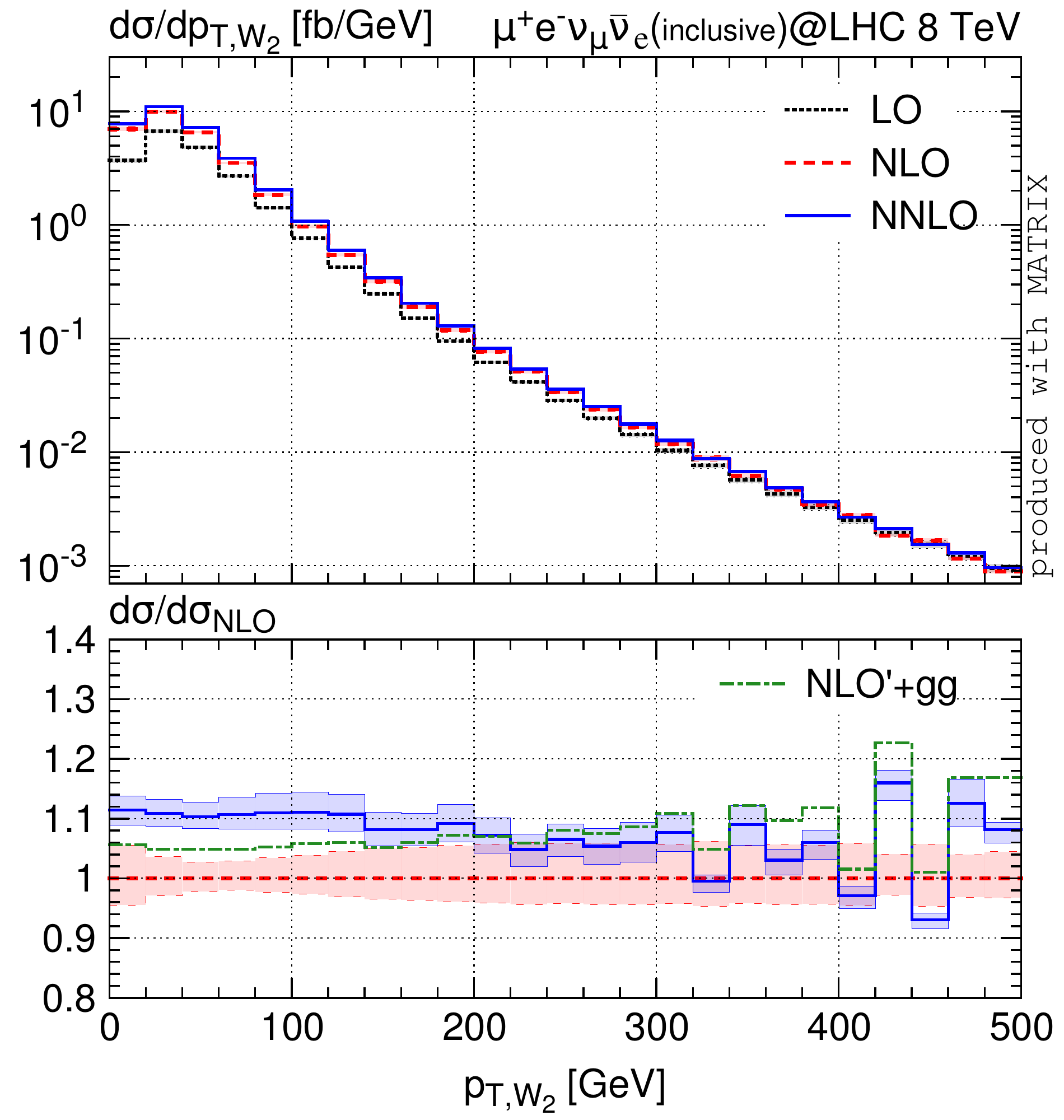} &
\includegraphics[trim = 7mm -7mm 0mm 0mm, width=.33\textheight]{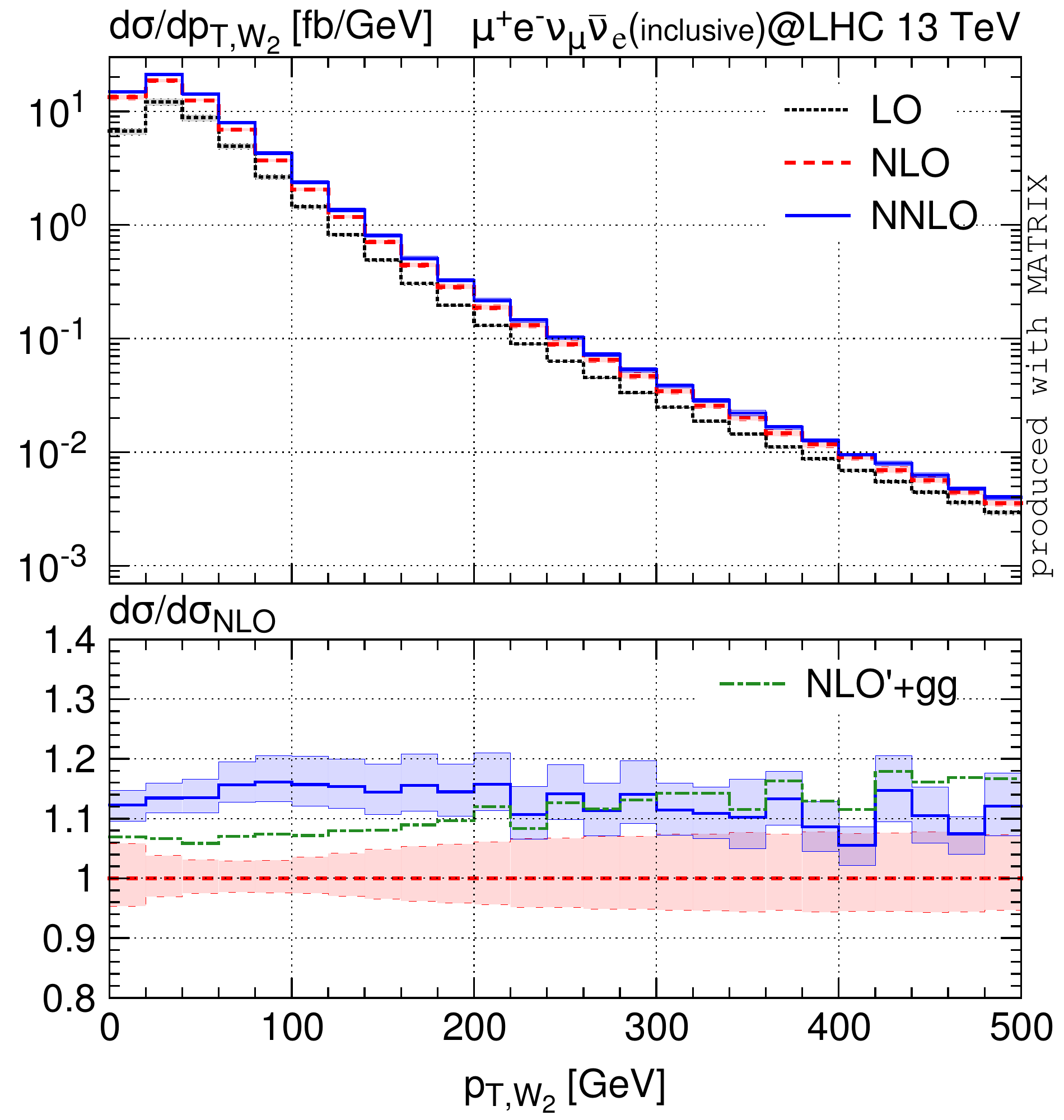} \\[-1em]
\hspace{0.6em} (a) & \hspace{1em}(b)
\end{tabular}
\caption[]{\label{fig:pTW2inclusive}{
Distribution in the transverse momentum of the softer reconstructed $W$ boson. 
No acceptance cuts are applied. 
Absolute predictions and relative corrections as in~\reffi{fig:mWWinclusive}.
}}
\end{center}
\end{figure}

In \reffitwo{fig:pTW1inclusive}{fig:pTW2inclusive} the transverse-momentum distributions of the 
harder $W$ boson, \ptwone{}, and the softer $W$ boson, \ptwtwo{}, are depicted.
The first eye-catching feature is the large NLO/LO correction in case of the harder $W$ boson,
which grows with \pt{} and leads to an enhancement by a factor of five at $\pt\approx 500$\,GeV, 
whereas such large corrections are absent for the softer $W$ boson.
This feature is due to the fact that the phase-space region with at least one hard $W$ boson
is dominantly populated by events with the NLO jet recoiling against this $W$ boson, 
while the other $W$ boson is relatively soft.
The LO-like nature of this dominant contribution for moderate and large values of \ptwone{} is reflected 
by the large NLO scale band.
The phase-space region where the softer $W$ boson has moderate or high transverse momentum as well 
is naturally dominated 
by topologies with the two $W$ bosons recoiling against each other.
Such topologies are present already at LO, and thus do not result in
exceptionally large corrections. 
Both for the leading and subleading $W$ boson, the NNLO corrections tend to exceed the NLO scale 
band at moderate transverse-momentum values.

For all distributions discussed so far, we find qualitatively the same
effects at $8$ and $13$\,TeV, essentially only differing by the larger
overall size of the NNLO corrections at the higher collider energy. 
Contributing only about one third of the total NNLO correction, the
\nloplusgg{} approximation does not provide a reliable description of the
full NNLO result.  Moreover, in general 
the loop-induced gluon--gluon channel alone cannot reproduce
the correct shapes of the full NNLO correction.

\subsection{Analysis of \bld{\muenn} production with \ww{} selection cuts}
\label{sec:results-ww}

In this Section we investigate the behaviour of radiative corrections in presence of 
acceptance cuts used in \ww{} measurements. The full set of cuts is summarized in~\refta{tablecuts}
and is inspired by the \ww{} analysis of \citere{Aad:2016wpd}\footnote{We do not apply any lepton-isolation criteria with respect to hadronic activity.}. 
Besides various restrictions on the leptonic degrees of freedom and the missing transverse momentum, 
this analysis implements a jet veto.

\renewcommand{\baselinestretch}{1.2}
\begin{table}[t]
\begin{center}
\begin{tabular}{|l|ccc|ccc|}
\hline
& \multicolumn{3}{c|}{$\sigmafid$(\ww{}-cuts)\,[fb]} &\multicolumn{3}{c|}{$\sigma/\sigma_{\rm NLO}-1$} \\
\hline
\multicolumn{1}{|c|}{$\sqrt{s}$} & 8\,TeV  & & 13\,TeV & 8\,TeV & & 13\,TeV  \\
\hline
LO       
& 147.23\phantom{0}(2)\,$^{+3.4\%}_{-4.4\%}$ 
& & 233.04(2)\,$^{+6.6\%}_{-7.6\%}$ 
& $-3.8\%$ 
& & $-\phantom{0}1.3\%$ \\
NLO      
& 153.07\phantom{0}(2)\,$^{+1.9\%}_{-1.6\%}$ 
& & 236.19(2)\,$^{+2.8\%}_{-2.4\%}$ 
& 0
& & 0 \\
\nloprime{}    
& 156.71\phantom{0}(3)\,$^{+1.8\%}_{-1.4\%}$ 
& & 243.82(4)\,$^{+2.6\%}_{-2.2\%}$ 
& $+2.4\%$ 
& & $+\phantom{0}3.2\%$ \\
\nloplusgg{} 
& 166.41\phantom{0}(3)\,$^{+1.3\%}_{-1.3\%}$ 
& & 267.31(4)\,$^{+1.5\%}_{-2.1\%}$ 
& $+8.7\%$ 
& & $+13.2\%$ \\ 
NNLO     
& 164.16(13)\,$^{+1.3\%}_{-0.8\%}$ 
& & 261.5(2)\phantom{0}\,$^{+1.9\%}_{-1.2\%}$ 
& $+7.2\%$ 
& & $+10.7\%$ \\
\hline
\end{tabular}
\end{center}
\renewcommand{\baselinestretch}{1.0}
\caption{\label{tableSignal} 
Cross sections with \ww{} fiducial cuts at different perturbative orders and 
relative differences with respect to NLO. Scale uncertainties and errors as in \refta{tableincl}.}

\renewcommand{\baselinestretch}{1.2}
\begin{center}
\begin{tabular}{|l|ccc|ccc|}
\hline
& \multicolumn{3}{c|}{$\efficiency=\sigmafid$(\ww{}-cuts)$/\sigmainc$} &\multicolumn{3}{c|}{$\efficiency/\efficiency_{\rm NLO}-1$} \\
\hline
\multicolumn{1}{|c|}{$\sqrt{s}$} & 8\,TeV  & & 13\,TeV & 8\,TeV & & 13\,TeV  \\
\hline
LO       
& 0.34608(7)$^{+0.6\%}_{-0.7\%}$
& & 0.29915(6)$^{+0.8\%}_{-1.0\%}$
& $+41.0\%$ 
& & $+52.6\%$ \\
NLO      
& 0.24552(5)$^{+4.4\%}_{-4.7\%}$
& & 0.19599(4)$^{+4.4\%}_{-4.7\%}$
& 0 
& & 0 \\
\nloplusgg{} 
& 0.25374(7)$^{+3.5\%}_{-3.7\%}$
& & 0.20773(5)$^{+3.2\%}_{-3.1\%}$
& $+\phantom{0}3.3\%$ 
& & $+\phantom{0}6.0\%$ \\
NNLO     
& 0.2378(4)\phantom{0}$^{+1.3\%}_{-0.9\%}$
& & 0.1907(3)\phantom{0}$^{+1.2\%}_{-0.9\%}$
& $-\phantom{0}3.2\%$ 
& & $-\phantom{0}2.7\%$ \\
\hline
\end{tabular}
\end{center}
\renewcommand{\baselinestretch}{1.0}
\caption{\label{tablewwacc} 
Efficiency of \ww{} acceptance cuts at different perturbative orders and 
relative differences with respect to NLO. Scale uncertainties and errors as in \refta{tableincl}.} 
\end{table}

\renewcommand{\baselinestretch}{1.0}

Predictions for fiducial cross sections at different perturbative orders are 
reported in \refta{tableSignal}. As a result of fiducial cuts, in particular 
the jet veto, radiative corrections behave very differently 
as compared to the inclusive case. 
The NLO corrections with respect to LO amount to only about $+4\%\,(+1\%)$ at $8\,(13)$\,TeV.
Neglecting the $+2\%\,(+3\%)$ shift due to the PDFs,
the NNLO corrections amount to $+5\%\,(+7\%)$. Their positive impact is, however, entirely due to the loop-induced
gluon-fusion contribution, which is not affected by the jet veto. In fact, comparing the NNLO and \nloplusgg{} 
predictions we see that
the genuine ${\cal O}(\as^2)$ corrections are negative and amount to roughly $-1\%\,(-2\%)$.

The reduction of the impact of radiative corrections when a jet veto is applied is a
well-known feature in perturbative QCD calculations~\cite{Catani:2001cr}.
A stringent veto on the radiation recoiling against the \ww{}
system 
tends to unbalance the cancellation
between positive real and negative virtual contributions, possibly leading to 
large logarithmic terms.
The resummation of such logarithms has been the subject of intense 
theoretical studies, especially in the important case of Higgs-boson 
production~\cite{Banfi:2012jm,Becher:2012qa,Stewart:2013faa},
and it has been recently addressed also for \ww{} 
production~\cite{Jaiswal:2014yba,Becher:2014aya}.
In the case at hand, the moderate size of radiative effects beyond NLO suggests that,
similarly as for Higgs production, fixed-order NNLO predictions 
should provide a fairly reliable description of jet-vetoed fiducial cross sections and distributions.

The reduced impact of radiative effects in the presence of a jet veto is often accompanied
by a reduction of scale uncertainties in fixed-order perturbative calculations.
Comparing the results in \refta{tableSignal}
with those in \refta{tableincl} we indeed see that the size of the NNLO scale uncertainty 
is reduced when cuts, particularly the jet veto, are applied. 
Such a small scale dependence should be interpreted with caution
as it tends to underestimate the
true uncertainty due to missing higher-order 
perturbative contributions.

The effect of radiative corrections on the efficiency of \ww{} fiducial cuts,
\begin{equation}
\label{acceptance}
\efficiency=\sigmafid/\sigmainc\,,
\end{equation}
is illustrated in \refta{tablewwacc}, where numerator and denominator are
evaluated at the same 
perturbative order and both with $\muR=\muF=m_W$.
Due to the negative 
impact of the newly computed NNLO corrections on the fiducial cross section (see 
\refta{tableSignal}) and their positive impact on the 
inclusive cross section (see \refta{tableincl}),
 the \nnlo{} corrections on the cut efficiency are quite significant. In particular,
at $\sqrt{s}=8\,(13)$\,TeV the
full \nnlo{} prediction lies about $6\%\,(9\%)$ below 
the \nloplusgg{} result. The uncertainties quoted 
in \refta{tablewwacc} are obtained by varying
$\muR$ and $\muF$ in a fully correlated way
in the numerator and denominator of~\refeq{acceptance}. Clearly, there is a large 
correlation at \lo{}, which results in a particularly small uncertainty. At 
NNLO the uncertainties are comparable to those of the fiducial cross sections.

\begin{figure}
\begin{center}
\begin{tabular}{cc}
\hspace*{-0.17cm}
\includegraphics[trim = 7mm -7mm 0mm 0mm, width=.33\textheight]{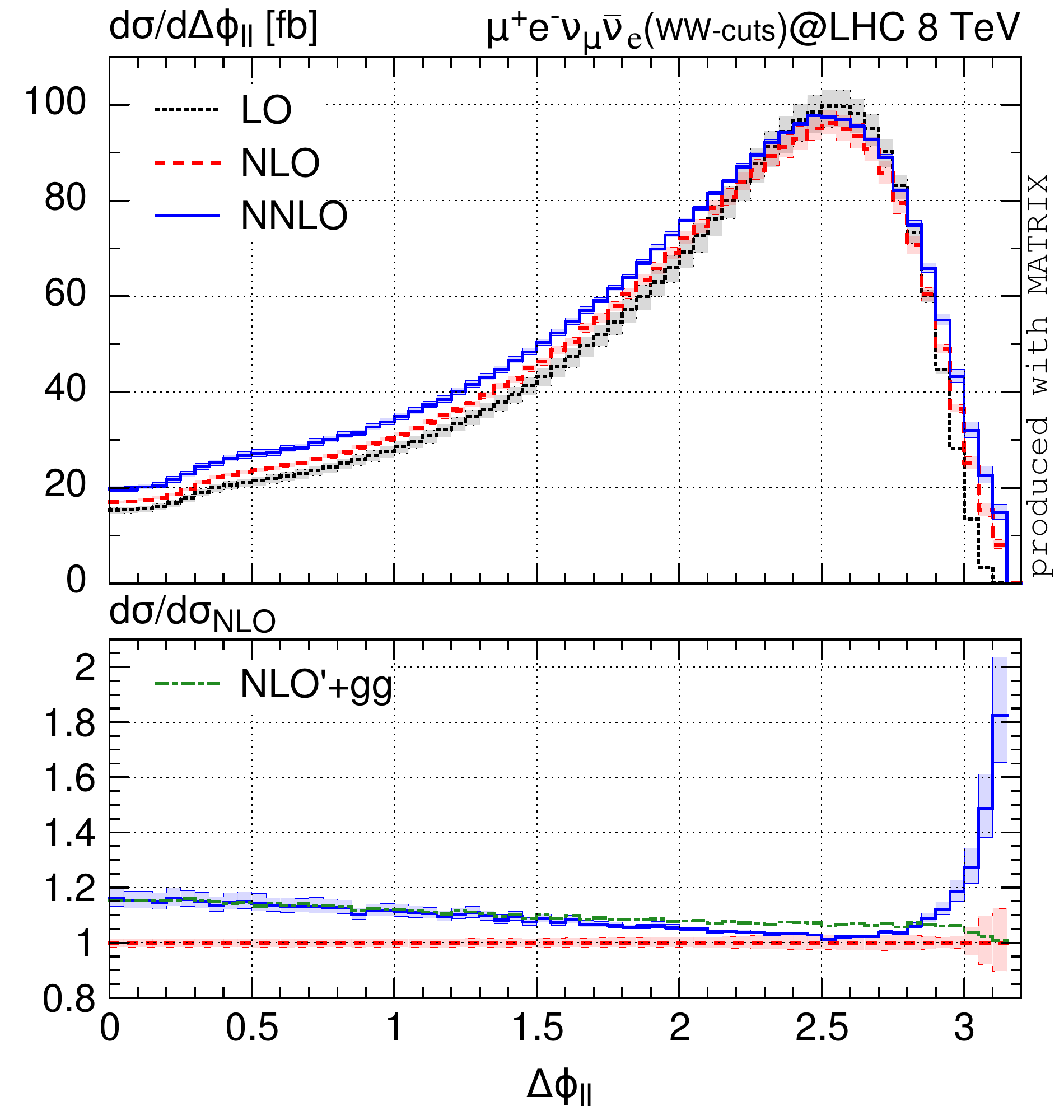} &
\includegraphics[trim = 7mm -7mm 0mm 0mm, width=.33\textheight]{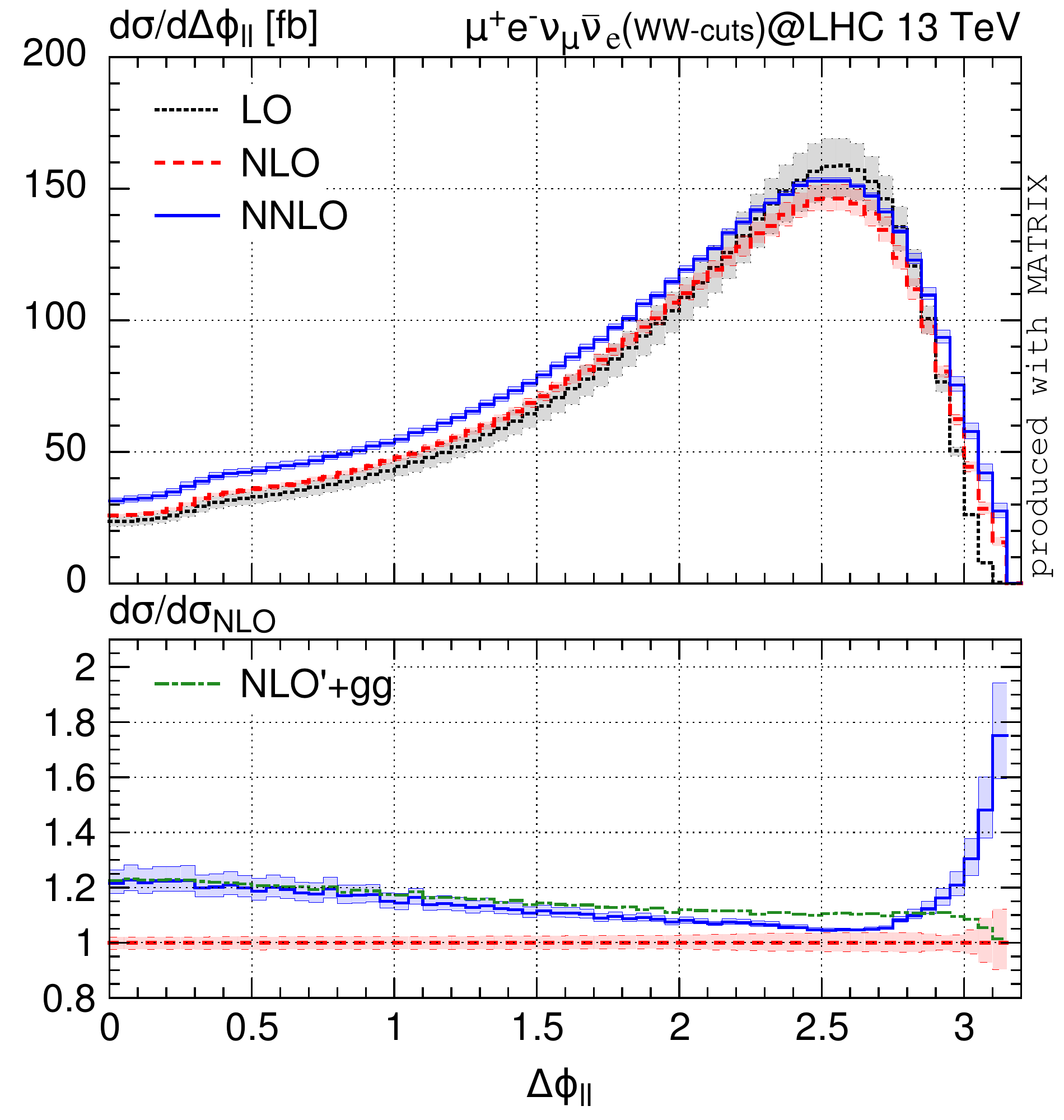} \\[-1em]
\hspace{0.6em} (a) & \hspace{1em}(b)
\end{tabular}
\caption[]{\label{fig:wwdphill}{
Distribution in the azimuthal separation of the charged leptons. \ww{} cuts are applied.  
Absolute predictions and relative corrections as in~\reffi{fig:mWWinclusive}.}}
\end{center}
\vspace{0.5cm}
\begin{center}
\begin{tabular}{cc}
\hspace*{-0.17cm}
\includegraphics[trim = 7mm -7mm 0mm 0mm, width=.33\textheight]{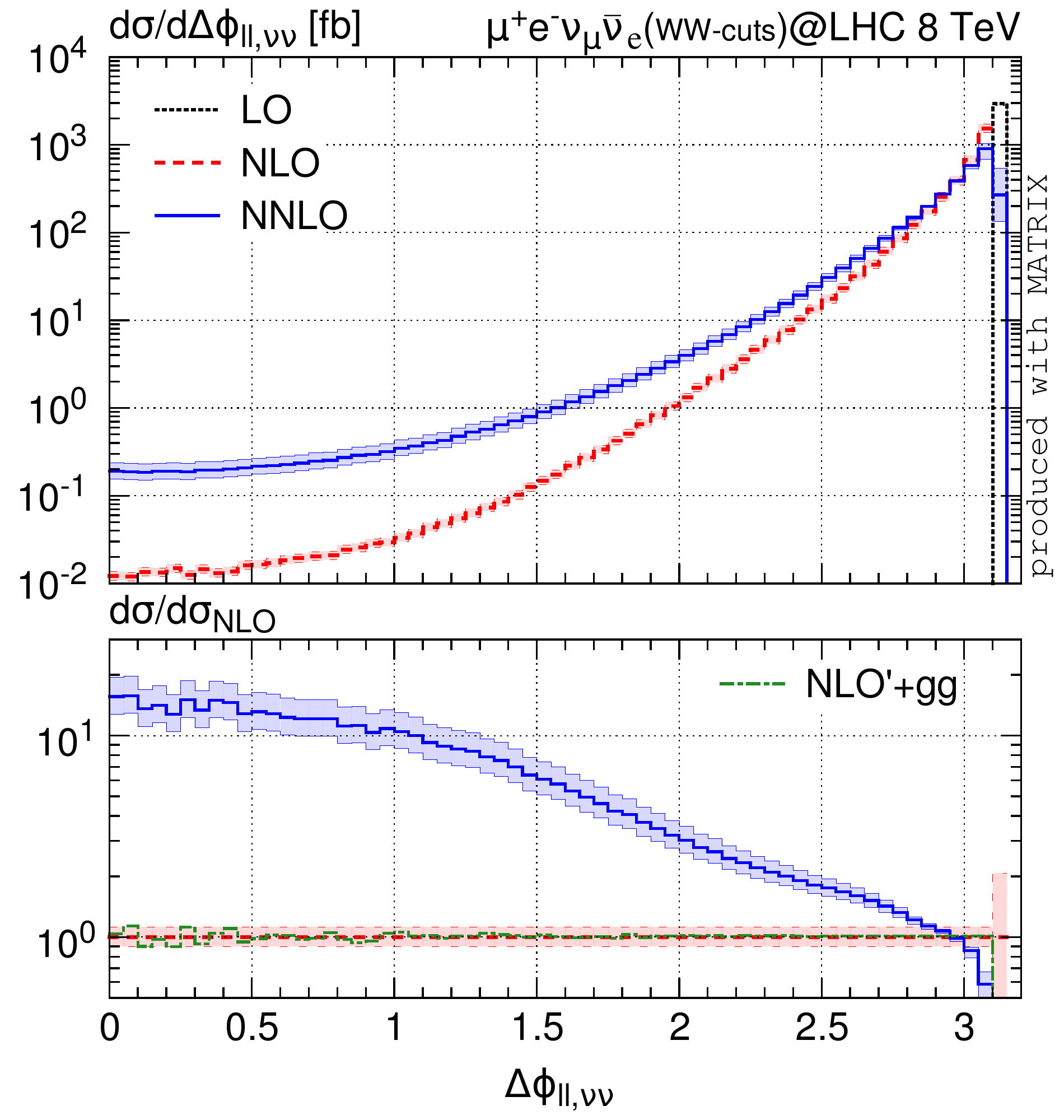} &
\includegraphics[trim = 7mm -7mm 0mm 0mm, width=.33\textheight]{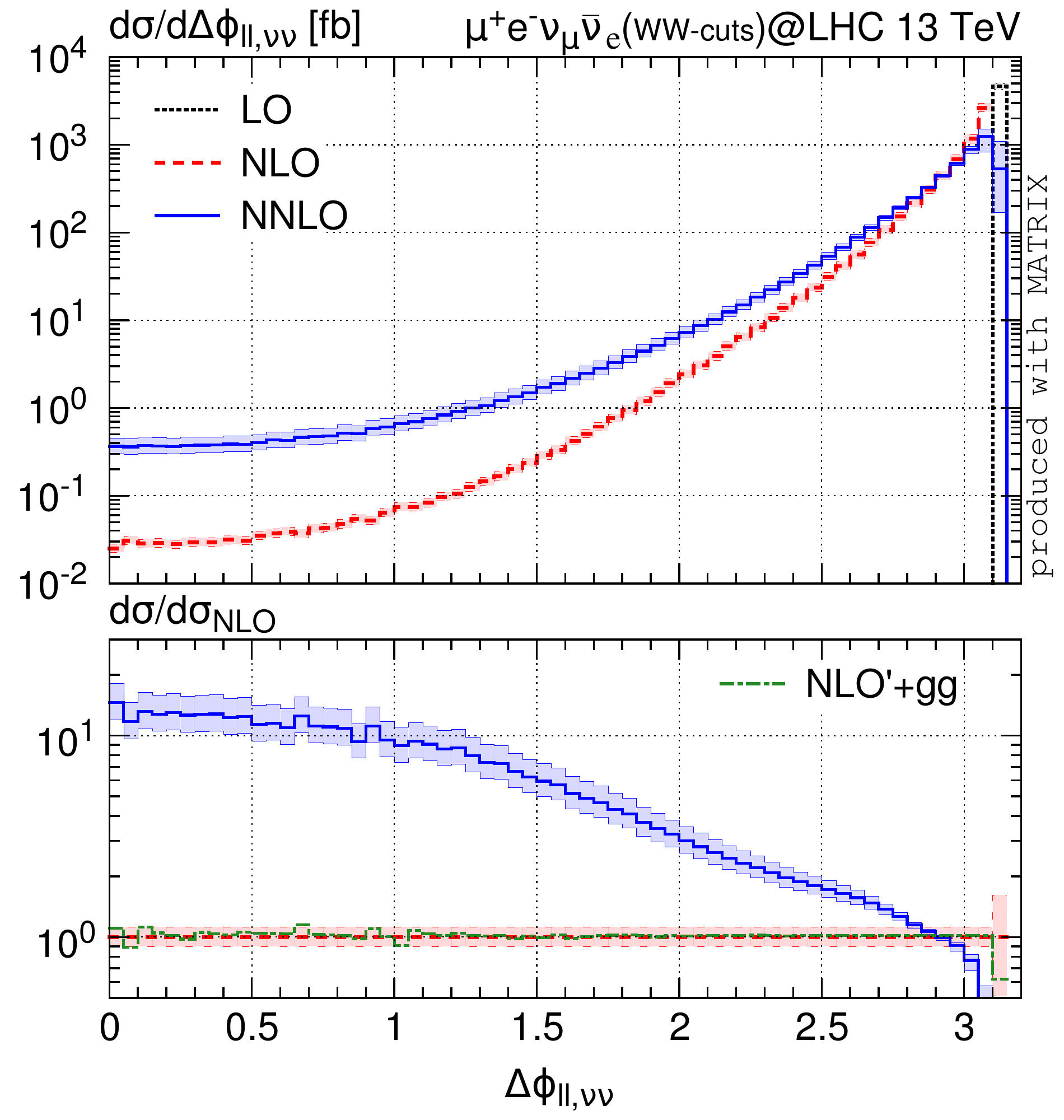} \\[-1em]
\hspace{0.6em} (a) & \hspace{1em}(b)
\end{tabular}
\caption[]{\label{fig:wwdphillnunu}{
Distribution in the azimuthal separation between the
transverse momentum of the dilepton system and 
the missing transverse momentum. \ww{} cuts are applied.  
Absolute predictions and relative corrections as in~\reffi{fig:mWWinclusive}.}}
\end{center}
\end{figure}

\begin{figure}[tp]
\begin{center}
\begin{tabular}{cc}
\hspace*{-0.17cm}
\includegraphics[trim = 7mm -7mm 0mm 0mm, width=.33\textheight]{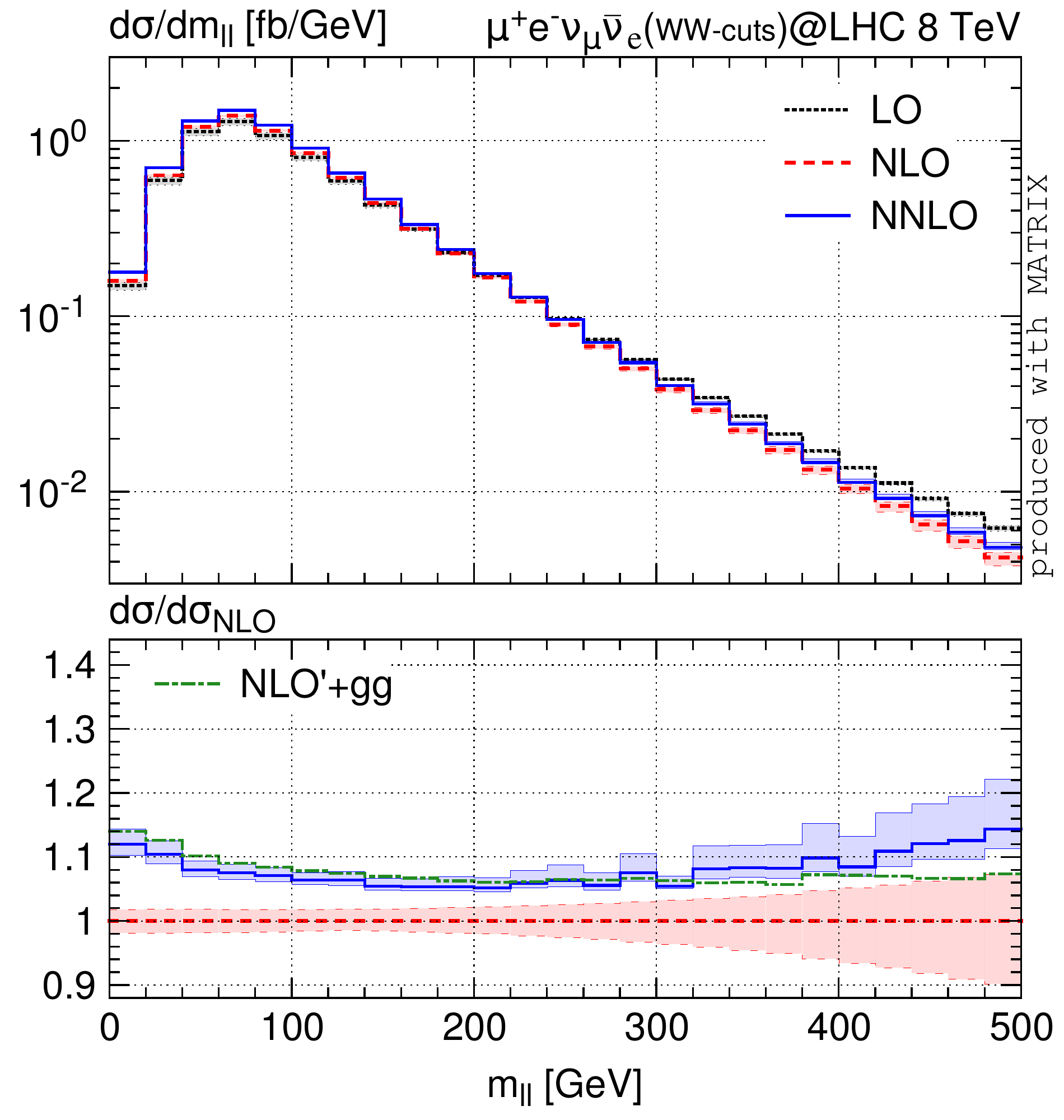} &
\includegraphics[trim = 7mm -7mm 0mm 0mm, width=.33\textheight]{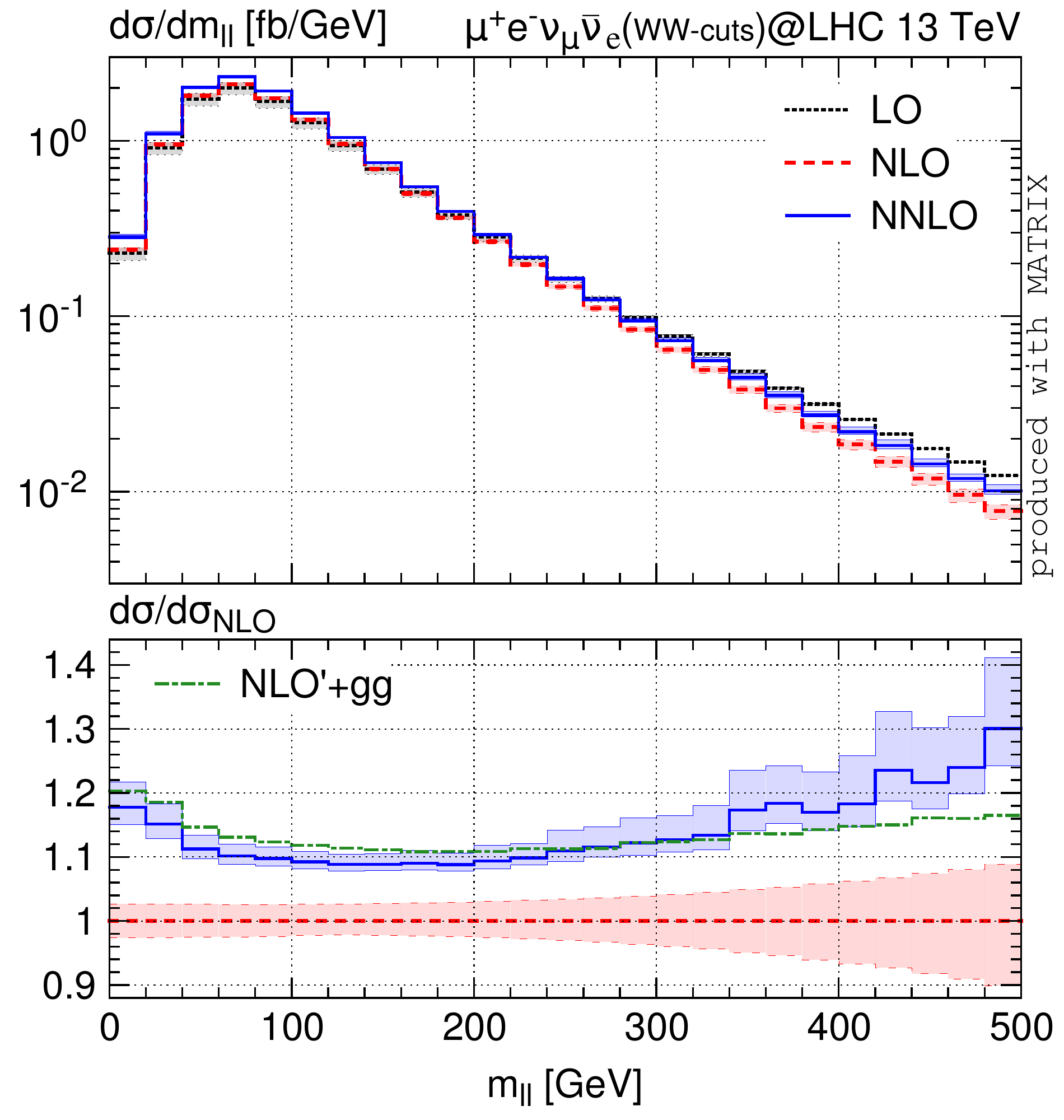} \\[-1em]
\hspace{0.6em} (a) & \hspace{1em}(b)
\end{tabular}
\caption[]{\label{fig:wwmll}{
Distribution in the  dilepton invariant mass. \ww{} cuts are applied.  
Absolute predictions and relative corrections as in~\reffi{fig:mWWinclusive}.}}
\end{center}
\end{figure}

As discussed in
\sct{subsec:top}, our default 4FS predictions are compared to 
an alternative top-subtracted computation in the \fs{5},
in order to assess the uncertainty related to the prescription for
the subtraction of the top contamination.
Without top subtraction we obtain 
$\sigma_{\rm NLO}=165.7(3)$\,fb and $\sigma_{\rm NNLO}=181.9(4)$\,fb
at $\sqrt{s}=8$\,TeV in the \fs{5}. Due to the jet veto, these 
fiducial cross sections feature a moderate top contamination of about
$8\%$ at NLO and $12\%$ at NNLO. Removing the top contributions, we find $\sigma_{\rm NLO}=153.4(4)$\,fb and 
$\sigma_{\rm NNLO}=162.5(3)$\,fb, which agree with the \fs{4} results within $1\%$.
At $\sqrt{s}=13$\,TeV, 
the top contamination in the \fs{5} is somewhat larger and 
amounts to $12\%\,(17\%)$ at \nlo{}\,(\nnlo{}). The top-subtracted fiducial cross sections, 
$\sigma_{\rm NLO}=238.3(6)$\,fb and $\sigma_{\rm NNLO}=265(2)$\,fb, on the other hand, are again in agreement with 
the \fs{4} results at the $1\%-2\%$ level.

Differential distributions in presence of \ww{} fiducial cuts are presented in 
\reffis{fig:wwdphill}{fig:wwptmiss}.
We first consider, in \reffi{fig:wwdphill}, 
the distribution in the azimuthal separation   of the charged leptons, $\dphill$.
The \nloplusgg{} approximation 
is in good agreement with full NNLO result at small $\dphill$, but
in the peak region the difference exceeds $5\%$, 
and the \nloplusgg{} result lies outside the \nnlo{} uncertainty band. The difference 
significantly increases in the large $\dphill$ region, where the cross 
section is strongly suppressed though.
The uncertainty bands of the NLO and NNLO predictions 
do not overlap. This feature is common to all distributions that are considered in the following. 
It is primarily caused by the loop-induced $gg$ contribution, which enters only at NNLO and is
not accounted for by the NLO scale variations. Ignoring the gluon-induced 
component, we observe a good perturbative convergence, apart from some peculiar phase-space corners.

In \reffi{fig:wwdphillnunu} we study the cross section as a function of
the azimuthal separation $\dphillnunu{}$ between the transverse momentum of the 
dilepton pair ($\mathbf\ptll$) and the missing transverse momentum ($\mathbf\ptmiss{}$).
Since $\dphillnunu{}=\pi$ at LO, the (N)NLO calculation is 
only (N)LO accurate at $\dphillnunu{}<\pi$. The NNLO corrections have a 
dramatic impact on the shape of the distribution:
The $\sigma_\nnlo/\sigma_\nlo$ $K$-factor grows with decreasing $\dphillnunu{}$ and reaches
up to  ${\cal O}(10)$ in the region $\dphillnunu{}\lesssim 1$, where the cross section is suppressed by more 
than three orders of magnitude.
This huge effect results from the 
interplay of the jet veto with the cuts on the $\pt$'s of the individual leptons and on \ptmiss{}.
At small $\dphillnunu{}$ the transverse momenta $\mathbf\ptll$ and 
$\mathbf\ptmiss{}$ must be balanced by recoiling QCD partons.
However, at NLO the emitted parton 
can deliver a sizeable recoil only in the 
region that is not subject to the jet veto, i.e.\ in the
strongly suppressed rapidity range $|y_j|>4.5$.
At NNLO, the presence of a second parton relaxes this 
restriction to some extent, thereby reducing the suppression 
by about one order of magnitude. 
The loop-induced $gg$ contribution does not involve any QCD radiation and
contributes only at $\dphillnunu{}=\pi$. As a consequence, the NLO and \nloplusgg{} predictions 
at $\dphillnunu{}<\pi$ are almost identical, apart from 
minor differences due to the PDFs.

\begin{figure}
\begin{center}
\begin{tabular}{cc}
\hspace*{-0.17cm}
\includegraphics[trim = 7mm -7mm 0mm 0mm, width=.33\textheight]{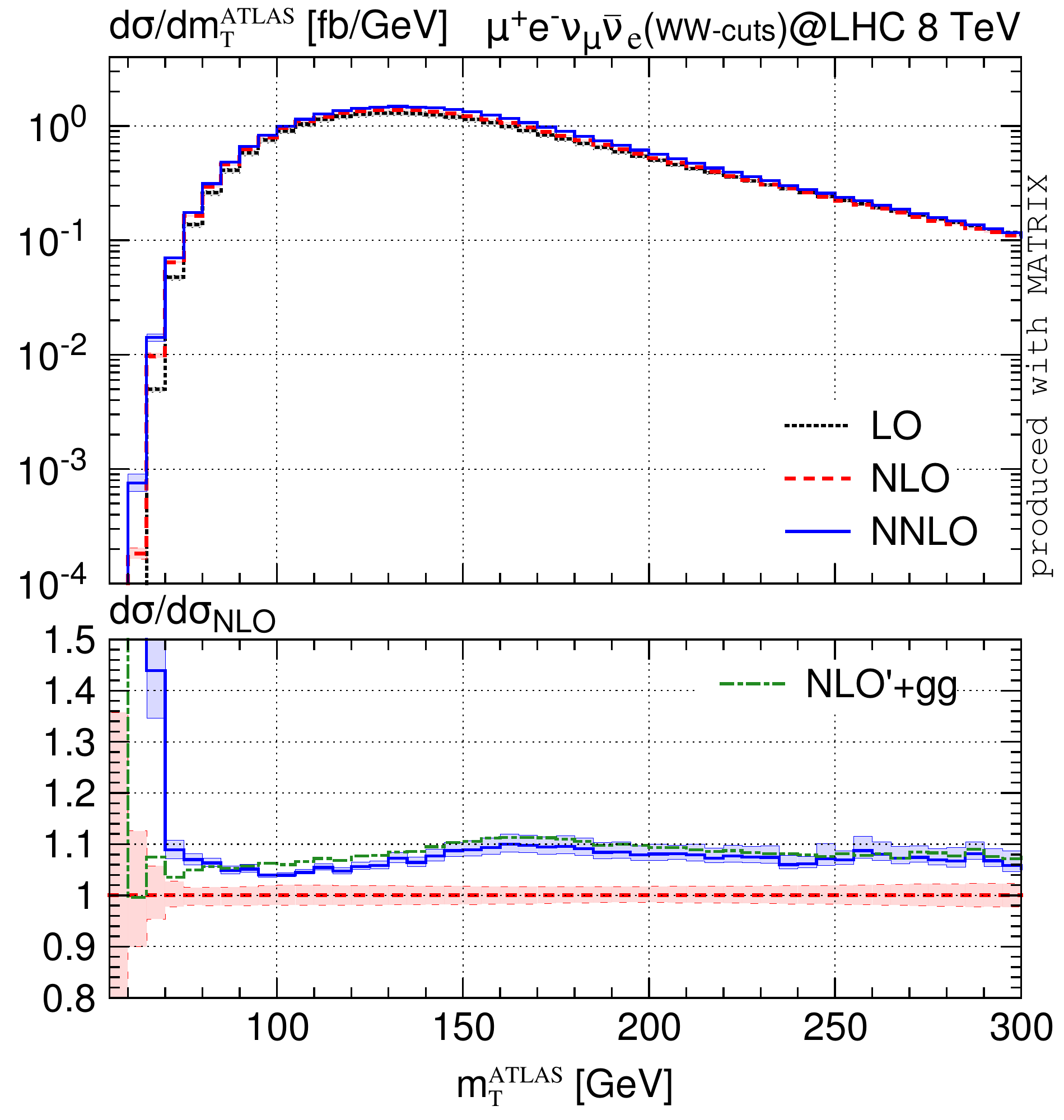} &
\includegraphics[trim = 7mm -7mm 0mm 0mm, width=.33\textheight]{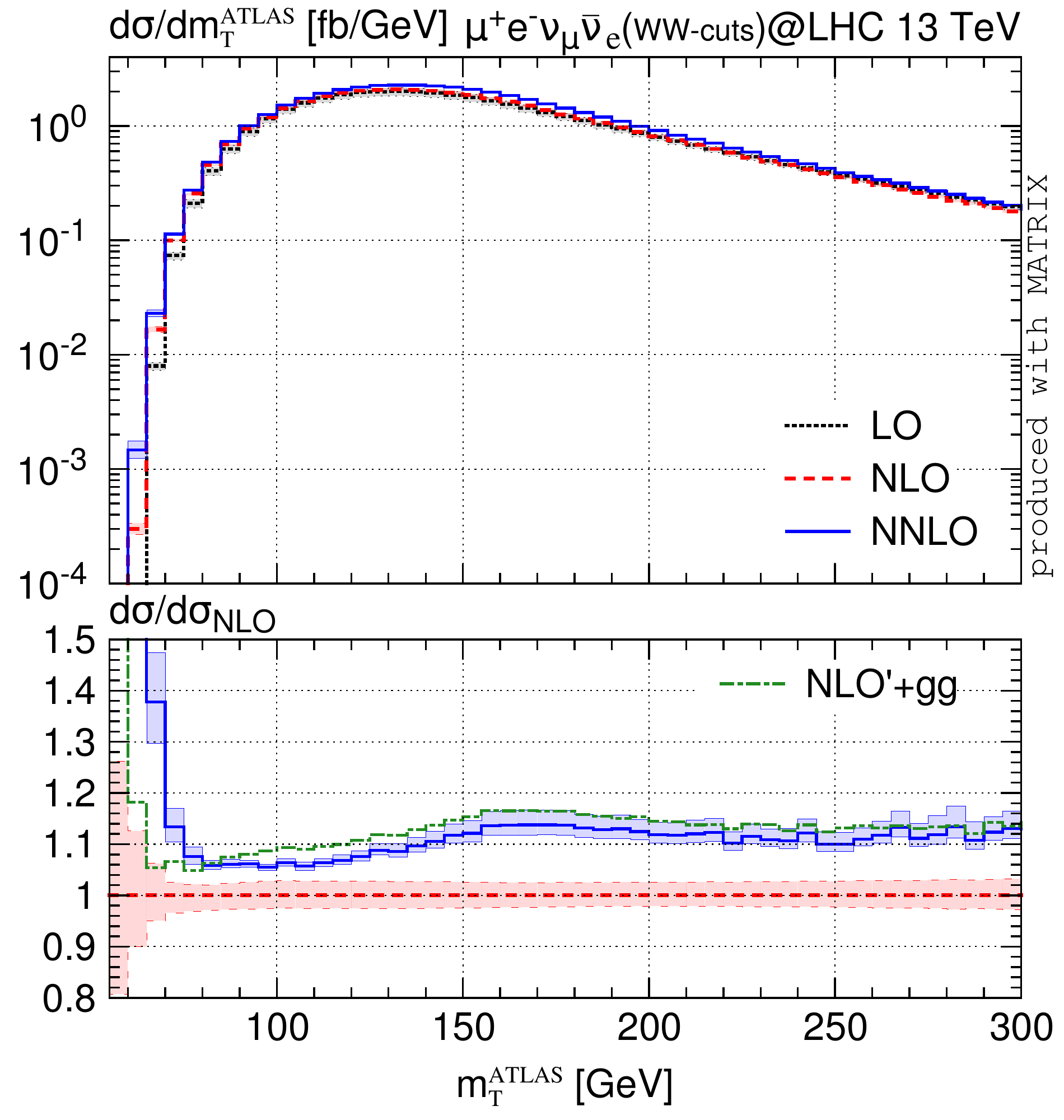} \\[-1em]
\hspace{0.6em} (a) & \hspace{1em}(b)
\end{tabular}
\caption[]{\label{fig:wwmt}{
Distribution in the \ww{} transverse mass. \ww{} cuts are applied.  
Absolute predictions and relative corrections as in~\reffi{fig:mWWinclusive}.}}
\end{center}
\end{figure}

The invariant-mass distribution of the dilepton pair is presented in \reffi{fig:wwmll}. On the one hand,
if one takes into account NNLO scale variations, the \nloplusgg{} result is by and large consistent with the \nnlo{} prediction.
On the other hand, the shapes of the \nloplusgg{} and \nnlo{} distributions
feature non-negligible differences, which range from $+5\%$ at low masses to $-5\%$ in the high-mass tail. 
Nevertheless, \nloplusgg{} provides a reasonable approximation 
of the full \nnlo{} result, in particular regarding the normalization.

The distribution in the \ww{} transverse mass,
\begin{equation}
  m_T^{\rm ATLAS}= \sqrt{\left(\Etlone+\Etltwo+\ptmiss\right)^2-\left({\bf p}_{T,l_1}+{\bf p}_{T,l_2}+
    {\ptmiss}\right)^2}\,,
\end{equation}
is displayed in \reffi{fig:wwmt}. 
Also in this case, apart from the strongly suppressed region of small $m_T^{\rm ATLAS}$,
the \nloplusgg{} approximation is in quite good agreement with the full \nnlo{} prediction.

\begin{figure}
\begin{center}
\begin{tabular}{cc}
\hspace*{-0.17cm}
\includegraphics[trim = 7mm -7mm 0mm 0mm, width=.33\textheight]{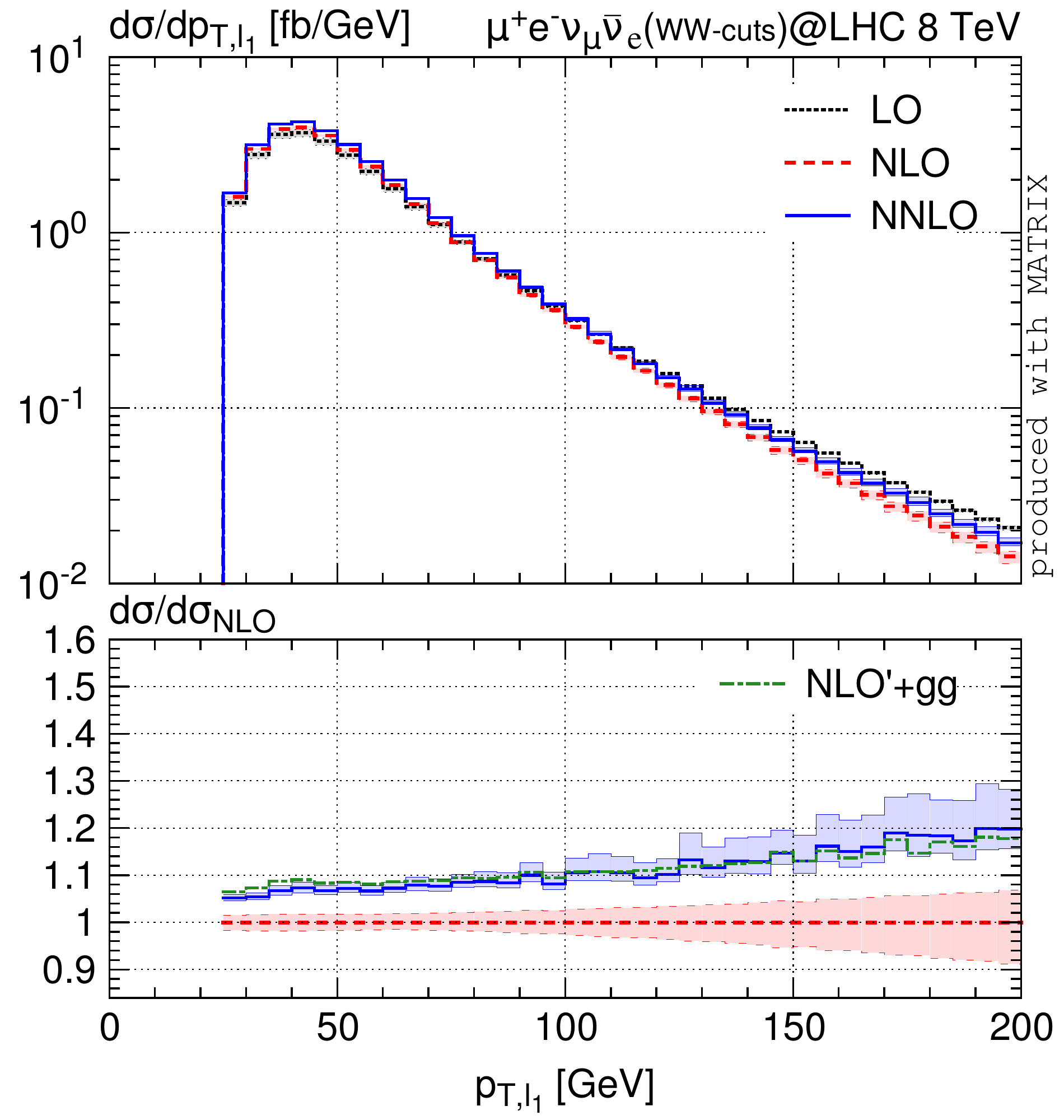} &
\includegraphics[trim = 7mm -7mm 0mm 0mm, width=.33\textheight]{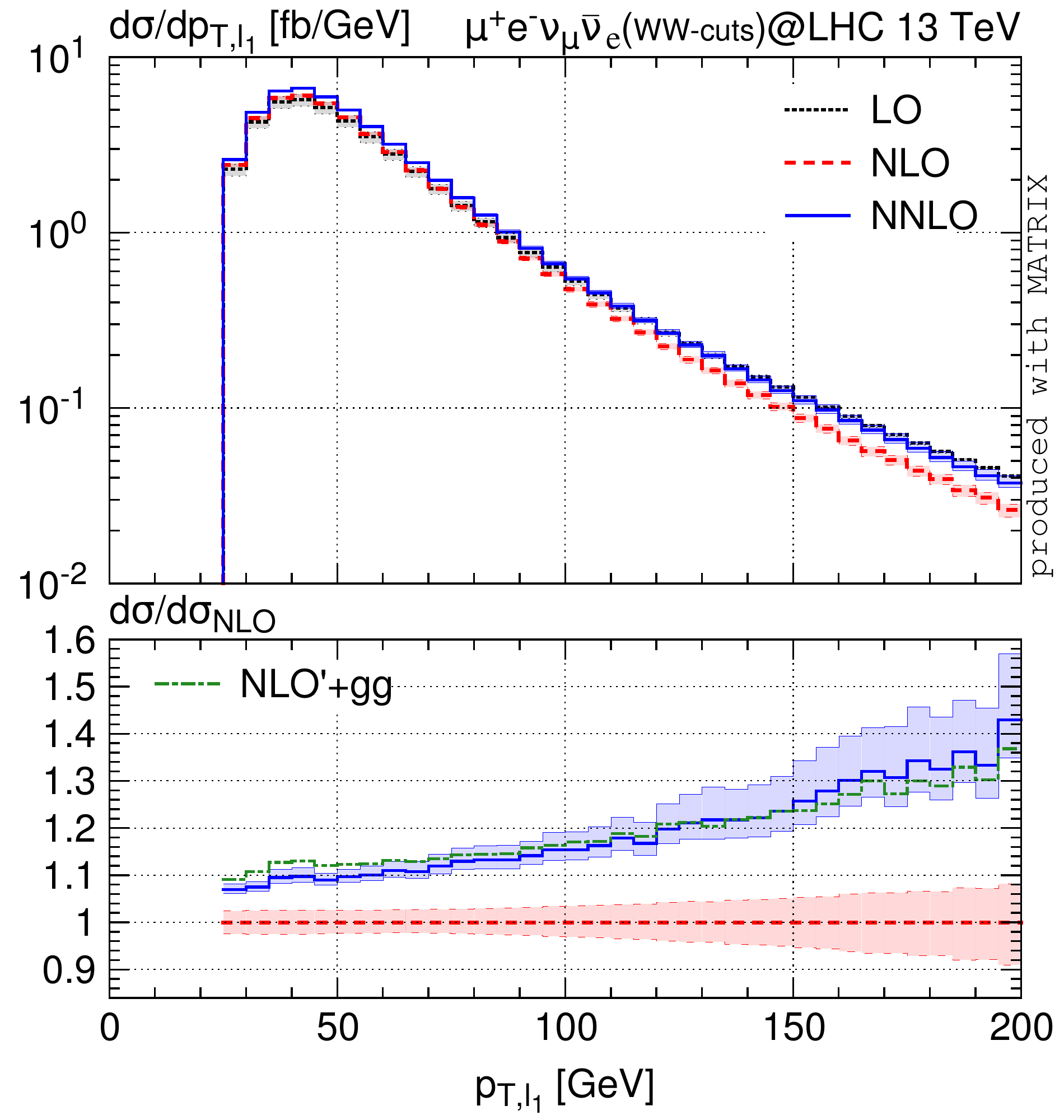} \\[-1em]
\hspace{0.6em} (a) & \hspace{1em}(b)
\end{tabular}
\caption[]{\label{fig:wwptl1}{
Distribution in the  $p_T$ of the leading lepton. \ww{} cuts are applied. 
Absolute predictions and relative corrections as in~\reffi{fig:mWWinclusive}.}}
\end{center}
\vspace{0.5cm}
\begin{center}
\begin{tabular}{cc}
\hspace*{-0.17cm}
\includegraphics[trim = 7mm -7mm 0mm 0mm, width=.33\textheight]{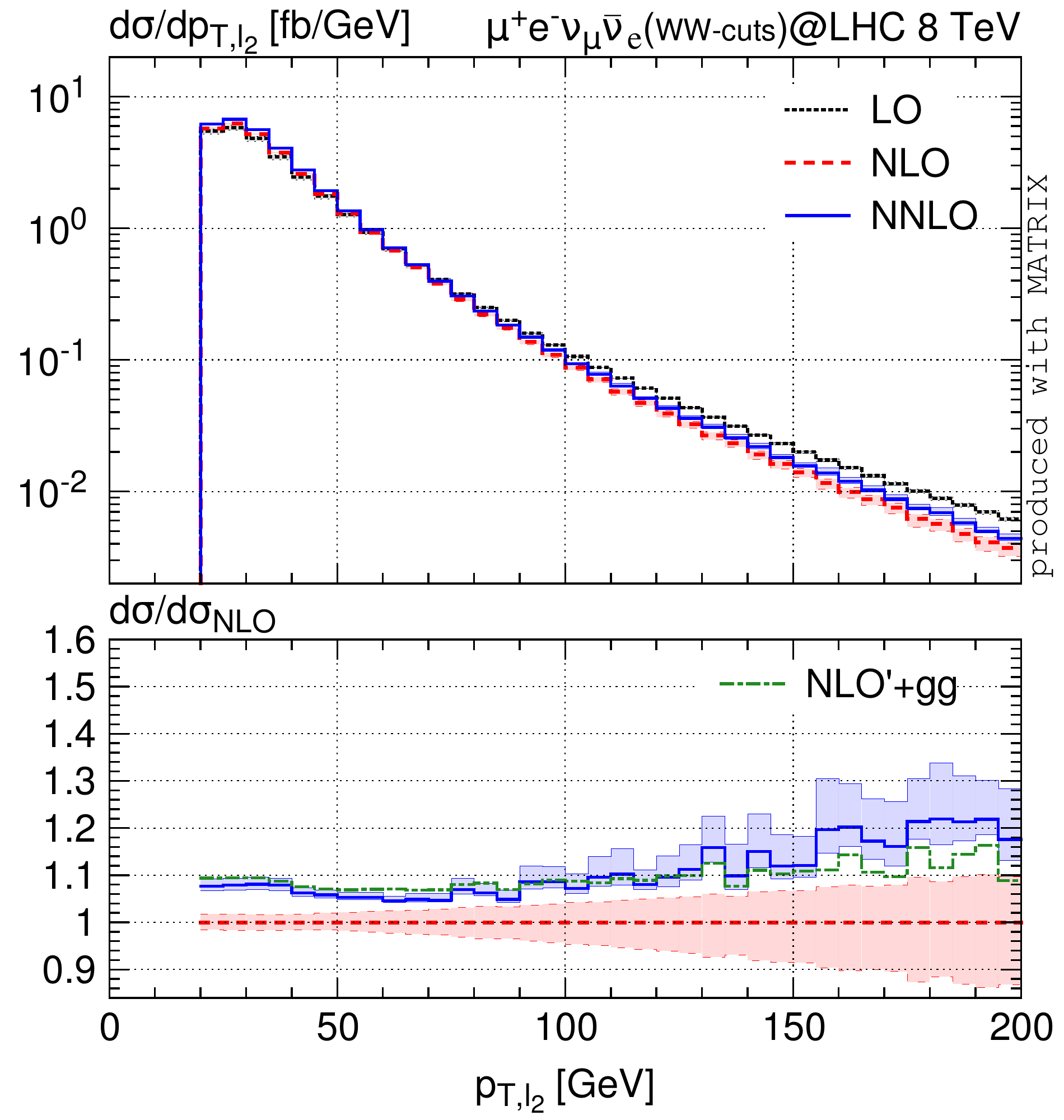} &
\includegraphics[trim = 7mm -7mm 0mm 0mm, width=.33\textheight]{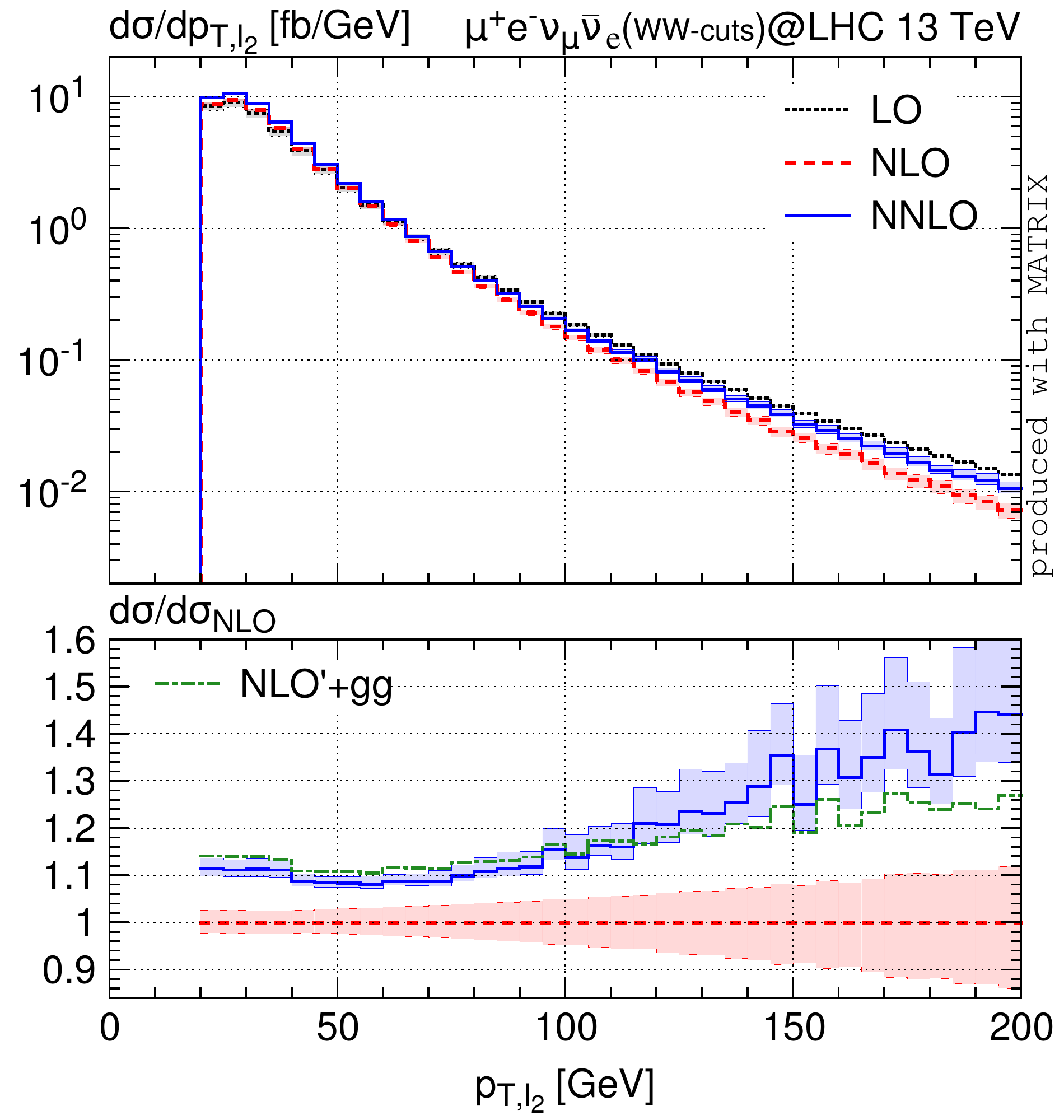} \\[-1em]
\hspace{0.6em} (a) & \hspace{1em}(b)
\end{tabular}
\caption[]{\label{fig:wwptl2}{
Distribution in the  $p_T$ of the subleading lepton. \ww{} cuts are applied.  
Absolute predictions and relative corrections as in~\reffi{fig:mWWinclusive}.}}
\end{center}
\end{figure}

\begin{figure}
\begin{center}
\begin{tabular}{cc}
\hspace*{-0.17cm}
\includegraphics[trim = 7mm -7mm 0mm 0mm, width=.33\textheight]{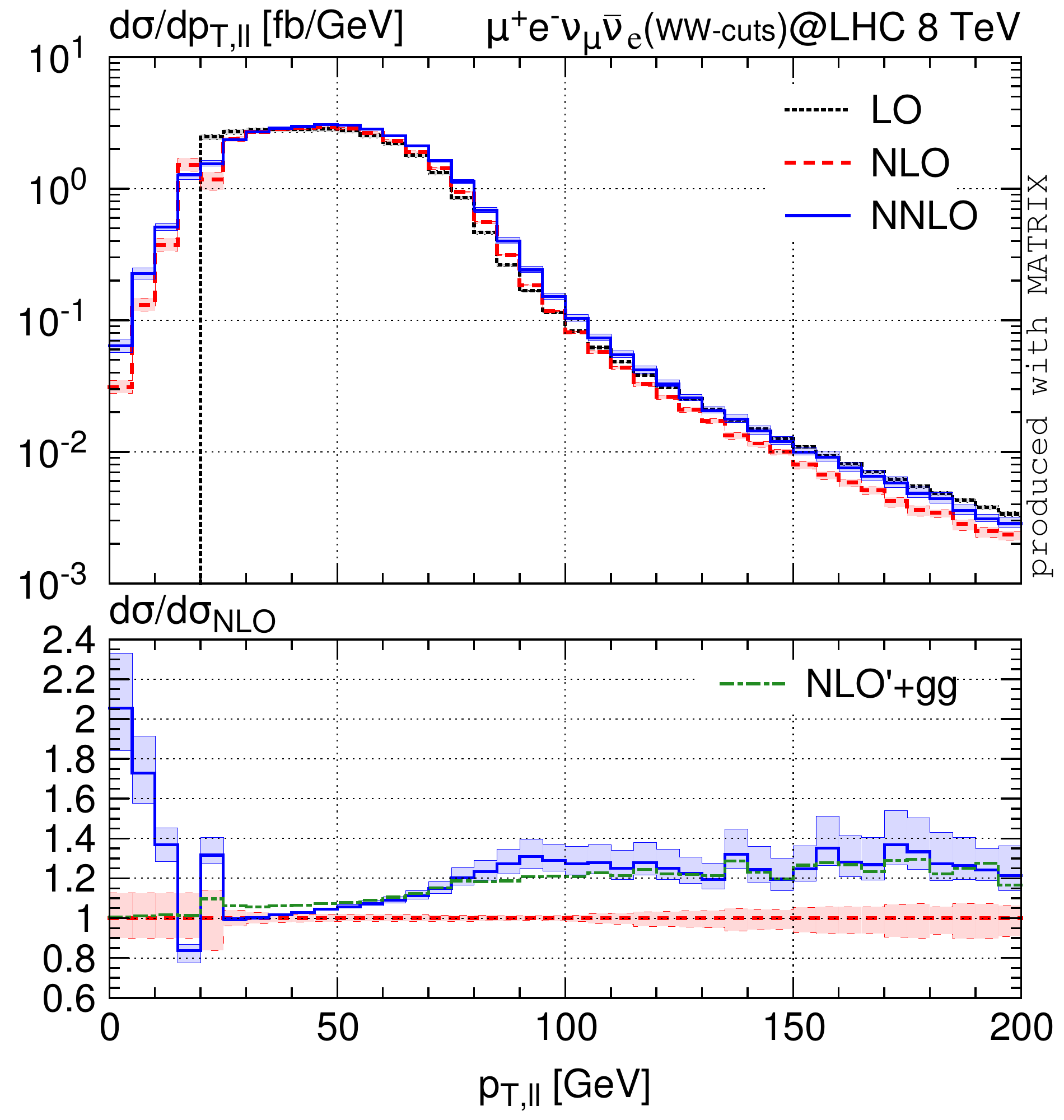} &
\includegraphics[trim = 7mm -7mm 0mm 0mm, width=.33\textheight]{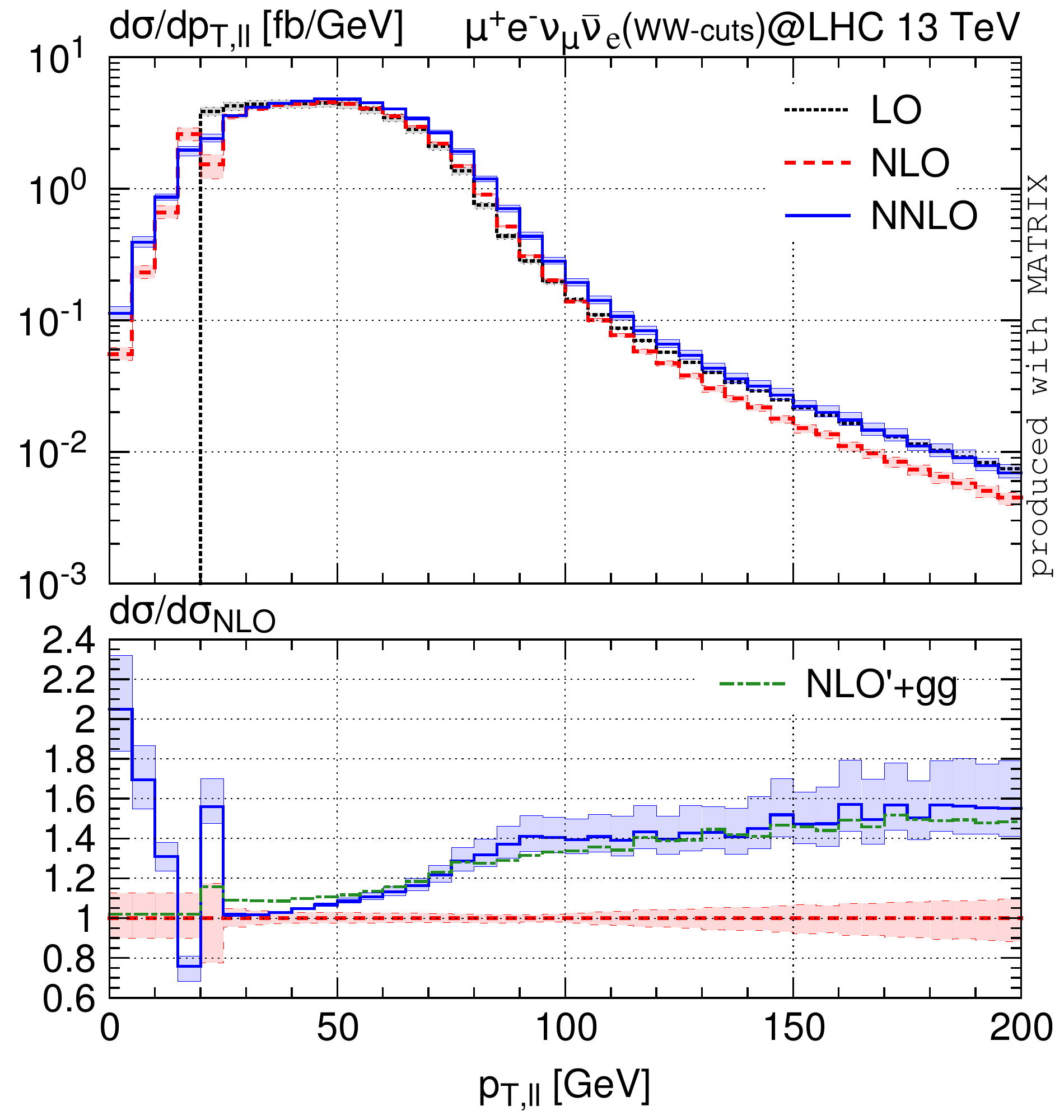} \\[-1em]
\hspace{0.6em} (a) & \hspace{1em}(b)
\end{tabular}
\caption[]{\label{fig:wwptll}{
Distribution in the  $p_T$ of the 
dilepton system. \ww{} cuts are applied.  
Absolute predictions and relative corrections as in~\reffi{fig:mWWinclusive}.}}
\end{center}
\vspace{0.5cm}
\begin{center}
\begin{tabular}{cc}
\hspace*{-0.17cm}
\includegraphics[trim = 7mm -7mm 0mm 0mm, width=.33\textheight]{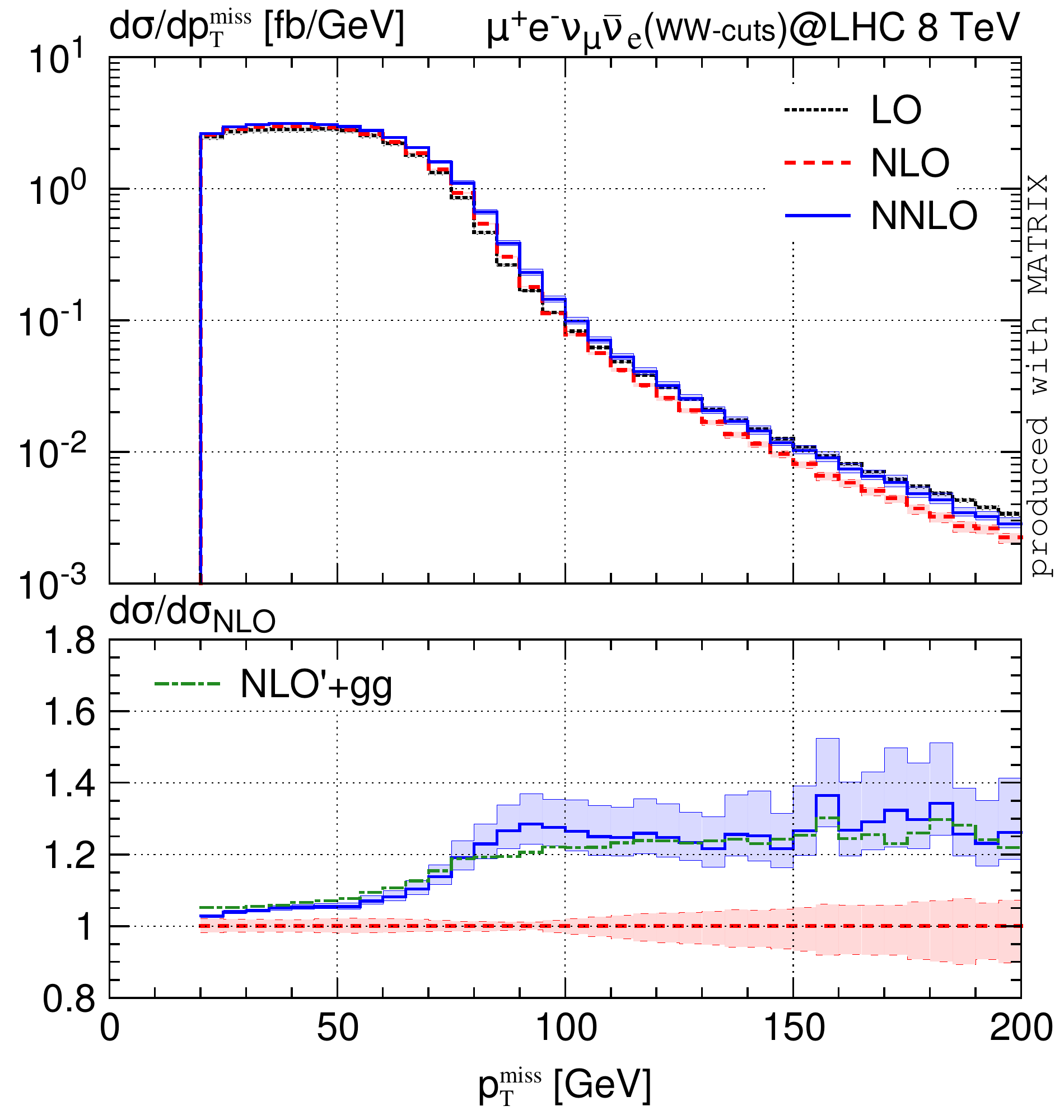} &
\includegraphics[trim = 7mm -7mm 0mm 0mm, width=.33\textheight]{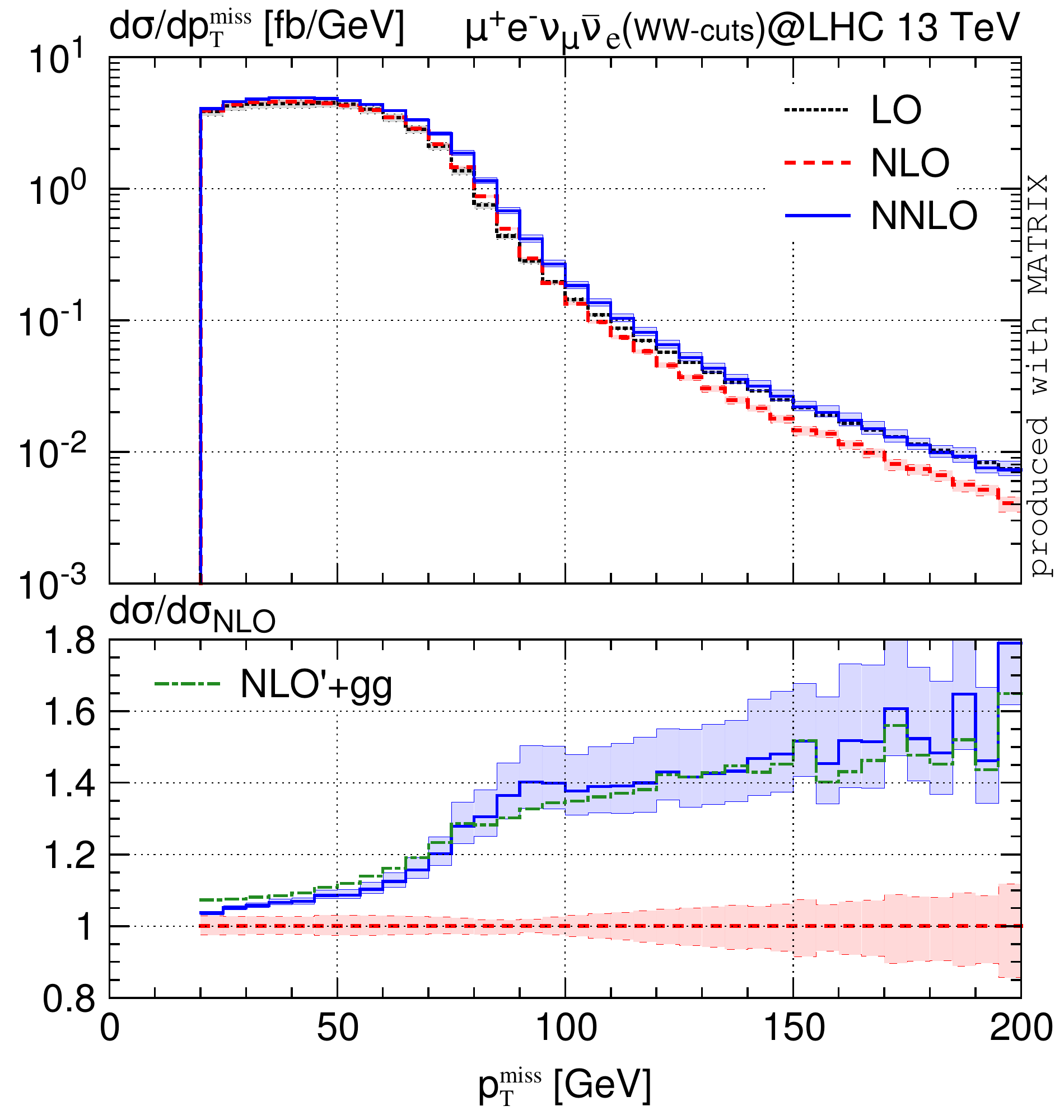} \\[-1em]
\hspace{0.6em} (a) & \hspace{1em}(b)
\end{tabular}
\caption[]{\label{fig:wwptmiss}{
Distribution in the missing transverse momentum. \ww{} cuts are applied.  
Absolute predictions and relative corrections as in~\reffi{fig:mWWinclusive}.}}
\end{center}
\end{figure}

In \reffitwo{fig:wwptl1}{fig:wwptl2} we show results for the 
$p_T$ distributions of the leading and subleading lepton, respectively. In both cases the 
impact of NNLO corrections grows with $p_T$. This is driven
by the gluon-induced contribution, which overshoots the complete NNLO result in the small-$p_T$ region 
and behaves in the opposite way
as \pt{} becomes large.
In the case of the subleading lepton, the genuine NNLO 
corrections are as large as ${\cal O}(10\%)$ around $p_{T,l_2}=200$\,GeV. Overall, 
there is a visible difference in shape between \nloplusgg{} and \nnlo{} for both 
the leading and subleading lepton transverse-momentum distributions.

The $p_T$ distribution of the dilepton pair is displayed in~\reffi{fig:wwptll}. 
This observable  has a kinematical boundary at LO, where the requirement $\ptmiss>20$ GeV implies that $\ptll>20$ GeV.
The region $\ptll<20$ GeV starts to be populated at NLO, but each perturbative
higher-order contribution (beyond LO) produces integrable logarithmic singularities
leading to perturbative instabilities 
at the boundary \cite{Catani:1997xc}.
This becomes particularly evident in the $d\sigma_\nnlo/d\sigma_\nlo$ ratio.
The loop-induced $gg$ contribution, having Born-like
kinematics, does not contribute to the region $\ptll<20$ GeV. In contrast, 
\nnlo{} corrections are huge, and the formal accuracy of NNLO predictions is only NLO in that region.
In the region of high $\ptll$ we observe significant NNLO corrections, and the
\nloplusgg{} approximation works rather well.
Similar features are 
observed in the \ptmiss{} distribution,
displayed in \reffi{fig:wwptmiss}, but without the perturbative instability at $\ptmiss=20$\,GeV, as the cut on $\ptmiss$ is explicit.

In general, radiative corrections behave in a rather similar way at $\sqrt{s}=8$\,TeV and $\sqrt{s}=13$\,TeV
in presence of \ww{} cuts.
Comparing the \nloplusgg{} approximation 
with the full \nnlo{} prediction, we find that the overall normalization is typically 
reproduced quite well, while genuine NNLO corrections can lead to significant shape differences
of up to $10\%$.
It does not come as a surprise that in kinematic regions that imply the presence of QCD radiation, 
loop-induced $gg$ contributions cannot provide 
a reasonable approximation of the full \nnlo{} correction.

\subsection{Analysis of \bld{\muenn} production with Higgs selection cuts}
\label{sec:results-higgs}

In this Section we repeat our study of radiative corrections
in presence of cuts that are designed for $H\to\ww{}$ studies at the LHC.
In this case, \ww{} production plays the role of irreducible background, and 
more stringent cuts are applied in order to minimize its impact on the
$H\to\ww{}$ signal.
The precise list of  cuts is specified in~\refta{tablecuts}
and corresponds to the $H\to\ww{}$ analysis of~\citere{ATLAS:2014aga}\footnote{In our analysis, we require $|\ymu|<2.4$ as for \ww{} signal cuts, in contrast to $|\ymu|<2.5$ in the ATLAS analysis. Moreover, we do not apply any lepton-isolation criteria with respect to hadronic activity.}. 
This selection implements a series of 
cuts similar to the ones used in \ww{} signal measurements, including 
a jet veto. The suppression of on-shell \ww{} production is achieved through
additional restrictions on $p_{T,ll}$, $m_{ll}$, $\Delta\phi_{ll}$ and
$\Delta\phi_{ll,\nu\nu}$.

\renewcommand{\baselinestretch}{1.2}
\begin{table}[t]
\begin{center}
\begin{tabular}{|l|ccc|ccc|}
\hline
& \multicolumn{3}{c|}{$\sigmafid$(H$-$cuts)\,[fb]} &\multicolumn{3}{c|}{$\sigma/\sigma_{\rm NLO}-1$} \\
\hline
\multicolumn{1}{|c|}{$\sqrt{s}$} & 8\,TeV  & & 13\,TeV & 8\,TeV & & 13\,TeV  \\
\hline
LO    
& 45.923(4)\,$^{+4.0\%}_{-5.0\%}$ 
& & 71.164\phantom{0}(7)\,$^{+7.2\%}_{-8.2\%}$ 
& $-\phantom{0}4.4\%$ 
& & $-\phantom{0}2.6\%$ \\
NLO   
& 48.045(5)\,$^{+1.9\%}_{-1.7\%}$ 
& & 73.085\phantom{0}(6)\,$^{+2.7\%}_{-2.4\%}$ 
& 0 
& & 0 \\
\nloprime{} 
& 49.318(7)\,$^{+1.7\%}_{-1.6\%}$ 
& & 75.578(11)\,$^{+2.5\%}_{-2.2\%}$ 
& $+\phantom{0}2.7\%$ 
& & $+\phantom{0}3.4\%$\\
\nloplusgg{} 
& 53.496(8)\,$^{+2.0\%}_{-1.5\%}$ 
& & 85.231(12)\,$^{+2.5\%}_{-2.5\%}$ 
& $+11.3\%$ 
& & $+16.6\%$ \\ 
NNLO  
& 52.30(4)\phantom{0}\,$^{+1.6\%}_{-1.0\%}$ 
& & 82.32(12)\phantom{0}\,$^{+2.4\%}_{-2.6\%}$ 
& $+\phantom{0}8.9\%$ 
& & $+12.6\%$\\
\hline
\end{tabular}
\end{center}
\renewcommand{\baselinestretch}{1.0}
\caption{\label{tableHiggs} 
Cross sections with Higgs fiducial cuts at different perturbative orders and 
relative differences with respect to NLO. Scale uncertainties and errors as in \refta{tableincl}.}

\renewcommand{\baselinestretch}{1.2}
\begin{center}
\begin{tabular}{|l|ccc|ccc|}
\hline
& \multicolumn{3}{c|}{$\efficiency=\sigmafid$(H$-$cuts)$/\sigmainc$} &\multicolumn{3}{c|}{$\efficiency/\efficiency_{\rm NLO}-1$} \\
\hline
\multicolumn{1}{|c|}{$\sqrt{s}$} & 8\,TeV  & & 13\,TeV & 8\,TeV & & 13\,TeV  \\
\hline
LO       
& 0.10795\phantom{0}(2)$^{+1.2\%}_{-1.4\%}$
& & 0.09135\phantom{0}(2)$^{+1.5\%}_{-1.7\%}$
& $+40.1\%$ 
& & $+50.6\%$ \\
NLO      
& 0.07706\phantom{0}(2)$^{+4.3\%}_{-4.6\%}$
& & 0.06065\phantom{0}(1)$^{+4.3\%}_{-4.5\%}$
& 0 
& & 0 \\
\nloplusgg{} 
& 0.08157\phantom{0}(2)$^{+3.1\%}_{-3.1\%}$
& & 0.06623\phantom{0}(2)$^{+2.7\%}_{-2.5\%}$
& $+\phantom{0}5.9\%$ 
& & $+\phantom{0}9.2\%$ \\
NNLO     
& 0.07575(11)$^{+1.2\%}_{-0.8\%}$
& & 0.06005(14)$^{+1.1\%}_{-0.9\%}$
& $-\phantom{0}1.7\%$ 
& & $-\phantom{0}1.0\%$ \\
\hline
\end{tabular}
\end{center}
\renewcommand{\baselinestretch}{1.0}
\caption{\label{tablehacc} 
Efficiency of Higgs acceptance cuts at different perturbative orders and 
relative differences with respect to NLO. Scale uncertainties and errors as in \refta{tableincl}.}
\end{table}

\renewcommand{\baselinestretch}{1.0}

In \refta{tableHiggs} we report predictions 
for fiducial  cross sections at different perturbative orders. 
The corresponding 
acceptance efficiencies, computed as in \refse{sec:results-ww}, are presented in \refta{tablehacc}. 
It turns out that Higgs cuts 
suppress the impact of QCD radiative effects in a similar way as 
\ww{} cuts. 
At $8\,(13)$\,TeV the \nlo{} and \nnlo{} corrections amount to $+5\%\,(+3\%)$
and to $+9\%\,(+13\%)$, respectively. The latter consist of a positive $+3\%$ shift due to NNLO PDFs,
a sizeable loop-induced $gg$ component of $+9\%\,(+13\%)$, 
and a rather small genuine $\order{\as^2}$ contribution of
$-2\%\,(-4\%)$.

We compare the \fs{4} predictions against the top-subtracted calculation in the \fs{5}:
At \mbox{$\sqrt{s}=8\,(13)$\,TeV} the latter yields 
$\sigma_{\rm NLO}=48.7\,(3)$\,fb ($\sigma_{\rm NLO}=73.4\,(2)$\,fb) 
and $\sigma_{\rm NNLO}=53.0\,(5)$\,fb ($\sigma_{\rm NNLO}=83.1\,(5)$\,fb), which corresponds to 
a $1\%-2\%$ agreement with the \fs{4} results. The size of the subtracted top 
contamination in the \fs{5} is slightly smaller than what was found for \ww{} cuts. It amounts to 
$5\%\,(9\%)$ at \nlo{} and $6\%\,(11\%)$ at \nnlo{}.

Similarly to the case of  \ww{} cuts, genuine 
${\cal O}(\as^2)$ corrections have a significant impact 
on the acceptance efficiency: At $\sqrt{s}=8\,(13)$\,TeV
the \nnlo{} prediction lies roughly $8\%\,(10\%)$ below the 
\nloplusgg{} result, which
exceeds the respective scale uncertainties. While the relative size of 
higher-order effects on the Higgs-cut efficiency 
is almost 
identical to the one found for \ww{} selection cuts, the absolute size 
of the acceptance efficiencies is much smaller. In the case of  Higgs cuts
it is almost a factor of three lower, primarily due to
the stringent cut on the invariant mass of the dilepton system.

\begin{figure}[tp]
\begin{center}
\begin{tabular}{cc}
\hspace*{-0.17cm}
\includegraphics[trim = 7mm -7mm 0mm 0mm, width=.33\textheight]{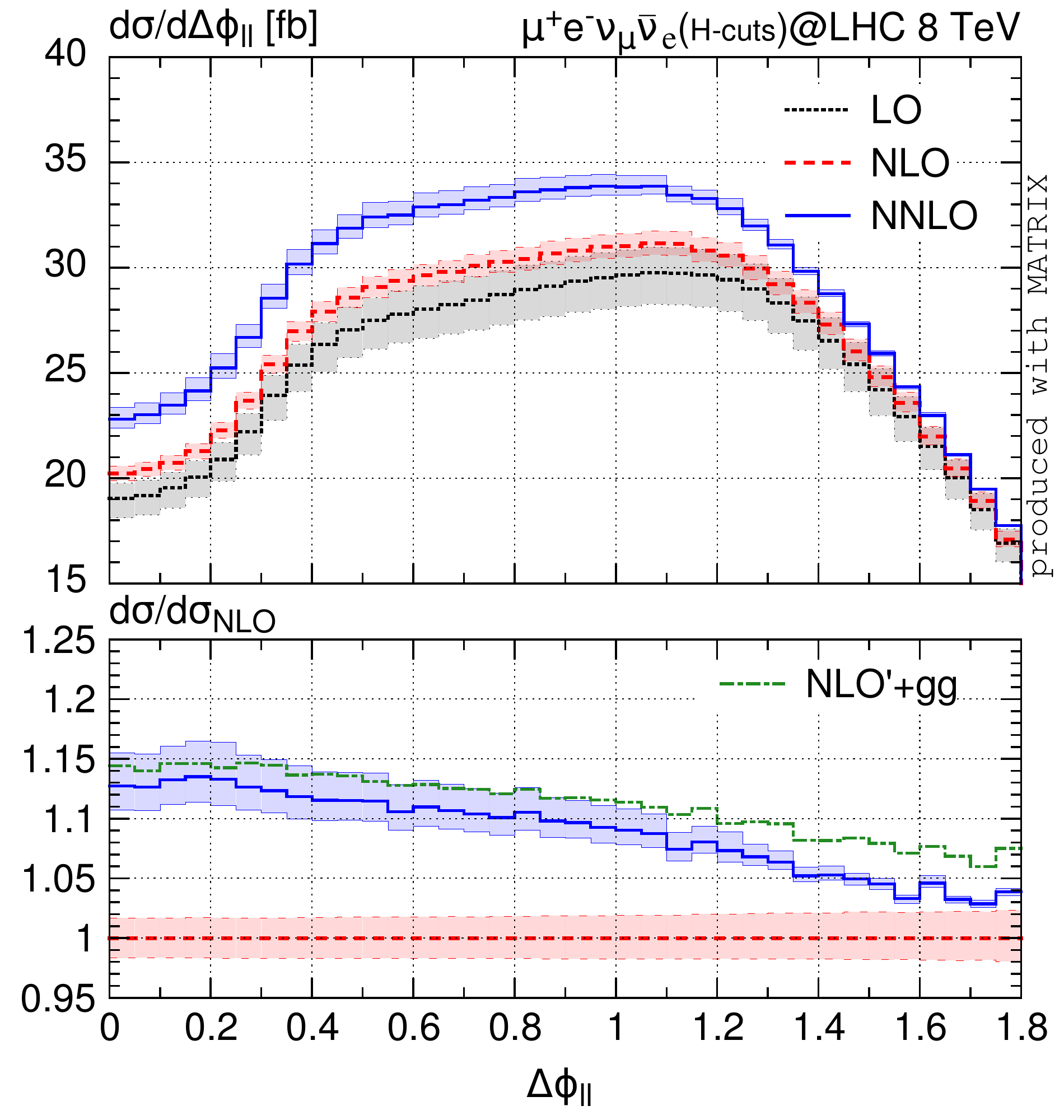} &
\includegraphics[trim = 7mm -7mm 0mm 0mm, width=.33\textheight]{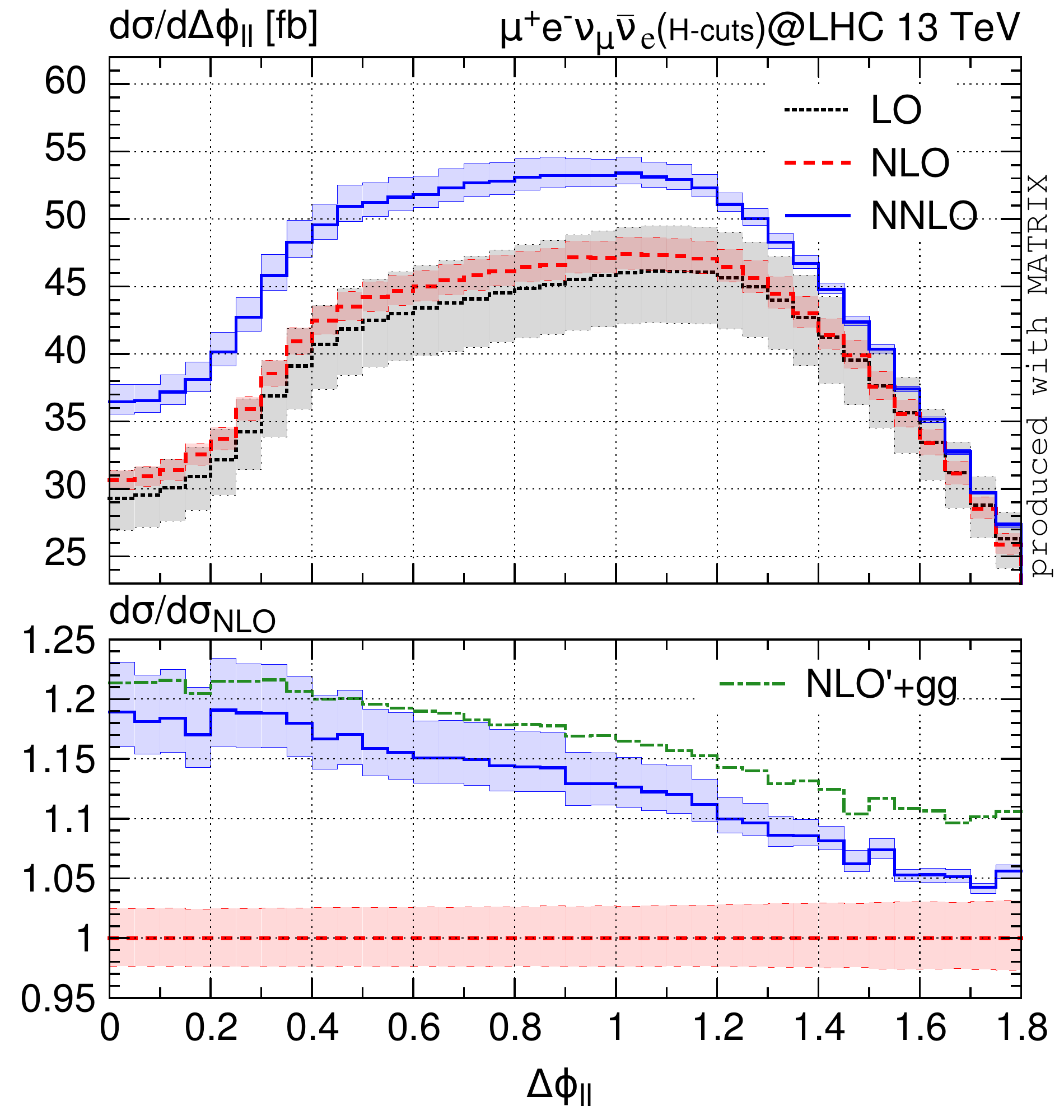} \\[-1em]
\hspace{0.6em} (a) & \hspace{1em}(b)
\end{tabular}
\caption[]{\label{fig:hdphill}{
Distribution in the azimuthal separation of the charged leptons. Higgs cuts are applied.  
Absolute predictions and relative corrections as in~\reffi{fig:mWWinclusive}.}}
\end{center}
\vspace{0.5cm}
\begin{center}
\begin{tabular}{cc}
\hspace*{-0.17cm}
\includegraphics[trim = 7mm -7mm 0mm 0mm, width=.33\textheight]{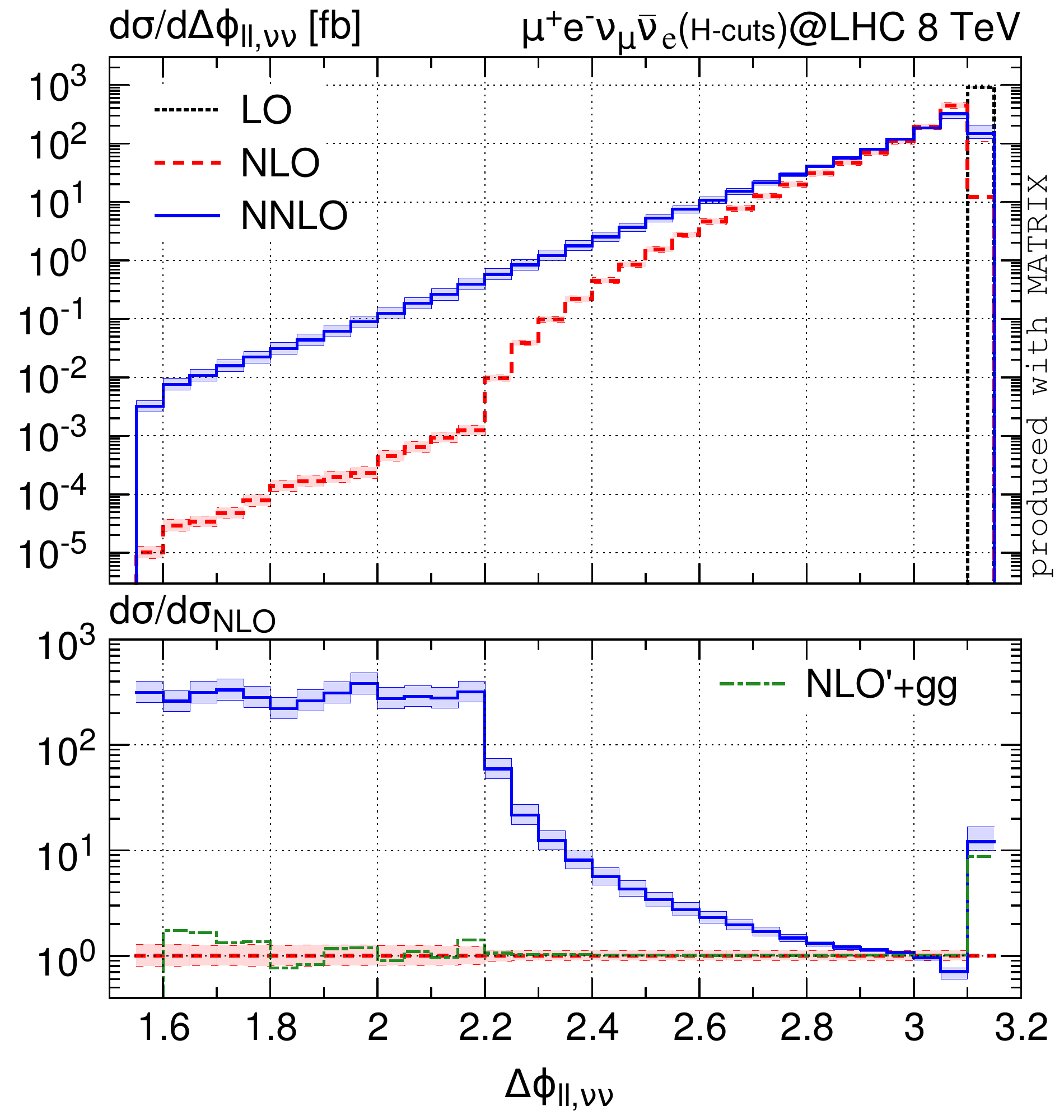} &
\includegraphics[trim = 7mm -7mm 0mm 0mm, width=.33\textheight]{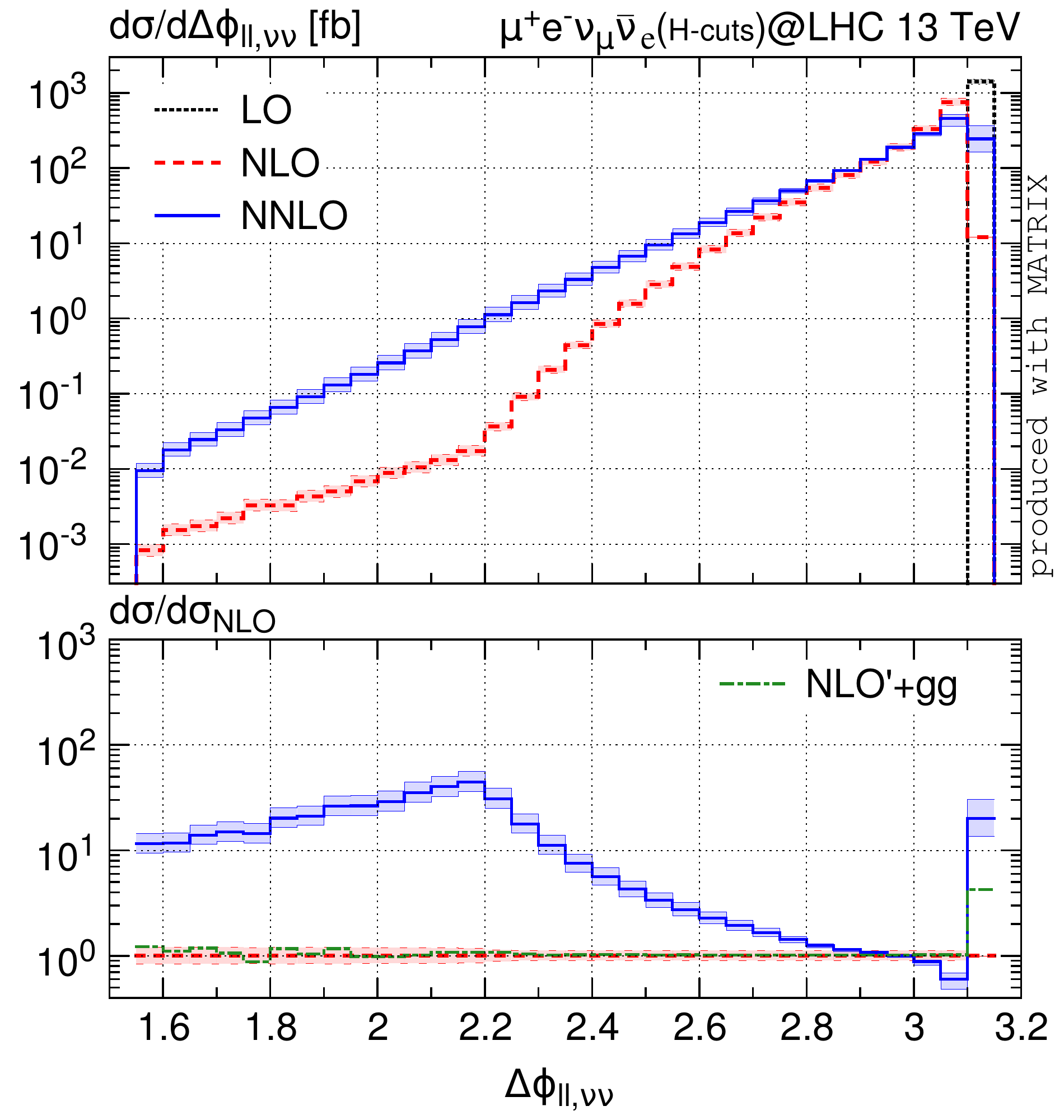} \\[-1em]
\hspace{0.6em} (a) & \hspace{1em}(b)
\end{tabular}
\caption[]{\label{fig:hdphillnunu}{Distribution in the
azimuthal separation between the
transverse momentum of the dilepton system and 
the missing transverse momentum. Higgs cuts are applied.  
Absolute predictions and relative corrections as in~\reffi{fig:mWWinclusive}.}}
\end{center}
\end{figure}

Differential distributions with Higgs cuts applied are presented
in~\reffis{fig:hdphill}{fig:hptmiss}. In general, they behave in a similar way as for the
case of \ww{} cuts discussed in \refse{sec:results-ww}. However, a few observables are quite sensitive 
to the additional cuts that are applied in the Higgs analysis.
Most notably, the distribution in the azimuthal separation of the charged leptons 
in~\reffi{fig:hdphill} exhibits a completely different shape
as compared to \reffi{fig:wwdphill}. In particular, it features an approximate 
plateau in the region $0.4\le\dphill\le 1.2$. The \nnlo{} corrections with respect to the \nlo{}
distribution at $\sqrt{s}=8\,(13)$\,TeV range from about $+13\%\,(+18\%)$
at small $\dphill$ to roughly $+2\%\,(+5\%)$ at separations close to 
the fiducial cut. The loop-induced $gg$ component provides a good approximation of the complete NNLO result for
small separations, but in the large $\dphill$ region it overshoots the complete NNLO result 
by about $5\%\,(7\%)$.

In the $\dphillnunu$ distribution, displayed in \reffi{fig:hdphillnunu}, we
observe that, similarly to the case of \ww{} cuts (see \reffi{fig:wwdphillnunu}), also
Higgs cuts lead to huge NNLO corrections at small
$\dphillnunu{}$. As discussed in \refse{sec:results-ww},
this behaviour is due to the fact that at small $\dphillnunu{}$
the leptonic and \ptmiss{} cuts require 
the presence of a sizeable QCD recoil, 
which is, however, strongly suppressed 
by the jet veto at NLO.
In the Higgs analysis, this suppression mechanism becomes even more powerful  
due to the additional
cut $\ptll>30$\,GeV, which forbids the two leptons to recoil against 
each other. This leads to the kink at
$\dphillnunu=2.2$ in the \nlo{} distribution and to the explosion of 
NNLO corrections below and slightly above this threshold.

\begin{figure}[tp]
\begin{center}
\begin{tabular}{cc}
\hspace*{-0.17cm}
\includegraphics[trim = 7mm -7mm 0mm 0mm, width=.33\textheight]{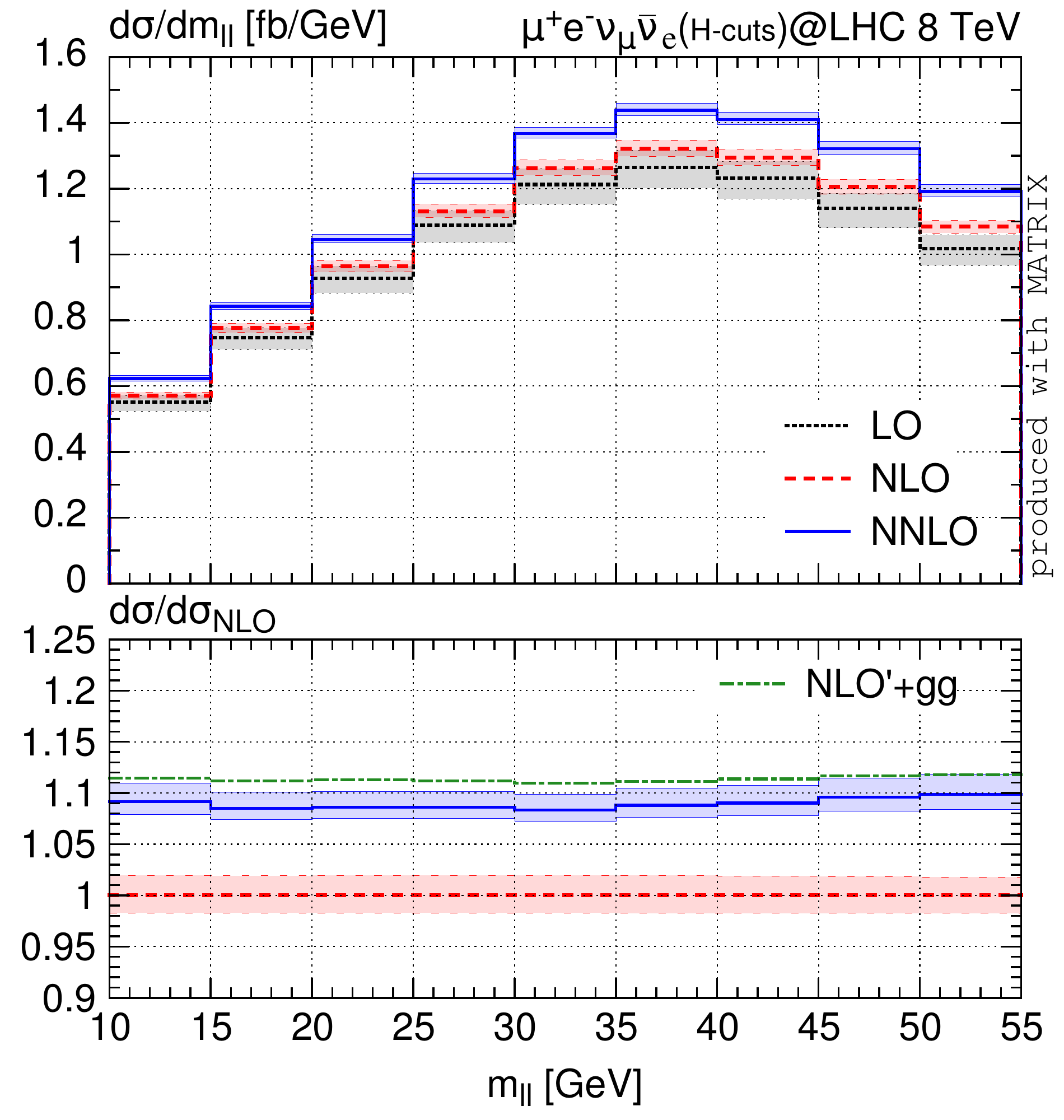} &
\includegraphics[trim = 7mm -7mm 0mm 0mm, width=.33\textheight]{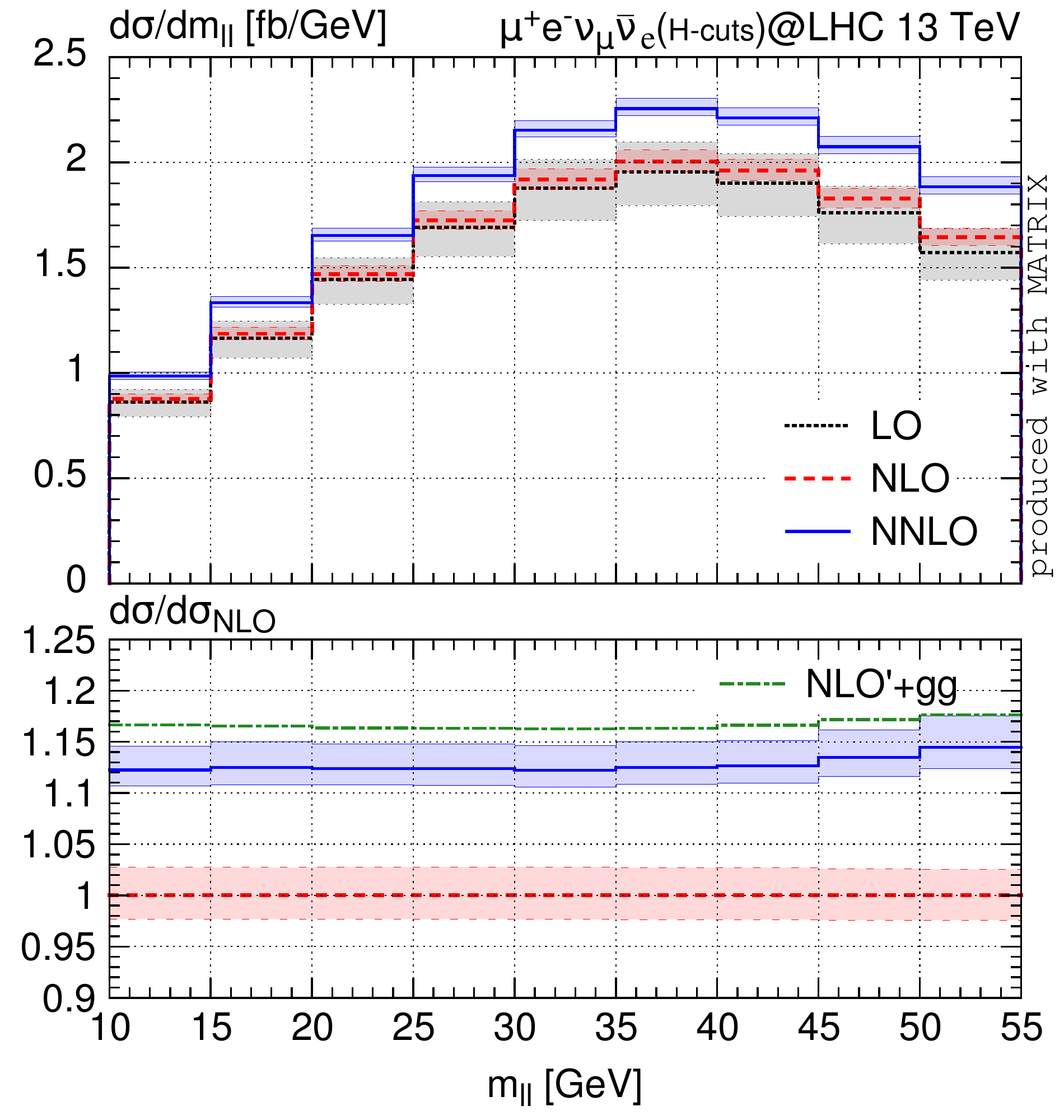} \\[-1em]
\hspace{0.6em} (a) & \hspace{1em}(b)
\end{tabular}
\caption[]{\label{fig:hmll}{Distribution in the
dilepton invariant mass. Higgs cuts are applied.  
Absolute predictions and relative corrections as in~\reffi{fig:mWWinclusive}.}}
\end{center}
\end{figure}

\begin{figure}[h]
\begin{center}
\begin{tabular}{cc}
\hspace*{-0.17cm}
\includegraphics[trim = 7mm -7mm 0mm 0mm, width=.33\textheight]{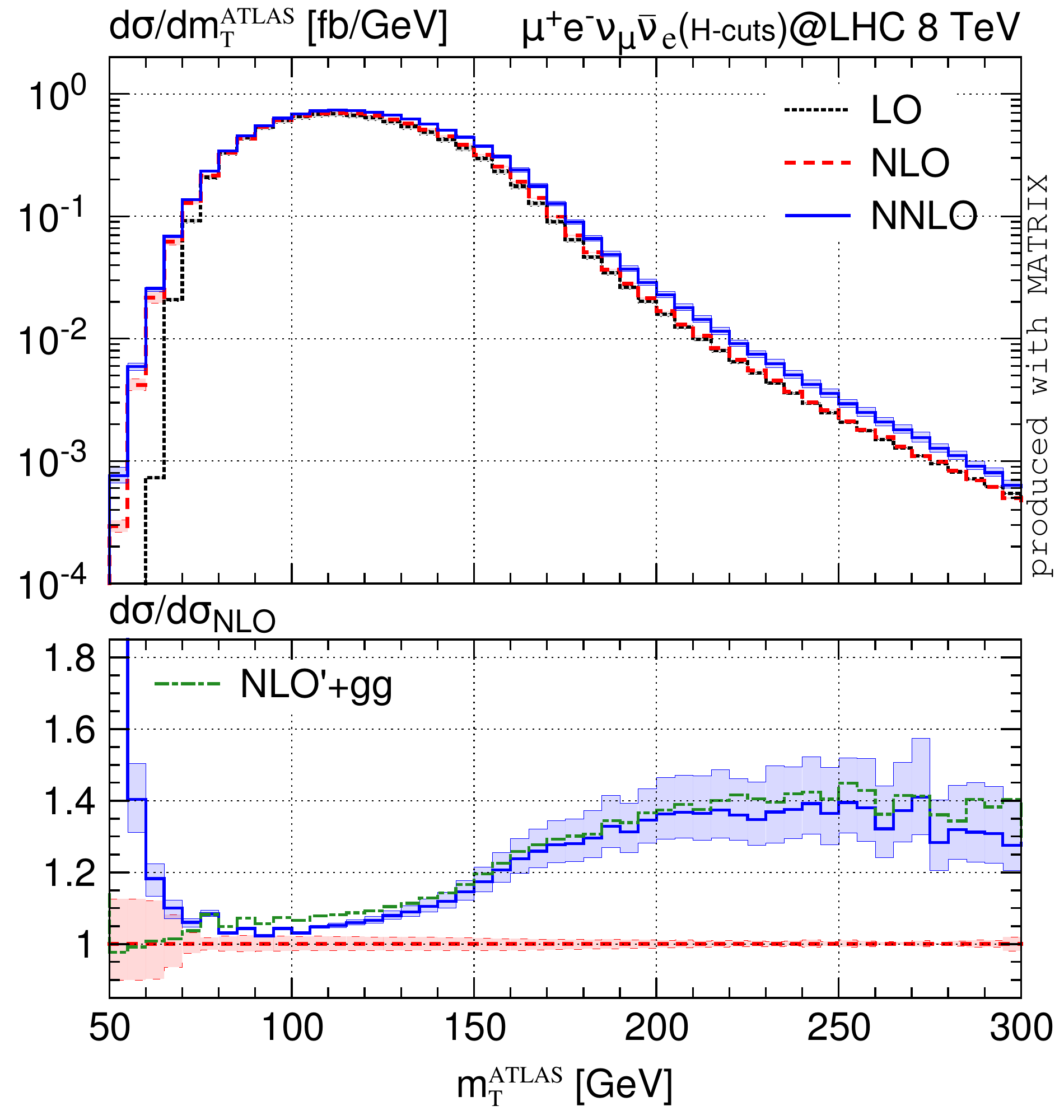} &
\includegraphics[trim = 7mm -7mm 0mm 0mm, width=.33\textheight]{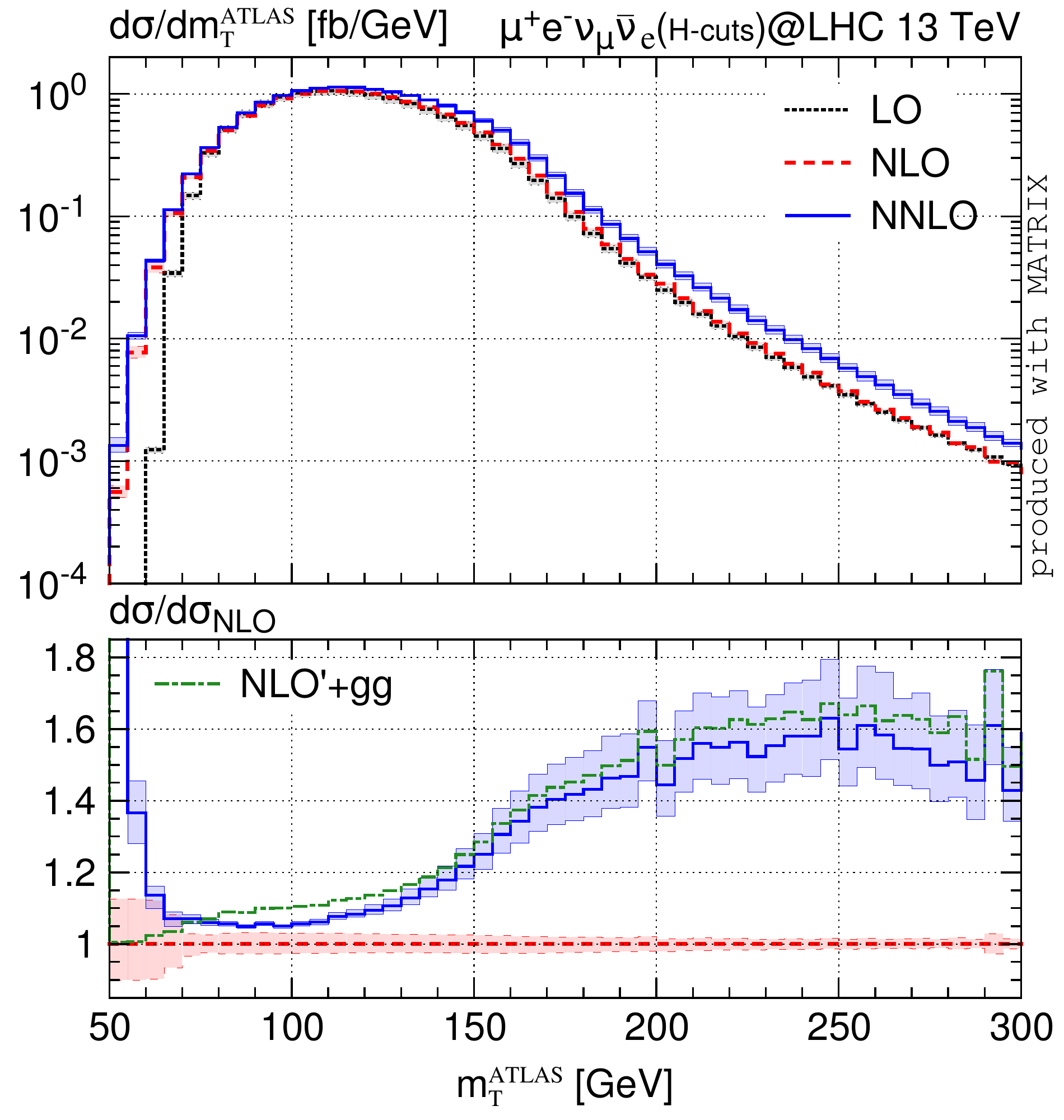} \\[-1em]
\hspace{0.6em} (a) & \hspace{1em}(b)
\end{tabular}
\caption[]{\label{fig:hmt}{Distribution in the \ww{} transverse mass. Higgs cuts are applied.  
Absolute predictions and relative corrections as in~\reffi{fig:mWWinclusive}.}}
\end{center}
\end{figure}

The invariant mass of the dilepton system, shown in~\reffi{fig:hmll},
is restricted to the region $10$\,GeV$\le \mll\le 55$\,GeV. The peak of 
the distribution is around $\mll=38$\,GeV, and the 
$\sigma_\nnlo/\sigma_\nlo$  $K$-factor 
is essentially flat. Also the \nloplusgg{} curve has a very similar shape
so that the radiative corrections precisely match those on the fiducial rates.

The distribution in $m_T^{\rm ATLAS}$ is presented in \reffi{fig:hmt}.
As compared to the \ww{} analysis (see \reffi{fig:wwmt}), 
we observe that the tail of the distribution drops significantly 
faster when Higgs cuts are applied. Moreover, in the high-$m_T^{\rm ATLAS}$ region the size of 
the loop-induced $gg$ corrections relative to \nlo{} and, hence, the size of 
the full \nnlo{} correction, is much larger than in the 
\ww{} analysis.
The NNLO corrections amount up to about $40\%\,(60\%)$
of the \nlo{} cross section at $\sqrt{s}=8\,(13)$\,TeV, while they hardly exceed $15\%$ when
\ww{} cuts are applied.

\begin{figure}[tp]
\begin{center}
\begin{tabular}{cc}
\hspace*{-0.17cm}
\includegraphics[trim = 7mm -7mm 0mm 0mm, width=.33\textheight]{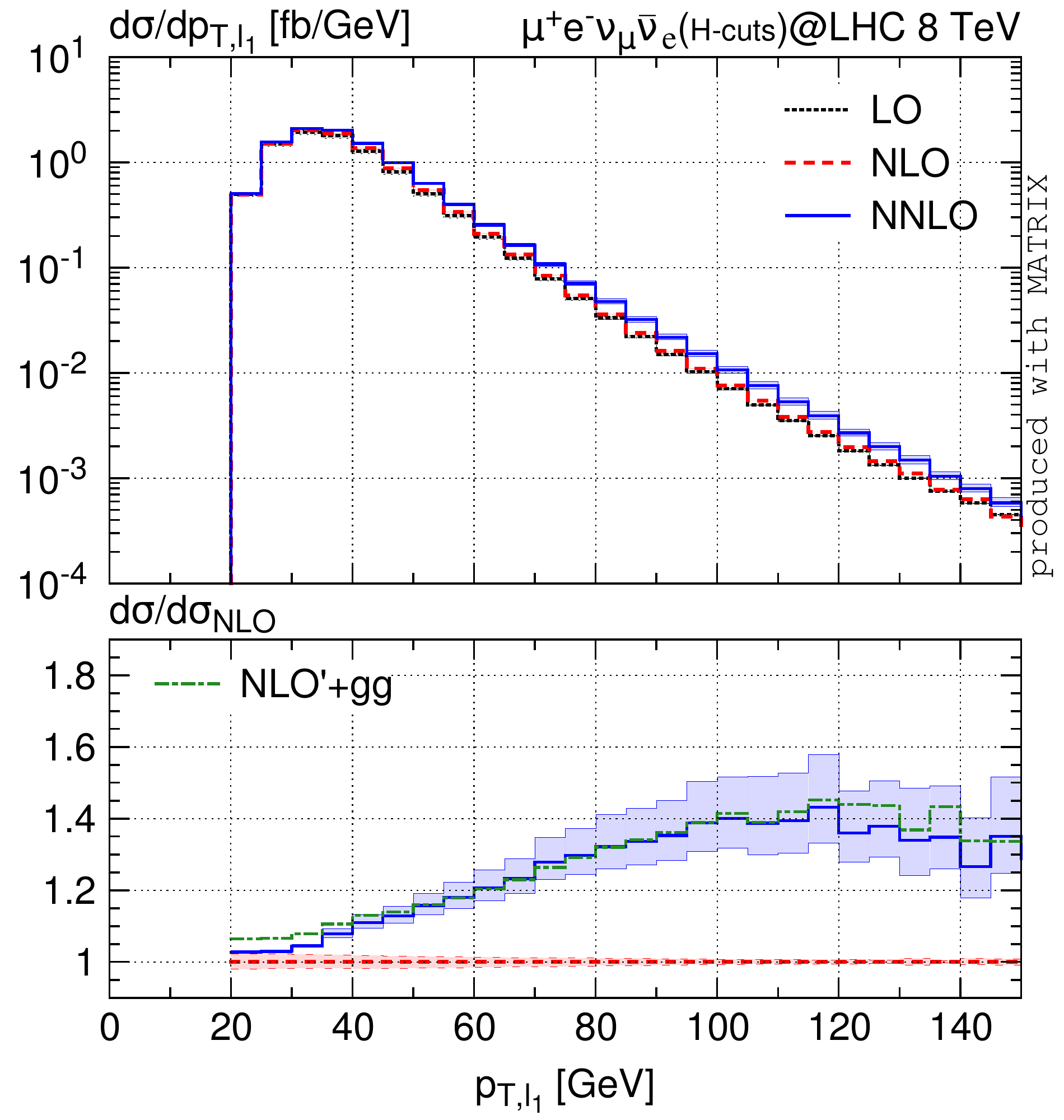} &
\includegraphics[trim = 7mm -7mm 0mm 0mm, width=.33\textheight]{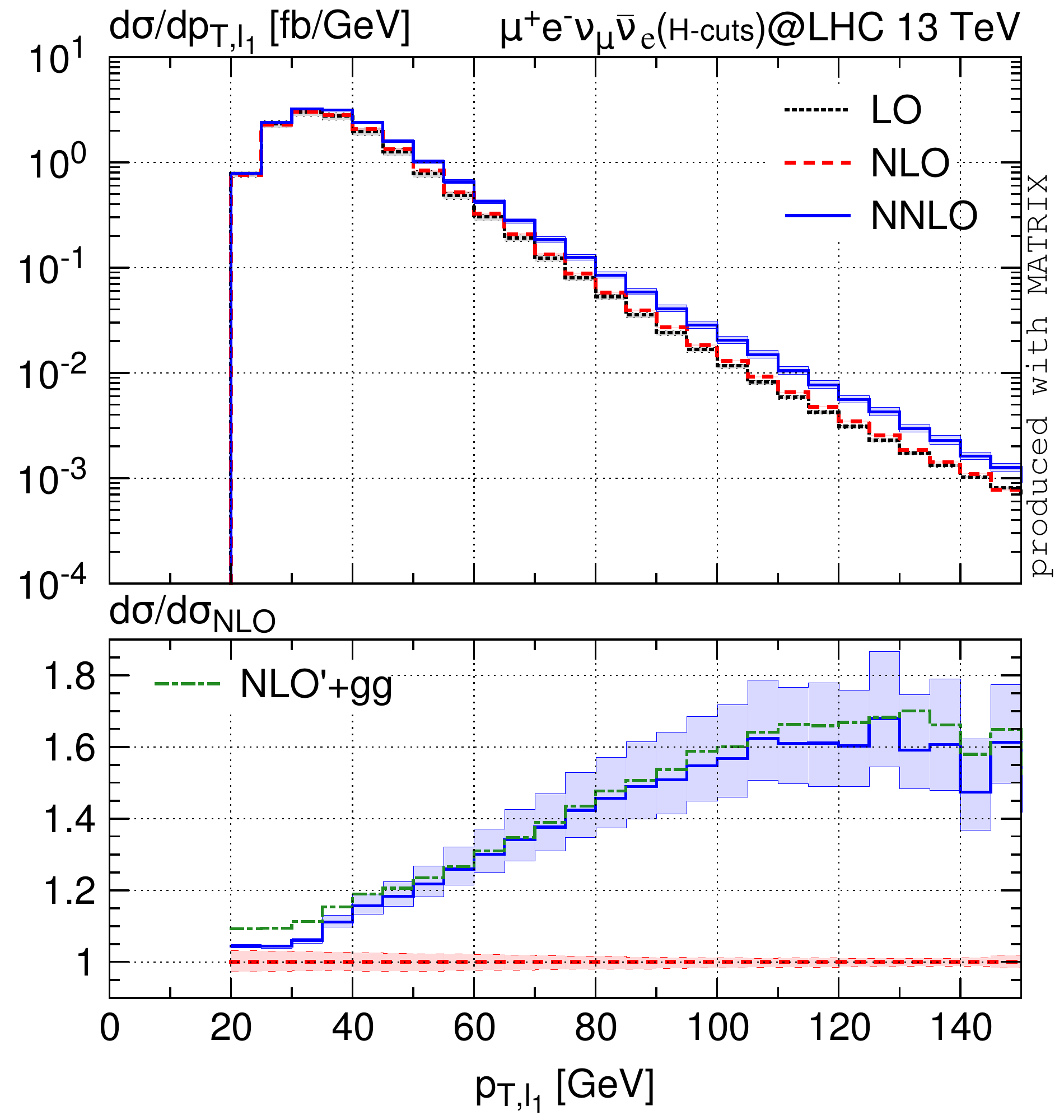} \\[-1em]
\hspace{0.6em} (a) & \hspace{1em}(b)
\end{tabular}
\caption[]{\label{fig:hptl1}{
Distribution in the 
$p_T$ of the leading lepton. Higgs cuts are applied.  
Absolute predictions and relative corrections as in~\reffi{fig:mWWinclusive}.}}
\end{center}
\vspace{1cm}
\begin{center}
\begin{tabular}{cc}
\hspace*{-0.17cm}
\includegraphics[trim = 7mm -7mm 0mm 0mm, width=.33\textheight]{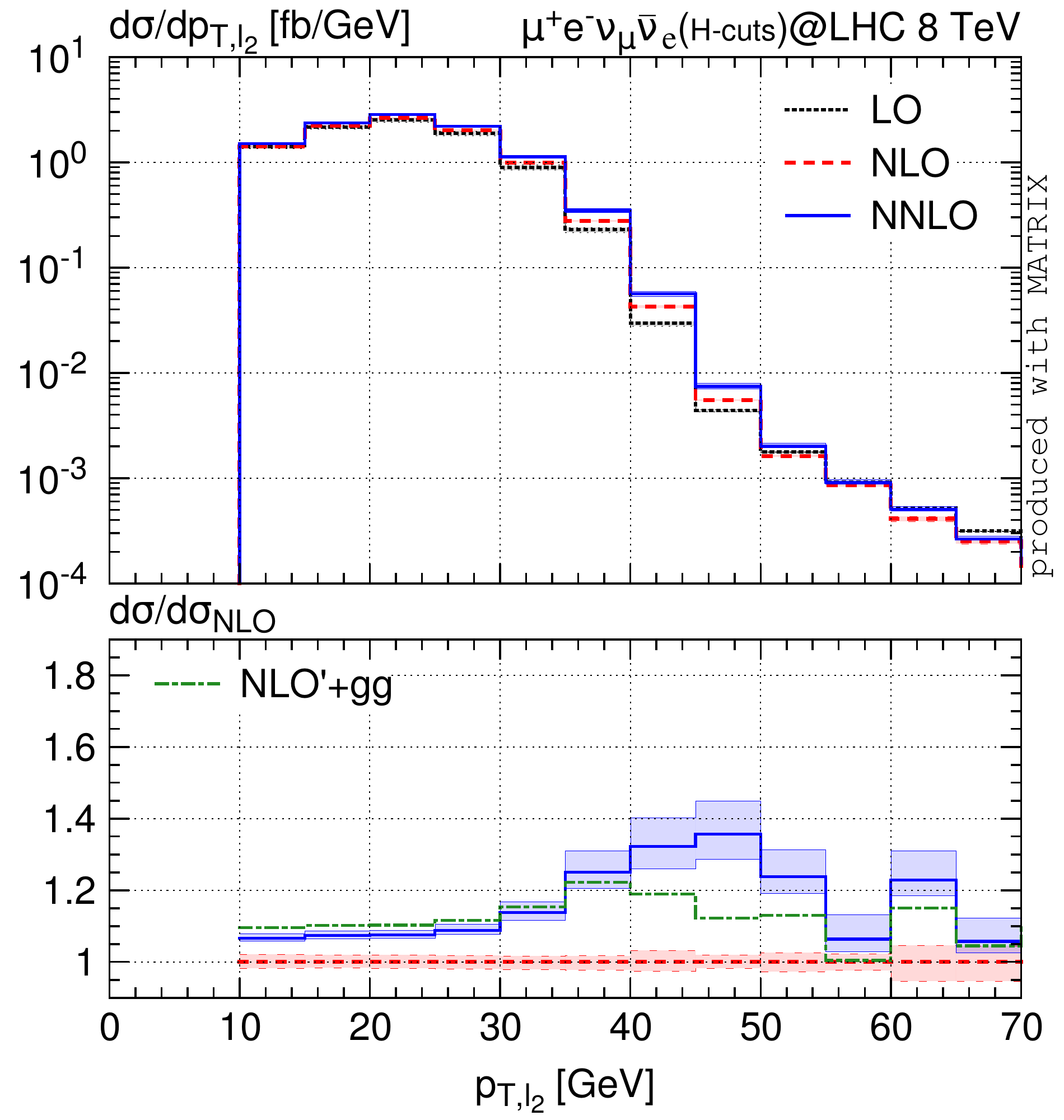} &
\includegraphics[trim = 7mm -7mm 0mm 0mm, width=.33\textheight]{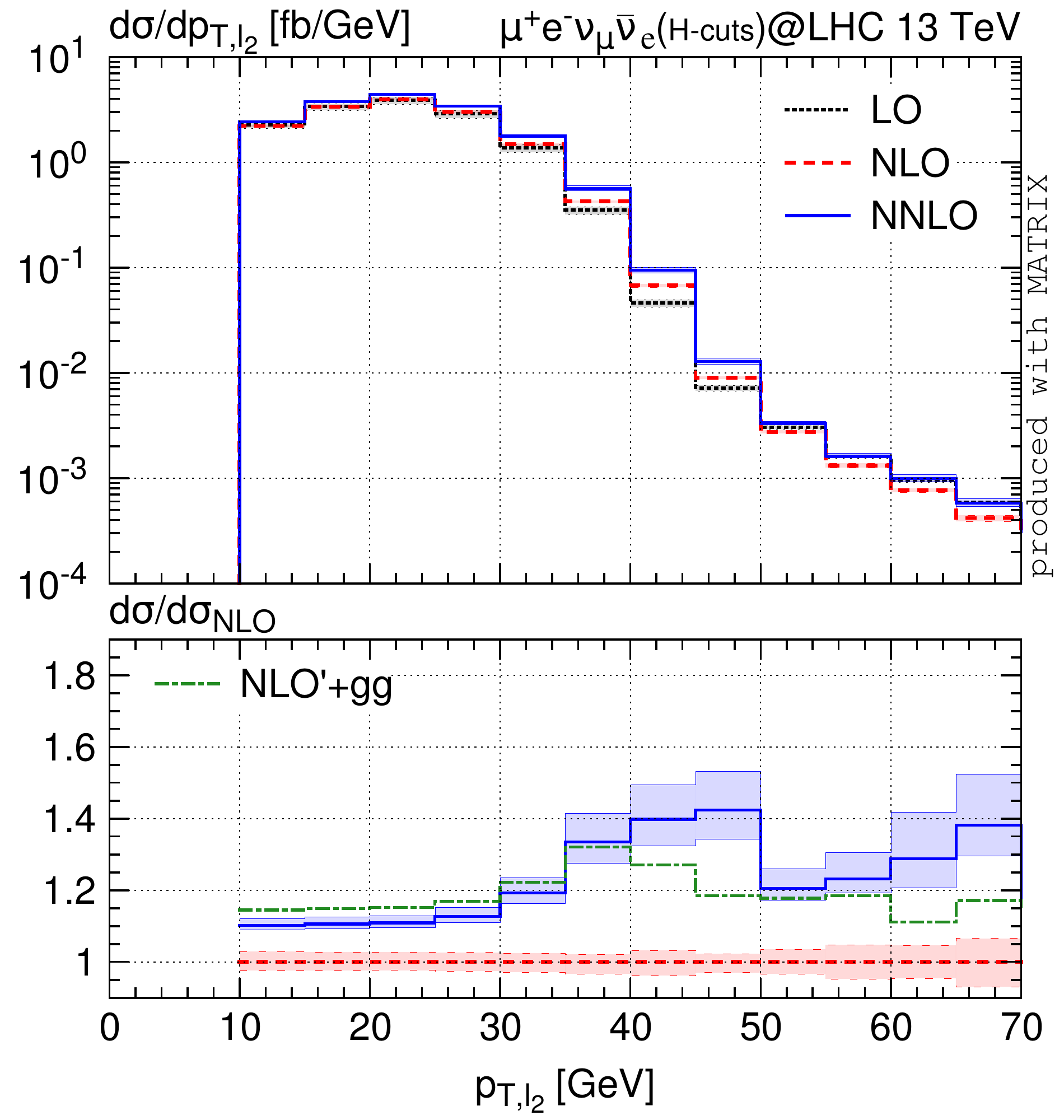} \\[-1em]
\hspace{0.6em} (a) & \hspace{1em}(b)
\end{tabular}
\caption[]{\label{fig:hptl2}{
Distribution in the 
$p_T$ of the subleading lepton. Higgs cuts are applied.  
Absolute predictions and relative corrections as in~\reffi{fig:mWWinclusive}.}}
\end{center}
\end{figure}

The distributions in the lepton $\pt$'s, depicted in
\reffitwo{fig:hptl1}{fig:hptl2}, behave in a similar way as
in~\reffitwo{fig:wwptl1}{fig:wwptl2}, apart from a steeper drop-off in the
tail and slightly larger corrections.

\begin{figure}[tp]
\begin{center}
\begin{tabular}{cc}
\hspace*{-0.17cm}
\includegraphics[trim = 7mm -7mm 0mm 0mm, width=.33\textheight]{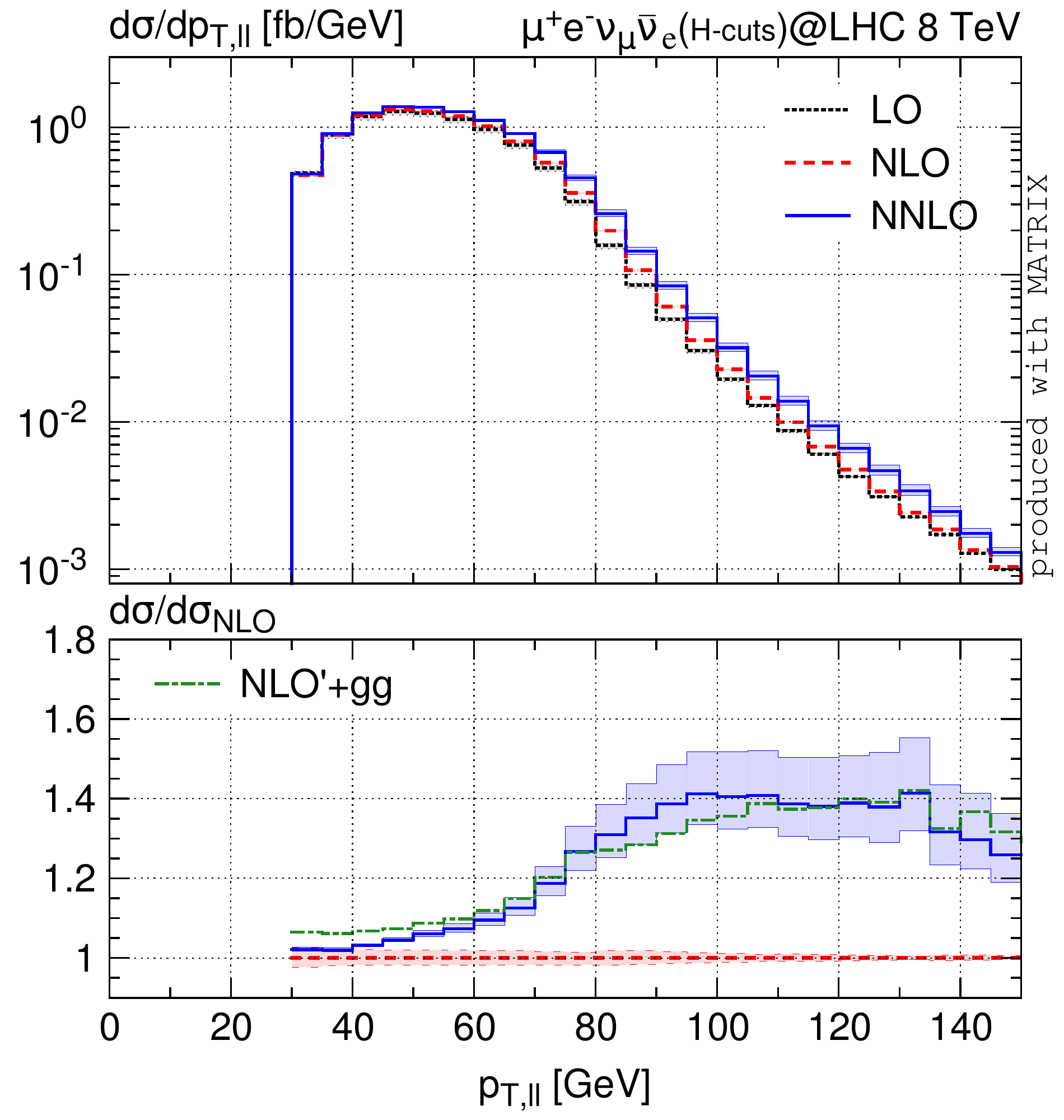} &
\includegraphics[trim = 7mm -7mm 0mm 0mm, width=.33\textheight]{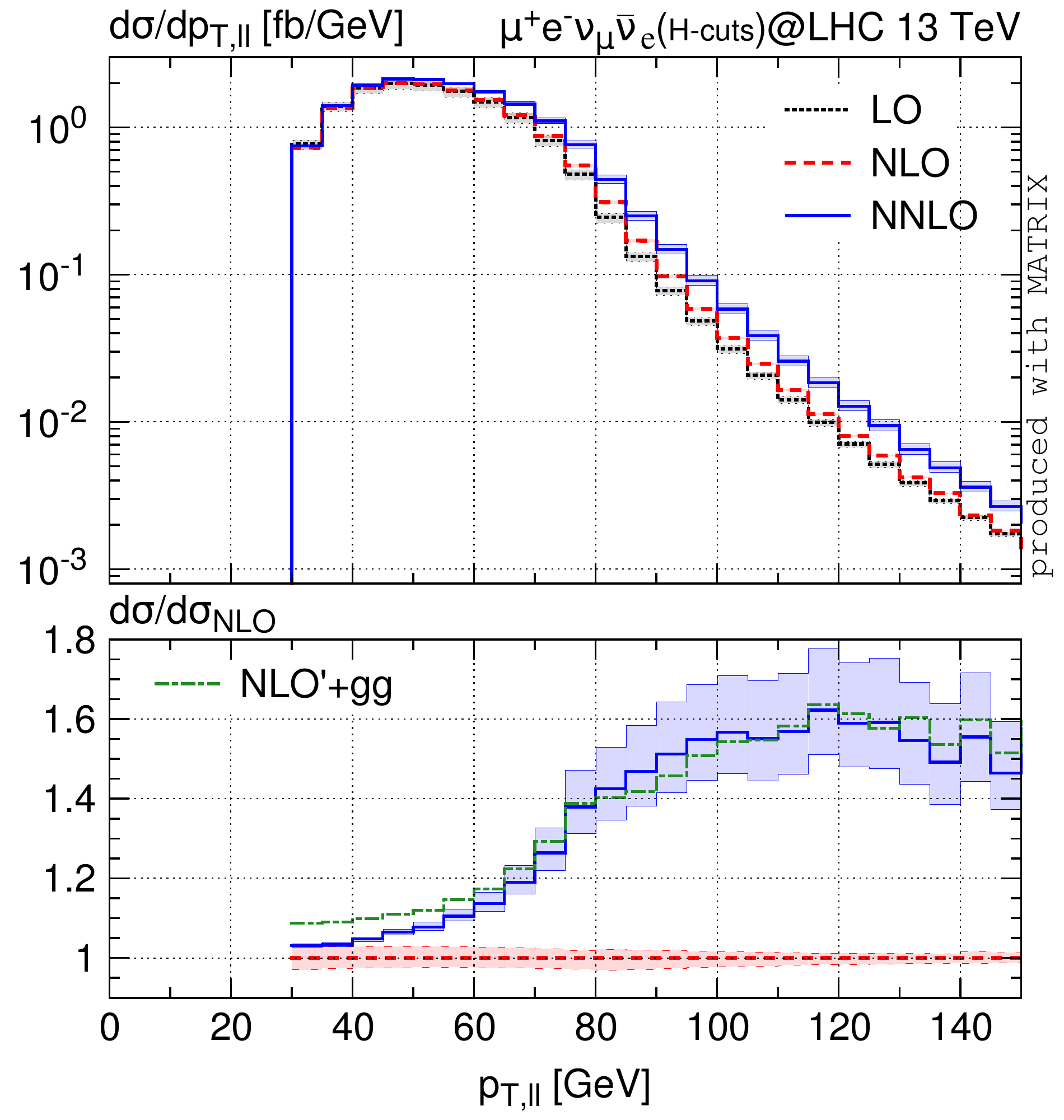} \\[-1em]
\hspace{0.6em} (a) & \hspace{1em}(b)
\end{tabular}
\caption[]{\label{fig:hptll}{Distribution in the 
$p_T$ of the dilepton system. Higgs cuts are applied.  
Absolute predictions and relative corrections as in~\reffi{fig:mWWinclusive}.}}
\end{center}
\vspace{1cm}
\begin{center}
\begin{tabular}{cc}
\hspace*{-0.17cm}
\includegraphics[trim = 7mm -7mm 0mm 0mm, width=.33\textheight]{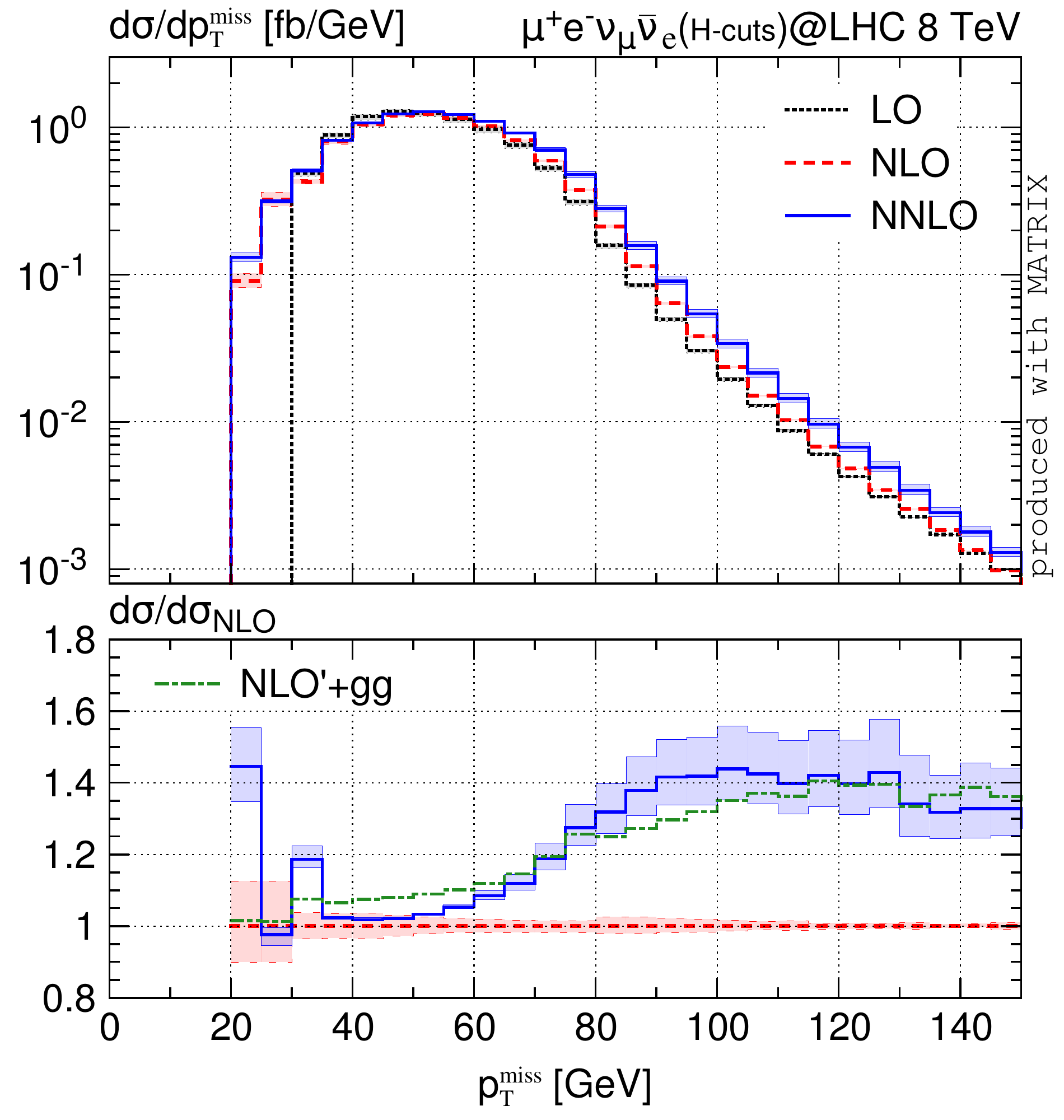} &
\includegraphics[trim = 7mm -7mm 0mm 0mm, width=.33\textheight]{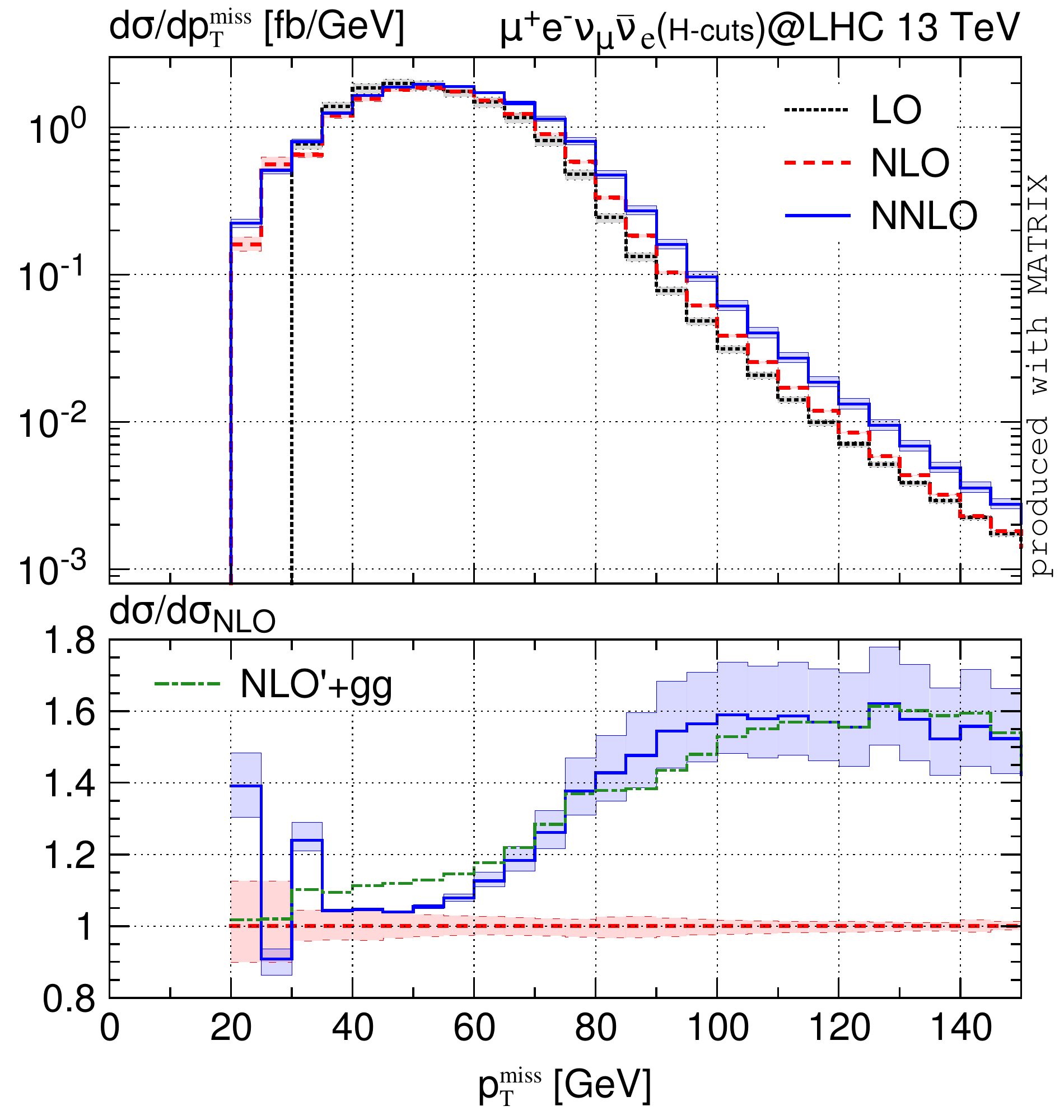} \\[-1em]
\hspace{0.6em} (a) & \hspace{1em}(b)
\end{tabular}
\caption[]{\label{fig:hptmiss}{
Distribution in the missing transverse momentum. Higgs cuts are applied.  
Absolute predictions and relative corrections as in~\reffi{fig:mWWinclusive}.}}
\end{center}
\end{figure}

For the distributions in the $p_T$ of the dilepton pair (see \reffi{fig:hptll})
and in $\ptmiss$ (see \reffi{fig:hptmiss}), we also find a
similar behaviour as in the case where \ww{} cuts are applied.
We note, however,
that the perturbative instability observed in the $\ptll$ distribution with
\ww{} cuts (see \reffi{fig:wwptll}) is removed by the 
explicit cut $\ptll>30$\,GeV in the Higgs analysis.
Accordingly, the \ptll{} cut implicitly vetoes events with $\ptmiss<30$\,GeV at Born level, which leads
to a perturbative instability in the \ptmiss{} distribution, particularly visible in the $\sigma_\nnlo/\sigma_\nlo$ ratio.
In fact, it is evident from~\reffi{fig:hptmiss}
that the phase-space region $\ptmiss<30$\,GeV 
is filled only upon inclusion of higher-order corrections.
Similarly to the case of \ww{} cuts, the behaviour of radiative effects 
is rather insensitive to the collider energy. 
Comparing \nloplusgg{} and full \nnlo{} predictions, in spite of the fairly good agreement
at the level of fiducial cross section, we observe
that the  genuine ${\cal O}(\as^2)$ corrections lead to significant shape distortions at the 10\% level.

\clearpage
\section{Summary}
\label{sec:summary}

We have presented the first fully differential calculation of the NNLO QCD
corrections to \ww{} production with decays at the LHC. Off-shell
effects and spin correlations, as well as all possible topologies
that lead to a final state with two
charged leptons and two neutrinos are consistently taken into account in the
complex-mass scheme. 
At higher orders in QCD perturbation theory, the inclusive \ww{} cross section is plagued by a
huge contamination from top-quark production processes, and the subtraction of 
top contributions is mandatory for a perturbatively stable
definition of the \ww{} rate.  In our calculation, any top contamination is 
avoided by 
excluding partonic channels with final-state bottom quarks in the \fs{4}, 
where the bottom-quark mass renders such contributions separately finite.
In order to quantify the sensitivity of the top-free \ww{} cross section on
the details of the top-subtraction prescription, our default predictions 
in the \fs{4} have been compared to an alternative calculation in the
\fs{5}.  In the latter case a numerical extrapolation in the narrow
top-width limit is used to separate contributions that involve top
resonances from genuine \ww{} production and its interference with $tW$ and
$t\bar t$ production.
The comparison of \fs{4} and  \fs{5} predictions for inclusive and fiducial
cross sections indicates that the dependence on the top-subtraction
prescription is at the $1\%-2\%$ level.

Numerical predictions
at $\sqrt{s}=8$ and $13$\,TeV have been discussed in detail for the
different-flavour channel $pp\to \muenn+X$.
As compared to the case of on-shell \ww{} production~\cite{Gehrmann:2014fva},
the inclusion of leptonic decays leads to a reduction of the total cross
section that corresponds to the effect of leptonic branching ratios plus an
additional correction of about $-2\%$ due to off-shell effects.
The influence of off-shell $W$-boson decays on the 
behaviour of (N)NLO QCD corrections is negligible.  
In fact,
apart from minor differences due to the employed PDFs, we find that the
relative impact of QCD corrections on the total cross sections
is the same as for on-shell \ww{} production~\cite{Gehrmann:2014fva}.
At $\sqrt{s}=8\,(13)$\,TeV,
ignoring the shift of $+2\%\,(+3\%)$ due to the difference
between NNLO and NLO PDFs, the overall NNLO correction is as large as
$+9\%\,(+11\%)$, while the loop-induced gluon--gluon contribution amounts to only $+3\%\,(+4\%)$; 
i.e., contrary to what was generally expected in the literature, the NNLO corrections
are dominated by genuine NNLO contributions to the $q\bar q$ channel, and the
loop-induced $gg$ contribution plays only a subdominant role.

The complete calculation of NNLO QCD corrections allows us to provide a first realistic
estimate of theoretical uncertainties through scale variations:  As is well-known,
uncertainties from missing higher-order contributions obtained through scale variations
are completely unreliable at LO and still largely underestimated at NLO.
This is due to the fact that the
$qg$ (as well as $\bar qg$) and
$gg$ (as well as $qq^{(}\hspace{-0.1em}'\hspace{-0.1em}^{)}$, $\bar q\bar q^{(}\hspace{-0.1em}'\hspace{-0.1em}^{)}$ and $q{\bar q}'$) partonic channels 
do not contribute at LO and NLO, respectively.
In fact, NNLO is the first order at which all partonic
channels contribute.
Thus NNLO scale variations, which are at the level of $2\%-3\%$
for the inclusive cross sections, can be regarded as a reasonable estimate of
the theoretical uncertainty due to the truncation of the perturbative
series.  This is supported by the moderate impact of the recently
computed NLO corrections to the loop-induced $gg$
contribution~\cite{Caola:2015rqy}.

Imposing a jet veto has a strong influence on the size of NNLO
corrections and on the relative importance of NNLO contributions from the 
$q\bar q$ channel and the loop-induced $gg$ channel.  
This was studied in detail for the case of standard fiducial
cuts used in \ww{} and $H\to\ww$ analyses by the LHC experiments.  As a
result of the jet veto, such cuts significantly
suppress all (N)NLO contributions that involve QCD radiation, thereby
enhancing the relative importance of the loop-induced $gg$ channel at NNLO.  
More precisely, depending on the analysis and the collider energy, fiducial cuts
lift the loop-induced $gg$ contribution up to $6\%-13\%$ with respect to \nlo{}, whereas the genuine NNLO corrections
to the $q\bar q$ channel are negative and range between $-1\%$ and 
$-4\%$, while the NLO corrections vary between $+1\%$ and $+5\%$.
The reduction of the impact of radiative corrections is accompanied by a
reduction of scale uncertainties, which, for the NNLO fiducial cross
sections, are at the $1\%-2\%$ level.  This is a typical side-effect of jet
vetoes, and scale uncertainties are likely to underestimate unknown
higher-order effects in this situation.

As a result of the different behaviour of radiative corrections to the
inclusive and fiducial cross sections, their ratios, which determine the
efficiencies of acceptance cuts, turn out to be quite sensitive to
higher-order effects. More explicitly, the overall NNLO corrections 
to the cut efficiency are small and range
between $-3\%$ and $-1\%$.
However, they arise from a positive shift between $+3\%$ and $+9\%$
due to the loop-induced $gg$ channel, and a negative shift
between $-6\%$ and $-10\%$ from genuine NNLO corrections to the $q\bar q$ channel.
The NLO prediction supplemented by the 
loop-induced $gg$ channel, i.e.\ the ``best'' prediction before the complete 
NNLO corrections 
were known, would thus lead to a significant overestimation of the efficiency, by up to about $10\%$.
Similarly to the case of fiducial cross sections, 
the scale uncertainties of cut efficiencies are at the 1\% level, 
and further studies are needed in order to 
estimate unknown higher-order effects in 
a fully realistic way. This, in particular, involves a more accurate 
modelling of the jet veto, which is left for future work.

Our analysis of differential distributions demonstrates that, in absence of
fiducial cuts, genuine NNLO corrections to the $q\bar q$ channel can lead to
significant modifications in the shapes of observables that are sensitive to
QCD radiation, such as the transverse momentum of the leading $W$ boson or
of the \ww{} system.  On the other hand, in presence of fiducial cuts, NLO
predictions supplemented with the loop-induced $gg$ contribution yield a reasonably good
description of the shape of differential observables, such as dilepton
invariant masses and single-lepton transverse momenta. 
We find, however, that even for standard \ww{} and Higgs selection cuts,
which include a jet veto,
genuine NNLO corrections tend to distort such distributions
by up to about $10\%$.
In phase-space regions that imply the presence of QCD radiation,
loop-induced $gg$ 
contributions cannot approximate the shapes of full NNLO corrections.

The predictions presented in this paper have been
obtained with \Matrix{}, a widely automated and flexible framework that 
supports NNLO calculations for all processes of the class
\mbox{$pp\to \elle^+\elle^{\prime\, -}\nu_{\elle}{\bar\nu}_{\elle^\prime}+X$},
including in particular also the channels with equal lepton flavours, $\elle=\elle'$.
More generally, \Matrix{} is able to address fully exclusive NNLO computations for
all diboson {production processes at hadron colliders.

\noindent {\bf Acknowledgements.}
We thank A.~Denner, S.~Dittmaier and L.~Hofer for providing us
with the one-loop tensor-integral library \Collier{} well before publication, and we are grateful 
to P.~Maierh\"ofer and J.~Lindert for advice on technical 
aspects of \OpenLoops.  
This research was supported in part by the Swiss National Science Foundation
(SNF) under contracts 200020-141360, 200021-156585, CRSII2-141847,
BSCGI0-157722 and PP00P2-153027, and by the Kavli Institute for Theoretical
Physics through the National Science Foundation's Grant No. NSF PHY11-25915.

\renewcommand{\em}{}
\bibliographystyle{UTPstyle}
\bibliography{wwnnlo}

\begin{thebibliography}{100}
\bibitem{munich}
{\sc Munich} is the abbreviation of ``MUlti-chaNnel Integrator at Swiss (CH)
  precision''---an automated parton level NLO generator by S.~Kallweit. In
  preparation.
\end{thebibliography}

\end{document}